\documentclass[aps,prx,amsmath,twocolumn,amssymb,titlepage]{revtex4-1}

\usepackage{graphicx}
\usepackage{nicefrac}
\usepackage{amsfonts}

\usepackage{amssymb}
\usepackage{amsmath} 
\usepackage{hhline}
\usepackage{subfig}

\usepackage{multirow} 
\usepackage{tabularx} 
\usepackage{array}

\usepackage{units}

\usepackage{color} 
\usepackage{tensor} 
\usepackage{braket}

\usepackage{bm}
\usepackage{hyperref}

\newcommand{\grafe}[1]{\left\{ #1 \right\}}
\newcommand{\tonde}[1]{\left( #1 \right)}
\newcommand{\quadre}[1]{\left[ #1 \right]}

\makeatletter
\renewcommand\@makecaption[2]{%
  \par
  \vskip\abovecaptionskip
  \begingroup
   \small\rmfamily
    \begingroup
     \samepage
     \flushing
     \let\footnote\@footnotemark@gobble
     \@make@capt@title{#1}{#2}\par
    \endgroup
  \endgroup
  \vskip\belowcaptionskip
}
\makeatother

\begin{document}
\title{Complex energy landscapes in spiked-tensor and simple glassy models:\\ ruggedness, arrangements of local minima and phase transitions }
\author{Valentina Ros}
\affiliation{Institut de physique th\'eorique, Universit\'e Paris Saclay, CEA, CNRS, F-91191 Gif-sur-Yvette, France}
\author{Gerard Ben Arous}
\affiliation{Courant Institute of Mathematical Sciences, New York University, 251
Mercer Street, New York, NY 10012, USA}
\author{Giulio Biroli}
\affiliation{Institut de physique th\'eorique, Universit\'e Paris Saclay, CEA, CNRS, F-91191 Gif-sur-Yvette, France}
\affiliation{Laboratoire de Physique Statistique, Ecole Normale Sup\'erieure,
PSL Research University, 24 rue Lhomond, 75005 Paris, France}
\author{Chiara Cammarota}
\affiliation{King's College London, Department of Mathematics, Strand, London WC2R
2LS, United Kingdom}

\begin{abstract}
\noindent
We study rough high-dimensional landscapes in which an increasingly stronger preference for a given configuration emerges. Such energy landscapes arise in glass physics and inference. In particular we focus on random Gaussian functions, and on the spiked-tensor model and generalizations.  We thoroughly analyze the statistical properties of the corresponding landscapes and  
characterize the associated geometrical phase transitions. In order to perform our study, we develop a framework based on the Kac-Rice method that allows to compute the complexity of the landscape, i.e. 
the logarithm of the {\it typical} number of stationary points and their Hessian. This approach generalizes 
the one used to compute rigorously the annealed complexity of mean-field glass models. We discuss its advantages with respect to previous frameworks, in particular the thermodynamical replica method which is shown to lead to partially incorrect predictions.
\end{abstract}

\maketitle
\section{Introduction}
\noindent
Characterizing rough multi-dimensional energy landscapes is a challenging task that is central in many different fields from physics to computer science, high-dimensional statistics, machine learning and biology. 
In a nutshell this problem consists in analyzing the statistical properties of functions defined on very high dimensional spaces. Relevant information that one wants to obtain is for instance the number of minima at a given energy, and more generally of the critical points, and the spectral properties of their corresponding Hessian. This issue is crucial to understand the {\it dynamics} within these landscapes, in particular gradient descent which has many physical and practical applications. 
Depending on the context, the landscape can correspond to the energy of a physical system, to the loss-function 
of a machine learning algorithm, to the cost function of an optimization problem or to the fitness function of a biological system. \\
Pioneering works on this subject were done in physics, in the context of mean field spin-glasses, starting from the 80s \cite{moore,kurchan,crisantisommers,giardinacavagnaparisi}, see \cite{cavagnapedestrian} for a review. One of the essential results, besides the explicit computations in several models, was the understanding that the statistical properties 
of rough energy landscapes are the ones characteristic of two different physical systems: spin-glasses and glasses \cite{cavagnapedestrian}. The origin of this universality lies in replica theory: the properties of the landscape 
are actually encoded in the type of mean-field solution obtained by the replica method, respectively full replica symmetry breaking and one step replica symmetry breaking~\cite{footnote1}. Remarkably, it was also realized that pure systems can behave as disordered ones, as first found in long-range spin models \cite{bernasconi} ; accordingly energy functions of several complex
systems are qualitatively similar to random functions.\\
In mathematics, in particular in probability theory, there has been a recent and growing research activity aimed at developing rigorous analysis of rough energy landscapes. Starting from the seminal work \cite{fyodorov} the Kac-Rice method has emerged as the mathematical framework suited to do that \cite{braydean,auffingerbenaouscerny,auffingerbenaous,fyodorovnadal,eliran}. It allowed to put on a firmer basis previous results obtained in the physics literature, and it highlighted important relationships with random matrix theory. Moreover, it has been recently exploited to analyze landscape properties of machine learning and inference models \cite{tengyuma,MontanariBenArous}. \\
The recent results and questions concerning the statistical properties of rough landscapes make clear that what found for mean-field glassy systems 
represents only a facet of a much more general challenge.
There are several different directions in which further investigations are timely and interesting.  
One of them is the characterization of landscapes in current problems central in machine learning and high-dimensional 
statistics, such as the analysis of rough energy landscapes
and associated phase transitions when an increasingly stronger preference for a given configuration arises. This problem is central in data science (the signal versus noise problem) \cite{reviewkrzakalazdeborova}, as well as in biology and in physics, in cases where a specific ground state competes with many random ones (e.g. protein folding \cite{remprotein} and random pinning glass transition \cite{randompinning}). Another important and quite distinct research direction consists in studying the number of equilibria in non-conservative dynamical systems that arise in neuroscience \cite{sompolinsky1} and theoretical ecology \cite{bunin}. In this case, forces do not derive from a potential, hence there is no landscape to start with, but nevertheless information 
about the number of equilibria and their stability can be obtained by methods similar to the one used for the conservative 
case \cite{touboul,fyodorovmay}. \\
From the methodological point of view, the main open crucial issue is developing the Kac-Rice method to compute the {\it typical} number of critical points, related to the average of the logarithm of the number of critical points (called quenched entropy). 
Computing the logarithm of the average (called annealed entropy), as done until now, 
is correct in a few cases only \cite{eliran}; in general, the two computations lead to different results even at leading order. 
Physics methods based on replica theory and super-symmetry provided guidance and results in specific cases, 
but as we shall discuss in the following they suffer important limitations \cite{Monasson,Annibale,CLR1, CLR2,Rizzo1,Rizzo2, Aspelmeier0}.\\
Our work has a double valence.  One is conceptual: we present a general analysis of the properties and the phase transitions occurring in rough energy landscapes whenever an increasingly stronger preference for a given configuration arises,
an interesting and timely issue as discussed above. The other is methodological: we develop the sought generalization of the Kac-Rice method to compute the {\it typical} number of critical points and the corresponding quenched entropy, a theoretical framework 
expected to have multiple applications in several fields. Overall, our work opens the way to throughout analysis of the statistical properties of rough landscapes in topical problems relevant in several different fields, from physics to machine learning and biology.  \\
We focus on the $p$-spin spherical model and add to its Hamiltonian a term favoring all configurations that are close to a given one \cite{sherrington1}. This choice is natural from different points of view. 
First, the system without the additional extra term has already proven to be an instrumental paradigm for rough energy landscapes \cite{cavagnapedestrian,crisom92}, so it is a natural starting point to study the effect of a preferred configuration 
on a random landscape. Second, it is directly relevant
for very recent problems studied in the computer science literature; in fact a particular realization of 
it corresponds to the so called spiked-tensor model, which recently attracted a lot of attention \cite{montanari,krzakala,bandera, Chen, MontanariBenArous}. 
The thermodynamics of the system we focus on, that we henceforth call generalized spiked-tensor, has been originally introduced in Ref. \onlinecite{sherrington1} to study the effect of a ferromagnetic coupling on a $p$-spin spherical model. Here we investigate in detail its energy landscape. Depending on the functional form of the additional term, we generically find different scenarios and different types of {\it energy landscape (or geometric) phase transitions}. Although this model is certainly extremely simplified, we think that the lessons that can be learnt from its analysis provide instrumental guidelines and extend to more realistic cases. 
Moreover, because of its relation with the spiked-tensor model, our results are directly relevant to current issues investigated in high-dimensional statistics and inference.\\
As stressed above, one of the main outcome of our work is the construction of a 
general Kac-Rice method which allows one to analyze cases in which the so-called quenched entropy does not coincide with the annealed one, as it happens for the model we consider. Since we use replicas in a rather innocuous way---we remain at the replica symmetric level---transforming it from a theoretical physics technique to a fully rigorous one
 should be within reach in a not too distant future. \\
In the following two sections we present a summary of the main results. In Section \ref{sec:ReplicaAnalysis} we discuss the zero temperature thermodynamics of the model by means of the replica method. In Section \ref{sec:KacRice} we present the new Kac-Rice method for the quenched complexity, and we compare its findings with the ones obtained with the replica method in Section \ref{sec:Monasson}. After reviewing the implications of these findings for the special case of the spiked-tensor model in Section \ref{sec:Spiked}, we present our conclusions in Section \ref{sec:Conclusion}.
 \begin{figure*}[!htbp]\centering
 \includegraphics[width=2\columnwidth]{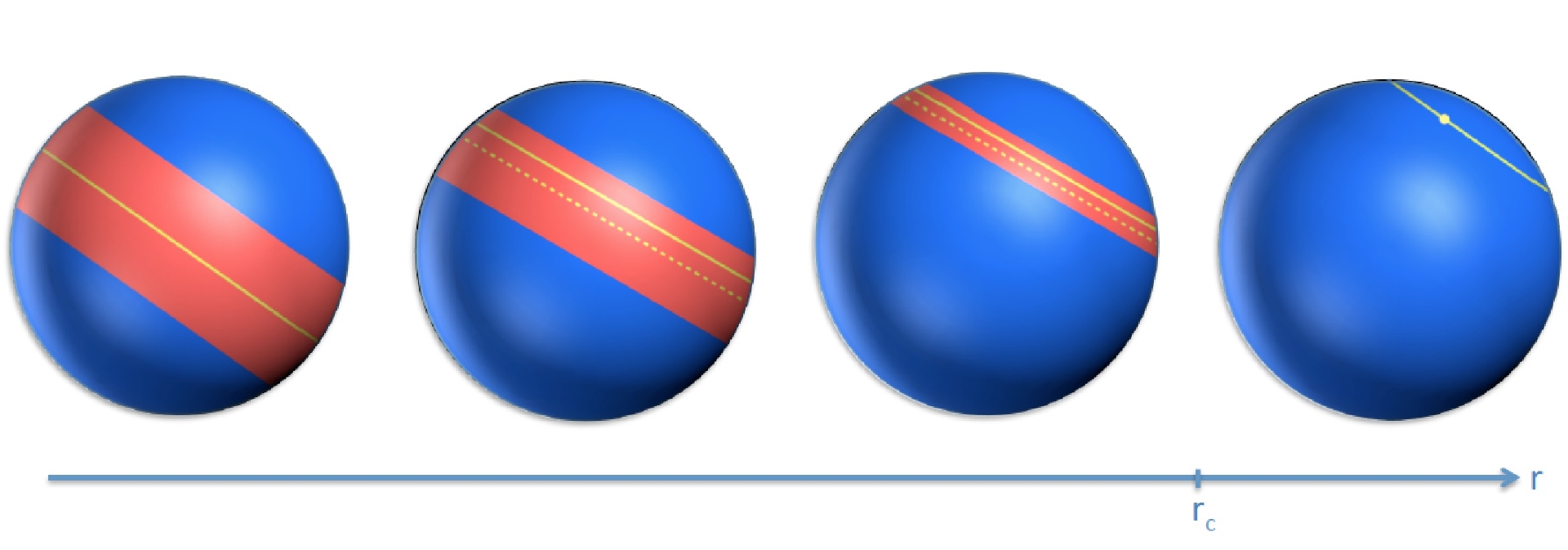}
 \caption{{\bf Evolution of the energy landscape in case I}. In this drawing we illustrate the evolution of the energy landscape due to the increase of $r$ in Case I ($k=1$). The red strip denotes the region on the sphere where minima lie in an exponential number. The continuous yellow line corresponds to the
 parallel where the deepest minima are located. The dashed yellow line corresponds to the
 parallel where the most numerous minima are located. At $r_c$ the energy landscape has a transition: for $r<r_c$ it is rough and full of minima, for $r>r_c$ it is smooth and only contains one minimum (represented by the yellow dot in the figure).}
 \label{Fig-k1}
\end{figure*}
   
\section{Definition of the model}
\noindent
We consider the Hamiltonian or energy functional:
$$
H_{p,k}(r)=-\hspace{-0.5cm}\sum_{\langle i_1,i_2,\dots,i_p\rangle}\hspace{-0.5cm}J_{i_1,i_2,\dots,i_p } s_{i_1} s_{i_2} \dots s_{i_p} -r N f_k\left(\frac{{\bf s}\cdot{\bf v_0}}{N}\right) 
$$
where the first sum is over all distinct $p$-uples and the subindices run from $1$ to $N$. 
The configuration space of the model is the sphere of radius $\sqrt{N}$, i.e. a given configuration is a vector ${\bf s}$ of $N$ components $\{s_1,s_2,\dots,s_N\}$ such that $\sum_i^N s_i^2=N$. The $N$-dimensional vector ${\bf v_0}$ points towards a specific direction, say ${\bf v_0}=\{1,1,\dots,1\}$ without loss of generality (we have imposed on ${\bf v_0}$ the same normalisation condition as ${\bf s}$).
In the following we are going to refer to this preferential direction of the model as the {\it North Pole}.\\
The first term of $H_{p,k}$ is the Hamiltonian of the standard spherical $p$-spin model~\cite{cavagnapedestrian} with random coupling $J_{i_1,i_2,\dots,i_p}$ normally distributed with zero mean and variance $\langle J^2 \rangle=p!/2N^{p-1}$ .\\
The second term represents an energetic gain when the system's configuration ${\bf s}$ is aligned with ${\bf v_0}$.
We generically describe this energetic gain by a function $f_k(x)$ of the scalar product $x=\sum_i^N s_iv_0^i/N$.
\begin{figure*}[!htbp]\centering
\includegraphics[width=2\columnwidth]{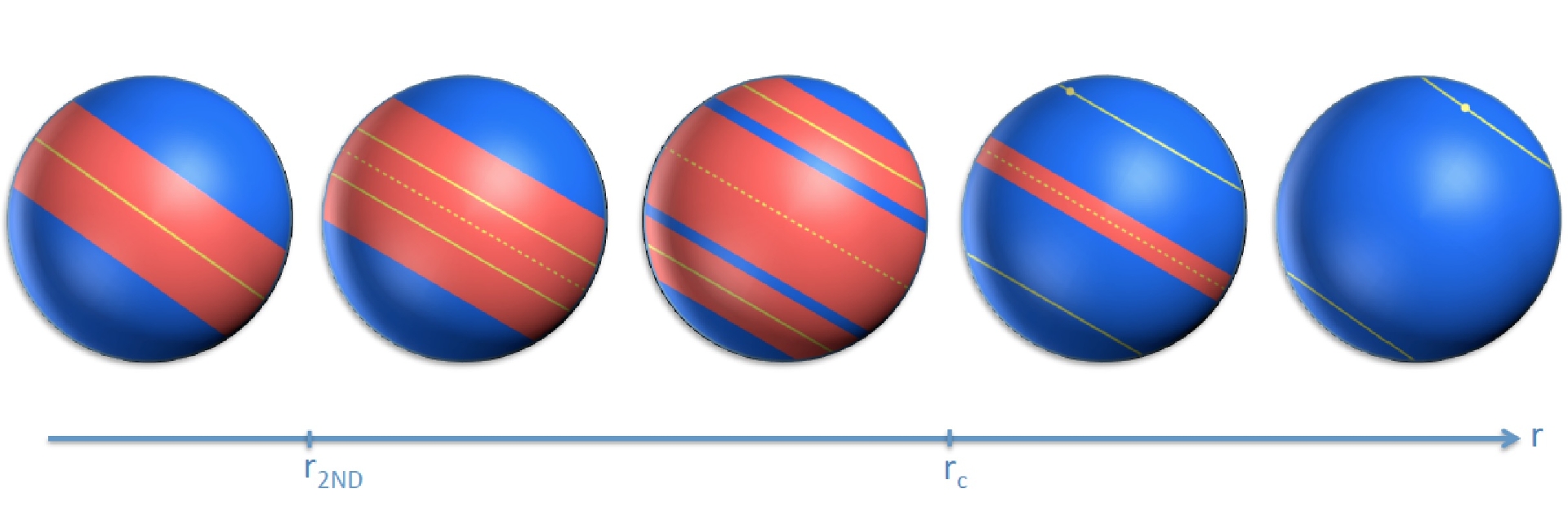}
 \caption{{\bf Evolution of the energy landscape in case II}. In this drawing we illustrate the evolution of the energy landscape due to the increase of $r$ in Case II ($k=2$). The red strip denotes the region on the sphere where minima lie in an exponential number. The continuous (dashed) yellow line corresponds to the
 parallel where the deepest (most numerous) minima are located. The energy landscape has several transitions. At $r_{2\rm{ND}}$ 
 the deepest minima are no longer on the equator and move toward the poles. Afterwards the band containing the exponential number of minima fractures in three parts, one around the equator and two symmetric ones closer to the poles.  At $r_c$ the bands closer to the pole implode and are replaced by two isolated global minima  (the one on the south hemisphere is not visible since it is on the back of the sphere) but the 
 band at the equator persists. Finally, for even larger values of $r$, the landscape becomes completely smooth with only two symmetric minima.}
 \label{Fig-k2}
\end{figure*}
Our aim is to use $f_k(x)$ as a template of a smooth function defined on the $N$ dimensional sphere, with a deep minimum in a specific direction. We found that the main relevant features of $f_k(x)$ are its derivatives in $x=0$: the sub-index $k$ indicates 
what is the first non zero derivative in $x=0$.  We assume that the function $f_k(x)$ reaches its highest value in $x=1$, is zero for $x = 0$ and is monotonously increasing in $[0,1]$. It also has a symmetric or an antisymmetric continuation for $x\le 0$ depending whether $k$ is even or odd respectively.\\
For concreteness, we shall often refer to the case $f_k(x)=x^k/k$, which was first introduced in Ref.~\onlinecite{sherrington1}. When $p=k$, the model corresponds to the so called spiked-tensor model 
which has been the focus of several recent studies in the computer science literature~\cite{montanari,krzakala,bandera,Chen, MontanariBenArous}. In particular, for this limiting case the calculation of the average number of stationary points has been very recently performed in~\cite{MontanariBenArous}.

\section{Summary of results}\label{sec:SummaryResults}
\noindent
The energy function $H_{p,k}$ contains two terms. The first is a random Gaussian function, whereas the second one is deterministic. These contributions are competing: the random fluctuations encoded in the former lead to an exponential (in $N$) number of critical points. On the sphere in very high dimensions the majority of the configurations are orthogonal to the \emph{North Pole}, thus it is on the equator that we expect the deepest minima created by the first term alone. Since there are exponentially less configurations in the direction of  ${\bf v_0}$, and the less so when the overlap with ${\bf v_0}$ is higher, the random fluctuations alone lead to minima of higher energy on parallels closer to the north pole. On the other hand,
the deterministic term energetically favors configurations aligned with ${\bf v_0}$. In consequence, depending on the relative strength of the two terms, that can be tuned by changing the value of $r$, and on the form of the function $f_k(x)$ the resulting rough energy landscape changes shape and the low lying energy minima change position and nature from many to a single one. As we shall see, all that corresponds to phase transitions in the
geometry of the landscape. Topological phase transitions, occurring when the landscape 
changes from being complex to simple, have been recently studied in \cite{FyodorovTopologyTrivialization,fyodledou,fyodalone} and dubbed \emph{topological trivialization}. 
The change in the global minima structure, which is directly accessible to a thermodynamic study, was already reported in Ref. \onlinecite{sherrington1}.  \\
In the following we present our main results on the evolution with $r$ of the full energy landscape. For the sake of the presentation we group the different scenarios in three classes, associated with the behavior of the global minima 
as a function of $r$.

\subsection{Case I: $f'(0)>0$}
\noindent
This case corresponds to functions $f_k(x)$ which are monotonically increasing and such that $f'(0)>0$. 
The simplest example, $f_k(x)=x$, corresponds to the $p$-spin spherical model in an external magnetic field (with $r$ playing the role of the field), and has been studied in \cite{crisantisommers,cavagnagarrahan,fyodledou}. 
In agreement with those analyses, we find that the energy landscape evolves as illustrated in Fig.~\ref{Fig-k1}. \\First, at $r=0$, there are an exponential number of minima located around the equator, i.e. for $\overline q \in [\overline{q}_m(0),\overline{q}_M(0)]$, where $\overline{q}_m(0)=-\overline{q}_M(0)$. This corresponds to the first sphere on the left, in which the presence of minima is indicated with a red strip. 
The deepest minima, not exponentially numerous, are at $\overline q=0$ and correspond to the continuous yellow line. 
The most numerous states, which are also the marginally stable ones since the density of states of their Hessian is a Wigner semicircle 
with left edge touching zero, are also at $\overline q=0$ for $r=0$ (and they are of course at higher energy).\\  
By increasing $r$ the strip containing all the minima 
moves toward the north pole, see the second sphere from the left in Fig. \ref{Fig-k1}. The deepest ones are on a parallel closer to the north pole as soon as $r>0$. 
The most numerous ones, always marginally stable, are now on a different parallel with smaller latitude, as it can be expected on general grounds since in order to have a lot of minima it is better to avoid too large latitudes at which less configurations are available (they are represented by a yellow dashed line in the figure).\\ 
By increasing $r$ the landscape becomes smoother due to a larger deterministic term and, accordingly, the number of minima and the strip where they are located shrink until reaching  a value $r_c$ above which only one minimum remains. This corresponds to a phase transition of the landscape, which is associated to recovering a replica symmetric solution for the global minimum within the replica method, and hence also to a phase transition 
in the thermodynamics (related to the structure of the global minima).  
\begin{figure*}[!htbp]\centering
\includegraphics[width=2\columnwidth]{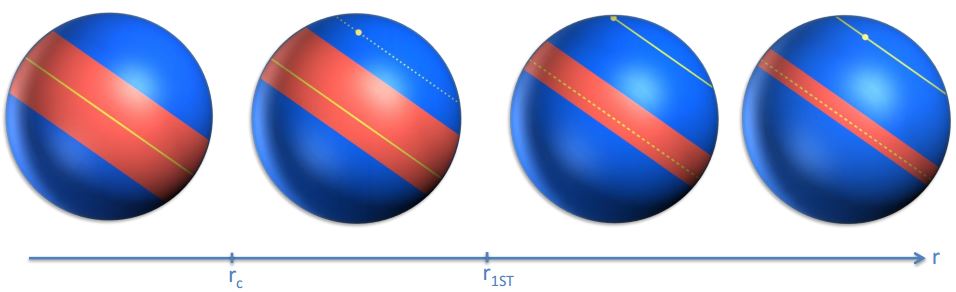}
 \caption{{\bf Evolution of the energy landscape in case III (option A)}. In this drawing we illustrate one of the  possible evolutions of the energy landscape due to the increase of $r$ in Case III. The red strips denote the regions on the sphere where minima lie in an exponential number. The continuous yellow lines corresponds to the
 parallel where the global minimum is located. At $r_{c}$ an isolated local minimum appears. The dotted yellow line denotes that it is not yet the global one.  At $r_{1\rm{ST}}>r_{c}$ 
 the deepest minimum is no longer on the equator and switches discontinuously to the isolated one close to the north-pole.  For larger values of $r$ the global minimum approaches the north pole and the band around the equator shrinks but does not disappear for any finite $r$. The most numerous states, denoted by a dashed line, are always located on the equator.}
 \label{Fig-k3A}
\end{figure*}
\begin{figure*}[htbp!]\centering
\includegraphics[width=2\columnwidth]{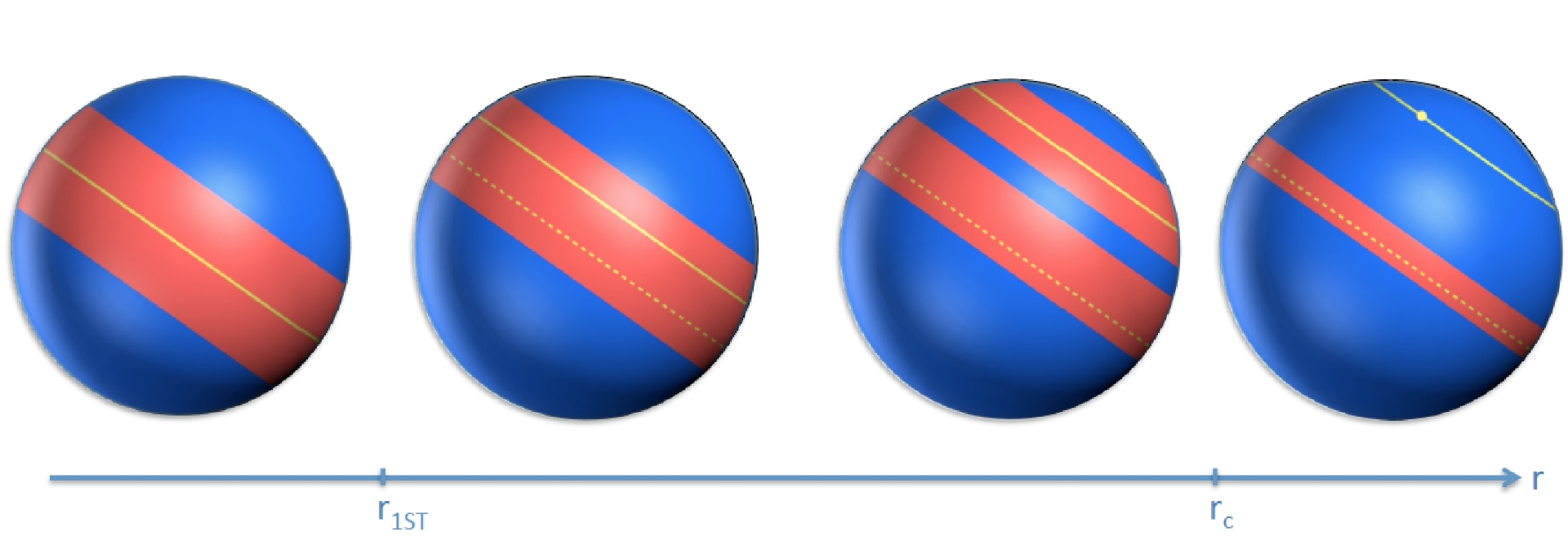}
 \caption{{\bf Evolution of the energy landscape in case III (option B)}. In this drawing we illustrate one of the  possible evolutions of the energy landscape due to the increase of $r$ in Case III. The red strips denote the regions on the sphere where minima lie in an exponential number.  At $r_{1\rm{ST}}$ 
 the deepest minimum is no longer on the equator and switches discontinuously to one at higher latitude.  
 For larger values of $r$ the band of local minima divides in two: one around the equator and one around the global minimum. For $r=r_c>r_{1\rm{ST}}$ the latter band shrinks to zero and the global minimum becomes isolated. The band around the equator shrinks but does not disappear for any finite $r$. The most numerous states, denoted by a dashed line, are always located on the equator.}
 \label{Fig-k3B}
\end{figure*}
For $r>r_c$ there is only one minimum in the energy landscape. In this case the random contribution due to the first term in the Hamiltonian is no longer strong enough to create a rugged landscape but still deforms it sufficiently to move the global minimum at a finite overlap with ${\bf v_0}$. This corresponds to the rightmost sphere in Fig. \ref{Fig-k1}. As we shall see in the following, a much richer 
energy landscape evolution is found for $k>1$. In these cases the behavior of the global minima is only a facet of a more general complex organization in configurations space.    
\subsection{Case II: $f'(0)=0$ and $f''(0)>0$}
\noindent
This regime corresponds to functions $f_k(x)$ which have vanishing derivative in $x=0$ but finite second 
derivative and are monotonically increasing from $x=0$ to $x=1$. In order to simplify the discussion we consider 
the symmetric case in which $f_k(-x)=f_k(x)$. The simplest example of such a function is $f_k(x)=x^2/2$. With this choice, $H_{p,k}$ corresponds to
a $p$-spin spherical model with an extra ferromagnetic interaction among spins ($r$ plays the role of the coupling). 
The evolution of the energy landscape is now different from Case I and it is illustrated in Fig. \ref{Fig-k2}. \\
The starting point at $r=0$ is the same. However, by increasing $r$ the strip containing all the minima widens 
and the deepest ones and the most numerous ones (always marginally stable) remain stuck on the equator. Actually they are exactly the same ones found for $r=0$ since 
$f_k(x)$ has no effect on the equation that determines the critical points on the equator (this is due to the vanishing of the first derivative in $x=0$).\\
This situation persists until $r=r_{2\rm{ND}}$, at which a second-order phase transition takes place at the bottom 
of the landscape, as already found in Ref. \onlinecite{sherrington1}. By increasing $r$ above $r_{2\rm{ND}}$ the deepest minima continuously  detach  from the equator, see the second sphere in Fig. \ref{Fig-k2} (due to the symmetry $x\rightarrow -x$ they are located both in the north and south hemispheres). \\
The behavior for larger $r$ is different from case I: there is first a transition in the structure of the energy landscape in which the strip separates in three bands, two closer to the north and south poles respectively, to which the deepest minima belong, and one around the equator where the most numerous ones are located, see the middle sphere in Fig. \ref{Fig-k2}. At $r=r_c$ there is another transition at which the two bands 
closer to the north and south poles containing an exponential number of minima shrink to zero and are replaced
by an isolated global minimum per hemisphere (fourth sphere from the left in Fig. \ref{Fig-k2}). This corresponds to recovering the RS solution in the thermodynamic treatment\cite{sherrington1}. Finally, at even larger $r$ all minima around the equator disappear and a final transition toward a fully smooth landscape characterized by only two minima takes place. This corresponds to the rightmost sphere in Fig. \ref{Fig-k2}.
The most numerous minima remain always at the equator for any value of $r$ until this final transition at which they disappear. However, they change nature when increasing $r$: at the beginning all the eigenvalues of their Hessian are distributed along a Wigner semi-circle whose left edge touches zero (so-called threshold states), whereas at large values of $r$ they are all distributed along a Wigner semi-circle whose support is strictly positive except for one eigenvalue, corresponding to an eigenvector oriented toward the north pole, which pops out from the semi-circle and is located exactly in zero. Thus, in both cases they are marginally stable but in a very different way.  \\
In conclusion, in the $k=2$ case in which the strength of the deterministic part is weaker in particular around the equator, the spurious local minima created by the random fluctuations are more stable. This results in a different evolution of the landscape, that 
before becoming fully smooth is characterized by isolated islands of ruggedness around the equator
and close to the global minima.  
\subsection{Case III: $f'(0)=f''(0)=0$}
\noindent
This regime corresponds to functions $f_k(x)$ which are monotonically increasing in $[0,1]$ and 
have vanishing first and second derivatives in $x=0$. For simplicity, we shall consider even and odd functions 
under $x\rightarrow -x$ when $k$ is odd and even respectively.  
The simplest example of such a function is $f_k(x)=x^k/k$ with $k\ge 3$, first introduced in Ref. \onlinecite{sherrington1}. With this choice and taking $p=k$, $H_{p,k}$ corresponds to the spiked-tensor model recently investigated in Refs. \onlinecite{montanari,krzakala,bandera,Chen, MontanariBenArous}. \\ 
The particularity of Case III is that the critical points on the equator are not affected at all by the deterministic perturbation, not even their Hessian (contrary to case II) since $f''(0)=0$: they remain stable and unperturbed for any finite value of $r$. In consequence, there is always a strip of minima around the equator. We have found that different evolution are possible in Case III depending on $p,k$. 
\subsubsection{Option A}
\noindent
This is the case found for example for spiked-tensor models such as $p=k=3$ and $p=k=4$. For concreteness we focus on $p=k=3$ ($p=k=4$ is analogous but one has to take into account that $f_k(x)$ is even instead of being odd). A band of minima, growing with $r$, is found around the equator. At a value $r_c$ an isolated minimum detaches from the top of the band, and for larger values of $r$ it moves to higher latitudes, while the rest of the band shrinks around the equator. The deepest minima are located on the equator and are the ones of the original (unperturbed) $p$-spin model until a value of $r$, that we call $r_{1\rm{ST}}$, is reached. When $r$ reaches the value $r_{1\rm{ST}}$ the global minimum switches from the equator to the single minimum outside the band and close to the north pole. 
Increasing $r$ further the isolated global minimum approaches the north pole and the band around the equator shrinks but never disappears for any finite $r$. The most numerous states are on the equator and are the threshold states 
of the unperturbed $p$-spin model. The evolution of the energy landscape and its transitions are illustrated in Fig. \ref{Fig-k3A}. 
\subsubsection{Option B} 
\noindent
This is the case found for example for $p=3$ and $k=4$. 
A band of minima, which first grows with $r$, is found around the equator. The deepest minima are located on the equator until $r_{1\rm{ST}}$ and are the ones of the original (unperturbed) $p$-spin model. When $r$ reaches the value $r_{1\rm{ST}}$ the global minimum switches discontinuously from the equator to another minimum inside the band, at higher latitude. Increasing $r$ further, the band divides in two: one closer to the equator and one around the global minimum. For $r=r_c$, the band around the global minimum shrinks to zero (this corresponds to recovering the RS solution in the thermodynamics treatment \cite{sherrington1}). For $r>r_c$ the global minimum is isolated. The remaining band around the equator shrinks but never disappears for any finite $r$. The most numerous states are on the equator and are the threshold states 
of the unperturbed $p$-spin model. The evolution of the energy landscape and its transitions are illustrated in Fig. \ref{Fig-k3B}.     \\\\
Two other options are possible: the discontinuous transition at $r_{1\rm{ST}}$ could take place after that the band has divided and, depending whether $r_{1\rm{ST}}$ is larger or smaller than $r_c$, 
it could take place when the global minimum is isolated (option C) or is still surrounded by many other local minima (option D). We did scan a few more (see Fig. \ref{pk}), but not all possible values of $p,k$, nor analyzed all possible functions $f_k(x)$ to search for these two behaviors but this can be easily (even though painfully) done if specific interest in these intermediate cases arises.

\subsection{Randomness versus deterministic contribution}
\noindent
A short conclusion of the results presented above is that the evolution of an energy landscape in which random fluctuations compete with a deterministic contribution favoring a single minimum depends on the behavior of the deterministic part on the portion of configuration space where the majority of minima created by randomness lie. 
If the deterministic part affects and deforms these minima then the evolution is quite simple: the number of minima 
decreases and they become more and more oriented toward the direction favoured by the deterministic part until a 
point at which only one isolated global minimum remains. A different behavior is instead found when the deterministic part does not deform the majority of minima created by randomness. In this case, the competition 
between random and deterministic contributions is resolved in two different ways: it deforms 
the landscape in the proximity of the configurations favoured by the deterministic part, which can even result in an island of ruggedness and many local minima, and it creates a very rugged landscape in the region where the deterministic part has no effect, where the majority of the configuration lie. As it can be easily guessed, this landscape structure can have crucial consequences on dynamical properties. \\
We shall discuss these issues and, more generally the implications and consequences of our results in the Conclusion. In the following we present 
the methods we used, and our findings in more details. We first recall the thermodynamic analysis of the model, focusing on the zero temperature limit, in order to discuss the behavior of the global minima of the landscape. Subsequently, we analyze the evolution of the full set of minima, encoded in the quenched complexity.

\section{Structure of global minima by the replica method}\label{sec:ReplicaAnalysis}
\noindent
Using the replica method, one can only partially characterize the energy landscape and its critical points. 
The aim of this section is to show and recall what kind of information can be gained in this way. 
The comparison with the Kac-Rice analysis is presented in Sec. \ref{sec:Monasson}. \\
Previous studies can be found in Ref.~\onlinecite{sherrington1} (see also Ref.~\onlinecite{CrisantiLeuzzi2013}).
Our motivation and perspective on the equilibrium results are different from those, since we focus on where, and to which extent, the recovery of a signal is thermodynamically favored against the noise dispersion. \\
The starting point of the thermodynamic analysis is the evaluation of the free energy $f(\beta,r)$, obtained by computing the $n$-times replicated partition function $\langle Z^n \rangle$:
\begin{equation}
f(\beta,r)=-\lim_{n\rightarrow 0, N\rightarrow\infty}\frac{1}{\beta N}\frac{\langle Z^n \rangle-1}{n} 
\end{equation}
where 
\begin{equation}\label{eq:PartFunc}
Z = \sum_{\{s_i\}}\exp[-\beta H_{p,k}(r)] \ ,
\end{equation}
and where the signal contribution to the Hamiltonian is represented by $f_k(x)=x^k/k$. To gain direct information on the energy landscape we focus on the zero temperature limit, when the equilibrium states dominating the partition function \eqref{eq:PartFunc} coincide with the absolute minima (or minimum) of the energy landscape. This thermodynamic analysis gives then access to the equilibrium transitions, which occur whenever these global minima detach from the equator and move at higher latitudes in the sphere, becoming correlated to the signal. Moreover, it allows to determine whether the bottom of the energy landscape is simple, i.e. just one global minimum, or has a more complicated structure, encoded in the Replica Symmetry Breaking (RSB) formalism.

\subsection{Energy at fixed overlap and the cases I,II,III}
\begin{figure}[htbp!]
\centering \textbf{MODELS PHASE DIAGRAM}\par\medskip
\begin{center}
\includegraphics[scale=0.4]{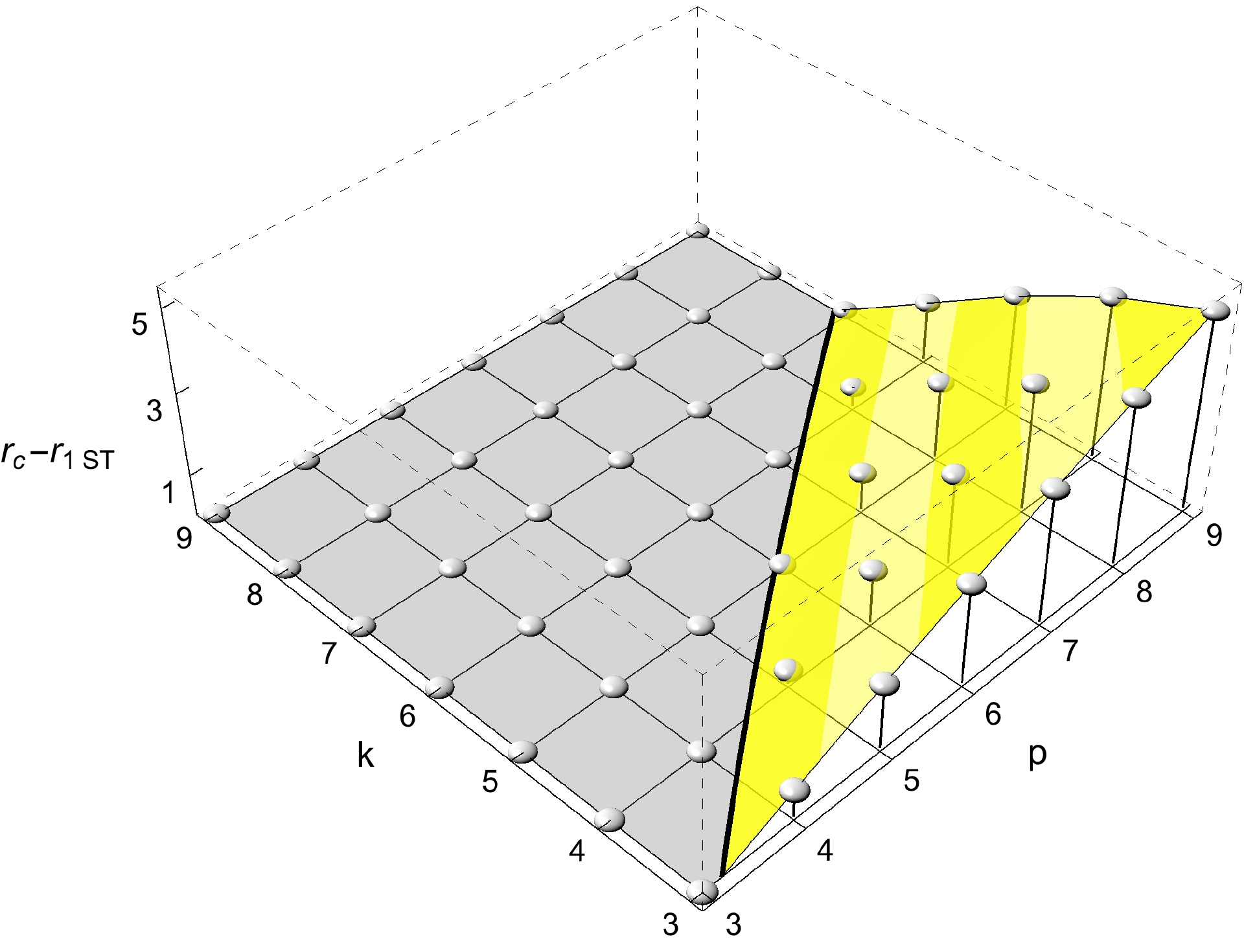}
\caption{Models (represented by the white spheres for integer $p$ and $k$) in which the 1RSB-RS transition takes place after (before) the discontinuous transition to the high overlap phase are to the right (left) of the black line. The yellow sheet represents the range of $r$, i.e. $r_c-r_{1ST}$, for which the $1$RSB phase of high $\overline q$ is globally stable for the first kind of models. Mesh lines are placed at $r_c-r_{1ST}=1,2,3,4$ as a reference.
}\label{pk} 
\end{center}
\end{figure}

\noindent
We first describe the main results of the replica analysis, the computation is shown later. 
One important remark is that the signal affects the model's solution only through the value of the typical overlap $\overline q$ with the north-pole.
To get the zero temperature solution, it is interesting then to focus on the intensive ground state energy of the original $p$-spin spherical model, i.e. without the function $f_k(x)$, for configurations constrained to have a fixed overlap $\overline q$. We denote this function $E(\overline q)$. 
As it is expected by the $\overline q \rightarrow -\overline q$ symmetry of the original $p$-spin problem, $E(\overline q)=E_{GS}+\frac{C_p}{2} \overline q^2+O(\overline q^4)$ for small $\overline q$, where $E_{GS}$ is the intensive ground state energy of the $p$-spin spherical model, and $C_p$ happens to be a positive constant.\\
This result already allows us to show the existence of the three regimes discussed in the previous section because now we can obtain and study the ground state energy of our model as $E(\overline q)-r\frac{\overline q^k}{k}$.
\begin{itemize}
\item {\bf Case I}: If $k=1$ then, no matter how small is $r$, the ground state is at $\overline q>0$ and increases when $r$ is augmented. This is the first scenario described in the previous section. 
\item {\bf Case II}: If $k=2$ then the ground state is at $\overline q=0$ for $r<r_{2{\rm ND}}=C_p$ and becomes continuously different from zero by increasing $r$ above $C_p$. This corresponds to a second-order like transition and to the second scenario discussed before.  
\item {\bf Case III}: if $k\ge 3$ then a discontinuous transition is bound to take place: for $r<r_{1{\rm ST}}$
the ground state is at $\overline q=0$, whereas for $r>r_{1{\rm ST}}$ it jumps to a finite value.  
This corresponds to a first-order like transition and to the third scenario discussed before.  \\
\end{itemize} 
An insight on the changes in the structure of the bottom of the landscape can be obtained along the same lines. At $r=0$ the replica solution is 1RSB. Proceeding as before, i.e. studying the $p$-spin spherical model at fixed $\overline q$, one can show that at fixed $p$ the solution always remains 1RSB until a given value of $\overline q_c$ is reached where the 1RSB-RS transition takes place. Moreover the replica structure is the same for identical values of $\overline q$.
Only the way in which $\overline q$ changes as a function of $r$ depends on the value of $k$. In consequence, 
when the ground state is at $\overline q=0$, there is a 1RSB structure of the energy landscape close to the global minimum (roughly speaking the energy landscape is rough close to the bottom). When the ground state is 
at  $\overline q>0$, for $r$ larger than a critical value $r_c$, the structure of the energy landscape close to the global minimum become RS (roughly speaking the energy landscape is convex close to the bottom). In case III 
there are two minima of $E(\overline q)$ close to the first order transition: one at $\overline q=0$ and one 
at $\overline q>0$. The high-overlap minimum
can become 1RSB before or after the discontinuous transition depending on the value of $r_c$ and $r_{1{\rm ST}}$. 
The replica analysis that we present below, see also Refs.~\onlinecite{sherrington1,CrisantiLeuzzi2013}, allows to find the models in which this happens, see Fig. \ref{pk}. The yellow sheet identifies (on its right) models 
that display a regime in which a rough landscape around the high-overlap global minimum
is present for $r_{1{\rm ST}}\le r \le r_c$. \\

\noindent We present below the replica computation. The following section will be also useful to fully understand the comparison with the Kac-Rice method discussed in Sec. \ref{sec:Monasson}. [Readers not interested 
in replica theory can skip Sec.\ref{sec:ReplicaEquations} and jump directly to Sec.\ref{sec:KacRice}]

\subsection{Replica Solution}\label{sec:ReplicaEquations}
\noindent
The standard replica computation \cite{cavagnapedestrian} for $f(\beta,r)$ leads to the following result 
$$
f=-\lim_{n\rightarrow 0, N\rightarrow\infty}\frac{1}{\beta N n }\int dQ_{\alpha,\beta}\exp(nNS[Q_{\alpha,\beta}]),
$$ 
where
$$
S=\frac{\beta^2}{4n}\sum_{a,b}Q_{ab}^p+\frac{1}{2n}\log \det Q_{\alpha\beta}+r\frac{\beta}{n}\sum_a f_k(Q_{0a})
$$
and $Q_{\alpha\beta}$ is an $(n+1)$x$(n+1)$ matrix ($\alpha, \beta \in [0,n]$) composed by $1$ on the diagonal, $Q_{0,a}$ on the $2n$ entries of the first line and column, and a matrix $Q_{a,b}$ with $a,b \in [1,n]$ on the remaining $n$x$n$ block.\\ The action has $3$ parts: the energy of the $p$-spin part of the original Hamiltonian, the energy due to the added potential $f_k$ controlled by the parameter $r$, and the entropy of a $N$-dimensional spherical system with one special direction.  \\
To proceed in the calculation, we use a RS {\it ansatz} on the entries $Q_{0,a}$, $Q_{0,a}=\overline{q}$, and
the usual RS or $1$RSB {\it ansatz} for the $Q_{a,b}$ matrix (no additional breaking of replica symmetry is expected).
The first case corresponds to $Q_{a,b}=q$ if $a\neq b$. In the second case the replicas are classified according to $n/m$ different blocks, $Q_{a,b}=q_1$ for $a\neq b$ with $a$ and $b$ in the same block of size $m$, and $Q_{a,b}=q_0$ when $a$ and $b$ belong to different blocks. 
The 1RSB ansatz contains the RS one: the second can be recovered by setting either $m=1$ or $q_1=q_0$. We thus only focus on the first. \\

\subsubsection{The 1RSB saddle point equations}
\noindent
The expression of the $1$RSB action $S_{\rm 1RSB}$ in the $N\rightarrow \infty$ and $n\rightarrow 0$ limit is reported in Appendix \ref{app:Replicas} for generic values of $\beta$. When $\beta\rightarrow\infty$ the action reads
\begin{eqnarray*}
\frac{S_{\rm 1RSB}}{\beta}=\frac{1}{4}[p\beta(1-q_1)+\beta m(1-q_0^p)]+r f_k(\overline{q})+ \\
	\frac{1}{2\beta m}\log\left(\frac{\beta(1-q_1)+\beta m(1-q_0)}{\beta(1-q_1)}\right)+ \nonumber \\
	\frac{1}{2}\frac{q_0-\overline{q}^2}{\beta(1-q_1)+\beta m(1-q_0)} \nonumber,
\end{eqnarray*}
where the parameters $\beta(1-q_1)$, $q_0$, $\beta m$, and $\overline{q}$ have to be determined by the following saddle point equations:
\begin{eqnarray*}
\frac{p}{2}=\frac{1}{\beta m}\left(\frac{1}{\beta (1-q_1)}-\frac{1}{\beta(1-q_1)+\beta m(1-q_0)}\right)+ \\
	\frac{q_0-\overline{q}^2}{[\beta(1-q_1)+\beta m(1-q_0)]^2} \nonumber,
\end{eqnarray*}
\begin{eqnarray*}
\frac{p}{2}q_0^{p-1}=\frac{q_0-\overline{q}^2}{[\beta(1-q_1)+\beta m(1-q_0)]^2},
\end{eqnarray*}
\begin{eqnarray}\label{spS}
\hspace{-.6cm}
\frac{1}{2} (1-q_0^p)+\frac{1}{(\beta m)^2}\log\left(\frac{\beta(1-q_1)}{\beta(1-q_1)+\beta m(1-q_0)}\right)+ \nonumber \\
	\frac{1}{\beta m}\frac{1-q_0}{\beta(1-q_1)+\beta m(1-q_0)}-\nonumber\\
	\frac{(1-q_0)(q_0-\overline{q}^2)}{[\beta(1-q_1)+\beta m(1-q_0)]^2}=0,
\end{eqnarray}
and
\begin{eqnarray*}
\overline{q}=r[\beta(1-q_1)+\beta m(1-q_0)]f'_k(\overline{q}) \ ,
\end{eqnarray*}
with $f'_k(x)=x^{k-1}$ being the first derivative of $f_k$. For each value of $r$, the value of $\overline{q}$ obtained solving the saddle point equations gives the latitude of the deepest minima of the landscape, while the function $-S_{\rm 1RSB}/\beta$ evaluated at the saddle point parameters gives their energy density.

\subsubsection{RS-$1$RSB transition for the high overlap phase}
\noindent
We now discuss the limit where the ground state energy ceases to be obtained by a $1$RSB solution and is instead determined by a RS one.
The transition between these two regimes signals a change in the structure at the bottom of the landscape from many low-lying minima to one single global minimum. We call $r_c$ the corresponding critical value of $r$ at which this occurs.\\ 
The first piece of information about this change of structure is obtained by expanding the four 1RSB saddle point equations~\cite{crisom92} for small $q_1-q_0$, and by keeping the lowest order non-zero terms. This gives four equations, see Appendix \ref{app:Replicas}. Applying them to the $1$RSB solution with high $\overline q$, we get that the critical point occurs at
\begin{equation}\label{eq:rCritico}
 r_c(p)=\sqrt{\frac{p(p-2)}{2}}f_k'\left(\sqrt{\frac{p-2}{p-1}}\right)^{-1},
\end{equation}
when 
$$
\beta(1-q_c)=\sqrt{\frac{2}{p(p-1)}}
$$
and 
\begin{equation}\label{eq:qc}
\overline{q}_c=\sqrt{\frac{p-2}{p-1}} \ .
\end{equation}
At this point the high $\overline q$ solution recovers a RS structure, i.e. it becomes a single minimum.
Still one has to consider whether this solution is a global minimum of the energy landscape or only a local one.
This piece of information is recovered by comparing the energy cost of the high $\overline q$ solution with the solution with $\overline q=0$, when this does still exist. A full account of all the possible models' solution obtained by using all the gathered information is presented in the next section.

\subsection{Results}
\noindent
From the numerical study of all the $T=0$ equations above we recover the three distinct scenarios accounted for in Sec. \ref{sec:SummaryResults}.  \\
{\bf Case $k=3$ and higher}.
For $k>2$, we find a stable $1$RSB $\overline q=0$ solution at every value of $r$. This solution is orthogonal to the signal and completely dominated by the noise represented by the $p$-spin part.
Beside this solution, when $r$ increases we find a second, high-$\overline q$ solution which undergoes a continuous transition between a $1$RSB phase and a RS phase at $r_c$. 
The high-$\overline q$ solution ($\overline q\neq 0$) contains at least partial information about the signal, the amount of this information being represented by the overlap $\overline q$. This solution is at first metastable compared to the $\overline q=0$ state, but it becomes stable at higher $r$.
This occurs through a first order transition at  $r_{1{\rm ST}}$.
If $r_{\rm 1ST}>r_c$ the first order-transition marks a thermodynamic discontinuity between a $1$RSB state (at $\overline q=0$) and a RS state (the high $\overline q$ one).
This scenario is generally found for $k\geq p$ as shown in the phase diagram in Fig.~\ref{pk}.
If $r_{\rm 1ST}<r_c$ instead two transitions are observed when $r$ increases. 
A first order transition will occur at lower $r$ showing the exchange of stability between the $\overline q=0$ and high-$\overline q$, $1$RSB, states. A continuos transition between the $1$RSB and the RS phase will follow within the high-$\overline q$ state at higher $r$.
An intermediate complex phase, related to a rugged landscape, already containing partial information on the signal, or North Pole, 
then emerges in this case. \\
{\bf Case $k=2$}.
The case $k=2$ is qualitatively different from $k=3$. 
The first order transition is replaced by a continuous, $2$nd order-like, transition between the $\overline q=0$ state and the high-$\overline q$ state before the last one becomes RS at $r_c$.
As explained before this can be rationalised thinking that the $1$RSB action is quadratic in $\overline q$. 
As such a term $f_k$ with higher power of $\overline q$ ($k>2$) cannot affect the local stability of the $\overline q=0$ state. When $k=2$ instead, $f_k$ can counterbalance the quadratic contribution of the $1$RSB action leading to the instability of the $\overline q=0$ solution at high enough $r_{2{\rm ND}}$. The 1RSB-RS transition happens for a strictly larger value of $r$ since it  takes place for a finite value of $\overline q$.\\
{\bf Case $k=1$}.
Finally the case $k=1$ has been extensively studied years ago~\cite{crisom92}, it corresponds to the 
$p$-spin spherical model in an external magnetic field. In this case there are no competing $1$RSB states at all. 
The linear field immediately shifts $\overline q$ of the $1$RSB phase away from zero until the continuous transition at $r_c$ brings the $1$RSB phase into the RS solution.\\
In order to show concrete examples, we report in Table~\ref{TBLtransitions} the different transition values 
for different $p$ and $k$, in particular for the spike-tensor model $p=k=3$.
\begin{table}[h!]
\centering
\begin{tabular}{c|c|c|c}
& $r_c$ & $r_{1{\rm ST}}$ & $r_{2{\rm ND}}$ \\
\hline
\begin{tabular}{c} $p=3$ \\ $k=1$ \end{tabular} & $1.225$ & n.a. & n.a. \\
\hline
\begin{tabular}{c} $p=3$ \\ $k=2$ \end{tabular} & 1.732 & n.a. & $ 0.7297$ \\
\hline
\begin{tabular}{c} $p=3$ \\ $k=3$ \end{tabular} & 2.449 & $2.559$ & n.a.  \\
\hline
\begin{tabular}{c} $p=4$ \\ $k=3$ \end{tabular} & 3. & $2.256$ & n.a.  \\
\hline
\begin{tabular}{c} $p=8$ \\ $k=3$ \end{tabular} & 5.715 & $1.594$ & n.a.  
\end{tabular}
\caption{Values of $r$ at the available transitions in a few instances of $p$ and $k$.}
\label{TBLtransitions}
\end{table}
As shown above, in the $k\ge 3$ case whether the discontinuous transition to the high $\overline q$ phase 
takes place before or after the 1RSB-RS transition depends on the model, i.e. on $p$ and $k$. 
In order to find a general criterion we evaluate the action of the 
high overlap phase at the 1RSB-RS transition:
$$
\frac{S^{Hc}_{RS}}{\beta}=\sqrt{\frac{p}{2(p-1)}}\frac{k+p-2}{k} \ ,
$$
see Eqs. (\ref{eq1t},\ref{eq2t},\ref{eq3t},\ref{eq4t}) in Appendix \ref{app:Replicas}. We then compare this action to the one of the 1RSB phase with $\overline{q}=0$: $S_{1RSB}^L$. 
There are two possible cases:
\begin{itemize}
\item $S^{Hc}_{RS}<S^{L}_{1RSB}$. In this case  the high-overlap phase becomes energetically favorable after the 1RSB-RS transition takes place. Thus, the rough energy landscape around the north pole described by the 1RSB phase does not contain the global minima of the landscape, but only some local (metastable) ones. 
 \item $S^{Hc}_{RS}>S^{L}_{1RSB}$. In this case the high-overlap phase becomes energetically favorable before  
 the 1RSB-RS transition takes place, hence there is a range of $r$ where the stable high-overlap phase is $1$RSB. This region extends up to $r_c$.
\end{itemize}
In Fig. \ref{pk} we show a diagram in the $p, k-$space, with the black line
representing the point where $S^{Hc}_{RS}/\beta=S^{L}_{1RSB}/\beta$ at zero temperature. 
To the right (respectively left) of the line lie models in which the $1$RSB-RS transition takes place after (respectively before) the discontinuous transition to the high overlap phase. Whenever the $1$RSB-RS transition takes place after the discontinuous transition, the complex phase with partial information about the signal contains ground state minima for a finite range of $r$. The height of the coloured sheet in Fig. \ref{pk} represents the range of $r$ for which this holds, i.e. $r_c-r_{1ST}$. Note that the range becomes larger and larger when $p$ increases for fixed $k$, or $k$ decreases at fixed $p$.\\
The two complex phases we have discussed present a multi-minima structure that is worth studying to get insights on the possibility to recover the signal through different sampling dynamics. We perform this study in the following section, making use of the Kac-Rice formalism. A comparison with the results obtained by means of the replica formalism is postponed to Sec. \ref{sec:Monasson}.

\section{Landscape analysis via replicated Kac-Rice formula}\label{sec:KacRice}
\noindent
In this section, we present the analysis of the energy landscape of $H_{p,k}(r)$ performed through the replicated version of the Kac-Rice method.\\
Our aim is to determine the number $\mathcal{N}_{N}(\epsilon, \overline{q})$ of local minima (or, more generally, of stationary points) of the energy functional, having a given energy density $\epsilon$ and a fixed overlap ${\bf s} \cdot {\bf v_0}=N \overline{q}$ with the special direction ${\bf v_0}$. The number $\mathcal{N}_{N}(\epsilon, \overline{q})$ is a random variable that, when the random fluctuations dominate over the signal, scales exponentially with $N$. This occurs over a finite range of energies; among the exponentially-many local minima, the lowest-energy ones dominate the thermodynamics of the model (described in detail in the previous section), while the higher-energy ones are expected to play a relevant role when discussing the dynamical evolution on the energy landscape.  \\
We are interested in determining the exponential scaling of the typical value of  $\mathcal{N}_{N}(\epsilon, \overline{q})$, that is, we aim at  computing the \emph{quenched} complexity $\Sigma_{p,k}(\epsilon, \overline{q}; r)$ defined as
\begin{equation}\label{eq:comp}
 \Sigma_{p,k}(\epsilon, \overline{q};r) \equiv \lim_{N \to \infty} \frac{\langle{\log \mathcal{N}_{N}(\epsilon, \overline{q})}\rangle}{N}.
\end{equation}
As anticipated, we perform the calculation making use of the Kac-Rice formula \cite{Kac, AdlerRandomFields}. This formalism has been recently exploited to characterize the topological properties of random landscapes associated to the pure and mixed $p$-spin models \cite{auffingerbenaouscerny, auffingerbenaous}, to the spiked-tensor model \cite{MontanariBenArous}, as well as to count the equilibria of dynamical systems modeling large ecosystems \cite{fyodorovmay, FyodorovTopologyTrivialization} and neural networks \cite{touboul}. In these contexts, results have been given for \emph{annealed} complexity, which governs the exponential scaling of the average number of stationary points, or equilibria. This corresponds to averaging $\mathcal{N}_N$ over the disorder realization before taking the logarithm, at variance with Eq. \eqref{eq:comp}.\\
For the Hamiltonian $H_{p,k}(r)$ with $r=0$, it is known that the quenched and annealed prescriptions give the same result for the complexity \cite{crisantisommers, eliran}. In presence of a signal, however, this equivalence does not hold (as we show below), so that the quenched calculation becomes necessary. We perform the latter by means of the replica trick, via the identity
\begin{equation}\label{eq:replica}
\Sigma_{p,k}(\epsilon, \overline{q}; r)= \lim_{N \to \infty}\lim_{n \to 0} \frac{\langle{\mathcal{N}^n_{N}(\epsilon, \overline{q})\rangle}-1}{N n},
\end{equation}
analytically continuing the expression for the higher moments of $\mathcal{N}_N$. The replicated version of the Kac-Rice formula allows us to obtain (to leading order in $N$) the moments $\langle \mathcal{N}_N^n\rangle$, for integers values of $n$.\\
As we show in the following, the expression for $\langle \mathcal{N}_N^n\rangle$ that we obtain involves $n$ critical points ${\bf s}^a$, $a=1, \cdots, n$, each with energy density $\epsilon$ and overlap $N \overline{q}$ with the \emph{North Pole}. Introducing their mutual overlaps ${\bf s}^a \cdot {\bf s}^b= N q_{ab}$, we find that we can parametrize the moments as:
\begin{equation}\label{eq:MomentsSaddlePOint}
 \langle \mathcal{N}_N^n (\epsilon, \overline{q})\rangle= \int \prod_{a<b} d q_{ab} \;  e^{N \mathcal{S}^{(n)}_{p,k}\tonde{\epsilon, \overline{q},r; \grafe{q_{ab}}}+ o(N)},
\end{equation}
where the integral can be computed with the saddle point approximation, optimizing over the order parameters $q_{ab}$. 
In consequence the action evaluated at the saddled point directly gives $\ln \langle \mathcal{N}_N^n (\epsilon, \overline{q})\rangle/N$ up to vanishing corrections in the large $N$ limit. 
To get the complexity, i.e the typical value of the number of critical points, we perform this calculation assuming replica symmetry, meaning that we set $q_{ab} \equiv q$ for $a \neq b$, and take the $n\rightarrow 0$ limit; we expect this to give accurate results, in view of the fact that $H_{p,k}(r)$ does not exhibit full-RSB but only 1-RSB in the statics. \\
Before entering into the details of the calculation, we collect the main resulting expressions in the following subsection.

\subsection{The main results: quenched complexity and mapping between the three cases}
\noindent
For arbitrary values of $n$ and assuming replica symmetry, we find that the action in Eq.  \eqref{eq:MomentsSaddlePOint} is given by
\begin{equation}\label{eq:ActionGeneraln}
 \mathcal{S}^{(n)}_{p,k}=  \frac{n}{2}\log [2 e (p-1)]+ n I[\beta(\epsilon, \overline{q})]- \tilde{Q}^{(n)}_{p,k}(\epsilon, \overline{q}; q),
\end{equation}
where $I(y)$ is an even function of its argument, equal to:
\begin{equation*}\label{eq:Inty}
\begin{split}
 I=\begin{cases} 
 \frac{y^2-1}{2} + \frac{y}{2} \sqrt{y^2-2}+ \log \tonde{\hspace{-0.05 cm} \frac{\hspace{-0.05 cm}-y+\hspace{-0.05 cm} \sqrt{y^2-2}}{2}\hspace{-0.05 cm}}  \hspace{-0.05 cm} \text{ if        } y \leq -\sqrt{2},\\
  \frac{1}{2} y^2 - \frac{1}{2} \tonde{1 + \log 2}  \text{  if        }-\sqrt{2}\leq y\leq 0,
 \end{cases}
 \end{split}
 \end{equation*}
while
 \begin{equation}\label{eq:beta}
  \beta(\epsilon, \overline{q})= \sqrt{\frac{p}{p-1}}  \epsilon + \frac{(p/k-1)}{ \sqrt{p(p-1)}}r \overline{q}^k.
     \end{equation}
The dependence on the overlap $q$ between the various replicas enters in the term $\tilde{Q}^{(n)}_{p,k}$, which reads:
\begin{equation*}\label{eq:QuGeneraln}
\begin{split}
 &\tilde{Q}^{(n)}_{p,k}=-\frac{1}{2}\log \tonde{\quadre{\frac{1-q}{1-q^{p-1}}}^{n-1} \frac{1-n \overline{q}^2+ (n-1)q}{1+(n-1) q^{p-1}}}\\
  &-n(n-1)r^2 \overline{q}^{2 k}\frac{(1-q)   q^{p} \quadre{1+(n-1) q^p-p (1-q)} }{p \left(1+(n-1) q^{p-1}\right) D(q)}\\
  &+n r^2 \overline{q}^{2 k-2}\left(1-\overline{q}^2\right) \frac{ 1}{p \left(1+(n-1) q^{p-1}\right)}\\
  &+2n (n-1) r \overline{q}^k  \left(\frac{r \overline{q}^k}{k}+\epsilon\right) \frac{q^{p+1} (1-q)   }{D(q)}\\
   &+n \left(\frac{r \overline{q}^k}{k}+ \epsilon\right)^2\frac{q^p-q^2-p q^p(1-q)(1+(n-1)q)}{D(q)},
   \end{split}
\end{equation*}
where we defined
\begin{equation}\label{eq:D(q)}
D(q)=(q^p-q^2)(1+(n-1)q^p)-pq^{p}(1-q)(1+(n-1)q).
\end{equation}
From this result, we can readily obtain the expression for the \emph{annealed} complexity, which is obtained setting $n=1$. In this case, the dependence on $q$ drops (as it is natural to expect, since there is only one replica and thus no overlap with any other one), and the action reduces to:
\begin{equation}\label{eq:FinalComplexityAnnealed}
\begin{split}
 & \Sigma^{(\text{ann})}_{p,k}(\epsilon, \overline{q}; r)= \frac{1}{2} \log \quadre{2 e (p-1)(1- \overline{q}^2)}+ I \quadre{\beta(\epsilon, \overline{q})}\\
 &- \tonde{\epsilon +  \frac{r \overline{q}^k}{k}}^2- \frac{1}{p}  r^2 \overline{q}^{2k-2} \tonde{1-\overline{q}^2}.
  \end{split}
\end{equation}
For $p=k$, this expression agrees with the results in Ref.~\cite{MontanariBenArous}. 
The annealed complexity is an upper bound to the quenched one. As we argue in Sec. \ref{sec:Results}, it captures correctly the properties of the energy landscape whenever this is smooth and has only one isolated minimum, i.e., in the regime $r>r_c$.\\
Our result at fixed $n$ provides all the integer moments of the number of critical points.  
To get the quenched complexity, the limit $n \to 0$ has to be performed, by analytically continuing \eqref{eq:ActionGeneraln}. The result is
\begin{equation}\label{eq:FinalComplexity}
\begin{split}
 &\Sigma_{p,k}(\epsilon, \overline{q}; r)= \frac{1}{2} \log \quadre{2 e (p-1)}+ I \quadre{\beta} -\tilde{Q}_{p,k}\tonde{\epsilon, \overline{q}; q_{\text{SP}}},
 \end{split}
\end{equation}
where $\tilde{Q}_{p,k}$ is defined from $\tilde{Q}^{(n)}_{p.k}= n \tilde{Q}_{p,k}+ O(n^2)$ and reads
\begin{equation}\label{eq:QTildeLinear}
\begin{split}
 &\tilde{Q}_{p,k}\tonde{\epsilon, \overline{q}; q}=\frac{1}{2}\quadre{\log \tonde{\frac{1-q^{p-1}}{1-q}}+\frac{q^{p-1}}{1-q^{p-1}}+\frac{\overline{q}^2-q}{1-q}}\\
 &+\frac{r^2 \overline{q}^{2k}}{p}\quadre{ \frac{(1-q^p) q^2 }{(q^p-q^2)(1-q^p)-p q^p(1-q)^2}+ \frac{\overline{q}^{-2}}{(1-q^{p-1})}}\\
 &-2 r \overline{q}^k \tonde{\frac{r \overline{q}^k}{k}+ \epsilon}\frac{q^{p+1}(1-q)}{(q^p-q^2)(1-q^p)-p q^p(1-q)^2}\\
  &+ \tonde{\frac{r \overline{q}^k}{k}+ \epsilon}^2\frac{q^p-q^2-p q^p(1-q)^2}{(q^p-q^2)(1-q^p)-p q^p(1-q)^2},\\
   \end{split}
\end{equation}
while $q_{\text{SP}}=q_{\text{SP}}(\epsilon, \overline{q})$ is the saddle point extremizing the function \eqref{eq:QTildeLinear}.\\
The evaluation of the quenched complexity therefore requires to compute a saddle point on $q$ for given values of the parameters $\overline{q}, \epsilon$. A substantial simplification comes from a general identity that we derive in Sec. \ref{sec:mapping} and which relates, for fixed $p$ and $\overline{q}$, the complexities $\Sigma_{p,k}(\epsilon, \overline{q}; r)$ for different values of $k$: 
\begin{equation}\label{eq:ExplicitMapping}
 \Sigma_{p,k}\tonde{\epsilon, \overline{q}; r}= \Sigma_{p,1}\tonde{\epsilon +  r  f_k -r\frac{ f'_k}{f'_1}f_1,\, \overline{q} \,;\, \frac{ f'_k}{f'_1} r}
\end{equation}
for $f_k \equiv f_k(\overline{q})$, meaning that all complexity curves for $k>1$ can be derived from the ones at $k=1$. This is convenient, as it allows us to solve the saddle point equations for $q$ in one single case. We remark however that not all the properties of the landscape at $k>1$ can be deduced from the case $k=1$: in particular, the analysis of the stability of the stationary points (i.e., of the spectrum of their hessian) has to be performed separately for any $k$, as we discuss in more detail in the following subsections.

\subsection{The replicated Kac-Rice formula}
\noindent
Here we present the replicated version of the Kac-Rice formula, and outline the main steps of the subsequent calculation. 
For convenience, we introduce the vectors ${\bm \sigma}= {\bf s}/\sqrt{N}$ and $ {\bf w_0}= {\bf v_0}/\sqrt{N}$ having unit norm, and we define the rescaled energy functional 
\begin{equation}\label{eq:EnegyDensity}
 h[{\bm \sigma}]= \sqrt{\frac{2}{N}} H_{p,k}(r) = h_{ps}[{\bm \sigma}]-\sqrt{2 N} r f_k\left({\bm \sigma}\cdot{\bf w_0}\right),
\end{equation}
with $h_{ps}[{\bm \sigma}]\equiv  -\sum_{\langle i_1,i_2,\dots,i_p\rangle} J'_{\bf i} \sigma_{i_1} \sigma_{i_2} \dots \sigma_{i_p}$ denoting the $p$-spin energy functional with rescaled coupling satisfying $\langle (J'_{\bf i})^2\rangle= p!$.
We count the stationary point ${\bm \sigma}$ of this functional satisfying $h \quadre{{\bm \sigma}}=\sqrt{2 N} \, \epsilon$ and ${\bm \sigma} \cdot {\bf w_0}=\overline{q}$, which are in one-to-one correspondence with the stationary points of $H_{p,k}(r)$ with energy density $\epsilon$ and ${\bf s} \cdot {\bf v_0}  =N \overline{q}$.\\
The Kac-Rice formula incorporates the spherical constraint, as it counts the number of stationary points of the functional $h[{\bm \sigma}]$ \emph{restricted} to the unit sphere; such points ${\bm \sigma}$ nullify the surface gradient of \eqref{eq:EnegyDensity}, which is a vector ${\bf g} [{\bm \sigma}]$ lying on the tangent plane to the sphere at the point ${\bm \sigma}$. Similarly, their stability is governed by the Hessian on the sphere, which we denote with $\mathcal{H}[{\bm \sigma}]$ (see Eq. \eqref{eq:relHessians} for a precise definition of this matrix). 
Given $n$ replicas  ${\bm \sigma}^a$, $a=1, \cdots, n$, we introduce the shorthand notation ${\bf g}^a\equiv {\bf g}[{\bm \sigma}^a]$,  $\mathcal{H}^a\equiv \mathcal{H}[{\bm \sigma}^a]$, $h^a\equiv h[{\bm \sigma}^a]$, and denote with $p_{\vec{{\bm \sigma}}}$ the joint density function of the $(N-1)n$ gradients components $g_{\alpha}^a$ and of the $n$ functionals $h^a$, induced by the distribution of the couplings $J'_{\bf i}$ in~\eqref{eq:EnegyDensity}. With this notation, the replicated Kac-Rice formula reads:
 \begin{equation}\label{eq:Full}
  \langle{\mathcal{N}^n_{N}(\epsilon, \overline{q})}\rangle=  \int \prod_{a=1}^n d{\bm \sigma}^a \, \delta \tonde{{\bm \sigma}^a \cdot {\bf w}_0- \overline{q}} \mathcal{E}_{\vec{{\bm \sigma}}}(\epsilon)\, p_{\vec{{\bm \sigma}}}({\bf 0}, \epsilon),
 \end{equation}
 with
 \begin{equation}\label{eq:Expectation}
  \mathcal{E}_{\vec{{\bm \sigma}}}(\epsilon)=  \Big\langle{\tonde{\prod_{a=1}^n \left| \text{det}\; \mathcal{H}^a\right|}  \Big| \grafe{h^a = \sqrt{2 N}\epsilon,  {\bm g}^a={\bf 0} \; \forall  a}}\Big\rangle.
 \end{equation}
 In \eqref{eq:Full} the integral is over $n$ replicas ${\bm \sigma}^a$ constrained to be in the unit sphere, at overlap $\overline{q}$ with the vector ${\bf w_0}$. The function $ p_{\vec{{\bm \sigma}}}({\bf 0}, \epsilon)$ is the joint density of gradients and energies evaluated at $g_{\alpha}^a=0$ and $h^a=\sqrt{2 N}\,\epsilon$ for any $a=1, \cdots, n$. The expectation value \eqref{eq:Expectation} is over the joint distribution of the Hessians $\mathcal{H}^a$, conditioned on each ${\bm \sigma}^a$ being a stationary point with rescaled energy $\sqrt{2 N} \epsilon$, and overlap $\overline{q}$ with ${\bf w_0}$. \\
 The computation of the moments \eqref{eq:Full} requires to determine, for each configuration of the replicas ${\bm \sigma}^a$, the joint distribution of the variables $\mathcal{H}^a_{\alpha  \beta}$, $g_{\alpha}^a$ and $h^a$, which are all mutually correlated and whose distribution depends, in principle, on the coordinates of all the replicas. For the simplest case ($n=1$) of a single replica ${\bm \sigma}$, it can be shown (see the discussion below, and Refs.~\cite{auffingerbenaouscerny,fyodorov}) that (i) the gradient ${\bf g}\quadre{{\bm \sigma}}$ is statistically independent from $h \quadre{{\bm \sigma}}$ and from the Hessian, and (ii) the distributions depend on ${\bm \sigma}$ only through its overlap $\overline{q}$ with the special direction ${\bf w_0}$ (in absence of the signal, the distribution turns out to be independent on $\bm{\sigma}$). These features make the computation of the annealed complexity feasible; in particular, (ii) is crucial, as it allows to integrate out the variable ${\bm \sigma}$ and get an expression for $\langle \mathcal{N}_N \rangle$ which depends only on few parameters. Moreover, it suggests that the distributions of the random vector  ${\bf g}\quadre{{\bm \sigma}}$ and of the random matrix $\mathcal{H} \quadre{{\bm \sigma}}$ satisfy some rotation invariant symmetry, hinting at the connection with the random matrix theory of invariant ensembles~\cite{auffingerbenaouscerny, fyodorov}. \\
   When the number of replicas is larger than one, the situation is more involved, as the random variables associated to different replicas are non-trivially correlated. However, it remains true that their joint distribution can be parametrized in terms of $\overline{q}$ and few additional order parameters, that are the overlaps $q_{ab}= {\bm \sigma}^{a} \cdot {\bm \sigma}^b$ between the different replicas~\cite{footnote2}. 
 In the following subsection, we discuss in more detail this structure, which allows us to re-express the moments \eqref{eq:Full} as an integral over the order parameters $q_{ab}$ of three terms scaling exponentially with $N$, see Eq.~\eqref{eq:Full3}. The first term is a volume factor, emerging when integrating over the variables ${\bm \sigma}^a$: this is evaluated with standard methods in Sec.~\ref{sec:PhaseSpaceFactor}.  The second terms is the joint distribution of gradients and energy fields; the difficulty in computing this term relies in the inversion of the correlation matrix of the gradients: we overcome it by realizing that it is sufficient to invert the projection of the matrix on a restricted portion of replica space, see Sec. \ref{sec:JointDensity}. Finally, the third term is the conditional expectation value of the product of determinants. We find that the conditioned Hessians of the various replicas are coupled, weakly perturbed GOE matrices, such that their mutual correlations can be neglected when computing the expectation value to leading order in $N$ (Sec.~\ref{sec:ConditionalLaw} and Sec.~\ref{sec:Factorization}). As a consequence, we find that this term contributes with a factor that is independent on $q_{ab}$, and which is governed by the properties of the GOE invariant ensembles, see Sec. \ref{sec:GOEcomputation}. We discuss the stability of the stationary points, which is encoded in the statistics of the spectrum of the Hessians, in Sec.~\ref{sec:Stability}. The final result of the calculation is Eq.~\eqref{eq:FinalComplexity}, where we remind that the integral over the order parameters has been performed within the saddle point approximation, assuming a replica-symmetric structure of the overlap matrix, $q_{ab} \equiv q$ for $a \neq b$.

 \subsection{Structure of covariances and order parameters}
 \noindent
  As a first step, we analyze the structure of the correlations between the random variables $\mathcal{H}^a$, $g^a$ and $h^a$: since they are Gaussian, their statistics is fully determined by their averages and mutual covariances, which turn out to depend only on $\overline{q}$ and on the overlaps $q_{ab}= {\bm \sigma}^{a} \cdot {\bm \sigma}^b$. \\
 To uncover this structure, we consider the gradients ${\bm \nabla} h^a \equiv {\bm \nabla} h[{ {\bm \sigma}^a}]$ and Hessian ${\bm \nabla}^2 h^a \equiv {\bm \nabla}^2 h[{\bm \sigma}^a]$ of the functional \eqref{eq:EnegyDensity} \emph{extended} to the whole $N$-dimensional space~\cite{footnote3}, and determine the covariances between their components along arbitrary directions in the $N$-dimensional space, given by some $N$-dimensional unit vectors ${\bf e}_i$. 
 From here, the correlations of the components ${\bf g}^a$ and $\mathcal{H}^a$ are easily determined setting ${\bf e}_i \to {\bf e}_\alpha^a$, where $\grafe{{\bf e}_\alpha^a}_{\alpha=1}^{N-1}$ is an arbitrarily chosen basis of the tangent plane at ${\bm \sigma}^a$. This follows from the fact that ${\bf g}^a$ is an $(N-1)$-dimensional vector with components ${g}^a_\alpha= {\bm \nabla} h^a \cdot {\bf e}_\alpha^a$, which is obtained from ${\bm \nabla} h^a$ by simply projecting it onto the tangent plane. 
 Similarly, $\mathcal{H}^a$ is an $(N-1) \times (N-1)$ matrix with components 
\begin{equation}\label{eq:relHessians}
 \mathcal{H}^a_{\alpha \beta}= {\bf e}_\alpha^a \cdot \tonde{{\bm \nabla}^2 h^a-\tonde{{\bm \nabla} h^a \cdot {\bm \sigma}^a}  \hat{1}} \cdot {\bf e}_\beta^a,
 \end{equation}
 as it follows from imposing the spherical constraint with a Lagrange multiplier~\cite{footnote4}.
  For arbitrary ${\bf e}_i$, taking the derivative of \eqref{eq:EnegyDensity} and computing the expectation value we find:
 \begin{equation}\label{eq:AvGrad}
 \begin{split}
&  \Big\langle {\bm \nabla} h^a \cdot {\bf e} \Big\rangle =-\sqrt{2 N} r f'\tonde{{\bm \sigma}^a \cdot {\bf w_0}}  \tonde{{\bf w_0} \cdot {\bf e}}\\
 & \Big\langle \tonde{{\bm \nabla} h^a \cdot {\bf e}} h^b  \Big\rangle_c = p ({\bm \sigma}^a \cdot {\bm \sigma}^b)^{p-1} \tonde{{\bf e} \cdot {\bm \sigma}^b},
  \end{split}
 \end{equation}
  and:
 \begin{equation}\label{eq:CovGrad}
 \begin{split}
 & \Big\langle \tonde{{\bm \nabla} h^a \cdot {\bf e}_1} \tonde{{\bm \nabla} h^b \cdot {\bf e}_2} \Big\rangle_c =p ({\bm \sigma}^a \cdot {\bm \sigma}^b)^{p-1} \tonde{{\bf e}_1 \cdot {\bf e}_2}+\\
  &p(p-1)({\bm \sigma}^a \cdot {\bm \sigma}^b)^{p-2} \tonde{{\bf e}_1 \cdot {\bm \sigma}^b} \tonde{{\bf e}_2 \cdot {\bm \sigma}^a},
  \end{split}
 \end{equation}
 where the subscript ``c'' indicates that the correlation function is connected. The same computation for the second derivatives gives:
\begin{equation*}\label{eq:CorrelationsHessianEn}
\begin{split}
 & \Big\langle  {\bf e}_1 \hspace{-0.05cm}  \cdot \hspace{-0.05cm} {\bm \nabla}^2 h^a \hspace{-0.05cm}\cdot \hspace{-0.05cm}{\bf e}_2 \Big\rangle\hspace{-0.05cm}=\hspace{-0.05cm}-\sqrt{2 N} r f''\tonde{\hspace{-0.05cm}{\bm \sigma}^a\hspace{-0.05cm} \cdot \hspace{-0.05cm}{\bf w_0}}  ({\bf w_0} \hspace{-0.05cm}\cdot \hspace{-0.05cm} {\bf e}_1 )({\bf w_0} \hspace{-0.05cm}\cdot \hspace{-0.05cm} {\bf e}_2 ),\\
 & \Big\langle \hspace{-0.1cm} \tonde{{\bf e}_1 \hspace{-0.1cm}\cdot \hspace{-0.1cm} {\bm \nabla}^2 h^a \hspace{-0.1cm} \cdot \hspace{-0.05cm} {\bf e}_2}  h^b \Big\rangle_c=p(p\hspace{-0.05cm}-\hspace{-0.05cm}1)({\bm \sigma}^a \hspace{-0.05cm}\cdot\hspace{-0.05cm} {\bm \sigma}^b)^{p-2}({\bf e}_1 \hspace{-0.05cm}\cdot\hspace{-0.05cm} {\bm \sigma}^b) ({\bm e}_2 \hspace{-0.05cm}\cdot\hspace{-0.05cm} {\bm \sigma}^b),
  \end{split}
\end{equation*}
and  
\begin{equation}\label{eq:HessTang}
\begin{split}
& \Big\langle  \tonde{ {\bf e}_1 \cdot  {\bm \nabla}^2 h^a \cdot {\bf e}_2} \tonde{ {\bf e}_3 \cdot {\bm \nabla}^2 h^b \cdot {\bf e}_4} \Big\rangle_c
=\\
&\frac{p! ({\bm \sigma}^a \cdot {\bm \sigma}^b)^{p-4}}{(p-4)!}({\bf e}_1 \cdot {\bm \sigma}^b) ({\bf e}_2\cdot {\bm \sigma}^b) ({\bf e}_3\cdot {\bm \sigma}^a)( {\bf e}_4 \cdot {\bm \sigma}^a)+\\
 &\frac{p!}{(p-3)!}({\bm \sigma}^a \cdot {\bm \sigma}^b)^{p-3} \,{({\bf e}_1 \cdot {\bf e}_4)( {\bf e}_2 \cdot {\bm \sigma}^b)( {\bf e}_3\cdot {\bm \sigma}^a)}+\\
  &\frac{p!}{(p-3)!}({\bm \sigma}^a \cdot {\bm \sigma}^b)^{p-3} \,{({\bf e}_2 \cdot {\bf e}_4)( {\bf e}_1  \cdot {\bm \sigma}^b)( {\bf e}_3 \cdot {\bm \sigma}^a)}+\\
  &\frac{p!}{(p-3)!}({\bm \sigma}^a \cdot {\bm \sigma}^b)^{p-3}\, {({\bf e}_1 \cdot {\bf e}_3)( {\bf e}_2 \cdot {\bm \sigma}^b)( {\bf e}_4 \cdot {\bm \sigma}^a)}+\\
   &\frac{p!}{(p-3)!}({\bm \sigma}^a \cdot {\bm \sigma}^b)^{p-3}\, {({\bf e}_2 \cdot {\bf e}_3)( {\bf e}_1 \cdot {\bm \sigma}^b)( {\bf e}_4\cdot {\bm \sigma}^a)}+\\
 &\frac{p! ({\bm \sigma}^a \cdot {\bm \sigma}^b)^{p-2} }{(p-2)!}\quadre{( {\bf e}_1\cdot  {\bf e}_3)( {\bf e}_2\cdot  {\bf e}_4)+( {\bf e}_1\cdot  {\bf e}_4)( {\bf e}_2\cdot  {\bf e}_3)}.
 \end{split}
\end{equation}
Finally, the correlations between Hessians and gradients read: 
\begin{equation}\label{eq:CorelationsHessianGrad}
 \begin{split}
& \Big\langle  \tonde{ {\bf e}_1 \cdot  {\bm \nabla}^2 h^a \cdot {\bf e}_2} \tonde{{\bm \nabla} h^b \cdot  {\bf e}_3 } \Big\rangle_c
 =\\
 &p(p-1)(p-2) ({\bm \sigma}^a \cdot {\bm \sigma}^b)^{p-3} ({\bf e}_1 \cdot {\bm \sigma}^b) ({\bf e}_2 \cdot {\bm \sigma}^b)  ({\bf e}_3\cdot {\bm \sigma}^a)+ \\
 & p(p-1) ({\bm \sigma}^a \cdot {\bm \sigma}^b)^{p-2} ({\bf e}_1 \cdot {\bf e}_3) ({\bf e}_2\cdot {\bm \sigma}^b)+ \\
 & p(p-1) ({\bm \sigma}^a \cdot {\bm \sigma}^b)^{p-2}({\bf e}_2 \cdot {\bf e}_3) ({\bf e}_1 \cdot {\bm \sigma}^b) .\\
 \end{split}
\end{equation}
Consider first the case of a single replica: choosing ${\bf e}_i \to {\bf e}_\alpha \quadre{ {\bm \sigma}}$ to be vectors in the tangent plane, using \eqref{eq:relHessians} and ${\bf e}_\alpha \quadre{ {\bm \sigma}} \cdot {\bm \sigma}=0$ one sees that ${\bf g}\quadre{ {\bm \sigma}}$ is uncorrelated from $\mathcal{H}\quadre{ {\bm \sigma}}$ and $h\quadre{ {\bm \sigma}}$; moreover, irrespectively of the choice of the basis in the tangent plane, the components of the gradient are independent Gaussian variables with variance $p$, while the Hessian is a GOE matrix with variance $p(p-1)$, shifted by a random diagonal matrix.\\
For more than one replica, correlations arise because of the non-zero overlaps between some directions ${\bf e}_\alpha^a$ in the tangent plane at ${\bm \sigma}^a$ and the other replicas ${\bm \sigma}^b$. However, the correlations of the components along directions that are orthogonal to ${\bf w}_0$ and to all the ${\bm \sigma}^a$ hugely simplify. To exploit this, it is convenient to separate the $N$-dimensional space embedding the sphere into the $(n+1)$-dimensional subspace $S$ spanned by the vectors ${\bf w}_0$ and  $\grafe{{\bm \sigma}^a}_{a=1}^n$, and its orthogonal complement $S^\perp$. The reference frame of the embedding space, which we denote with $\grafe{{\bf x}_i}_{i=1}^N$, can be chosen in such a way that the last $(n+1)$ vectors ${\bf x}_{N-n} \cdots, {\bf x}_N$ are a linear combination of ${\bf w_0}$ and of all the ${\bm \sigma}^a$, forming an orthonormal basis of $S$, while the remaining $N-n-1$ vectors ${\bf x}_1, \cdots, {\bf x}_{N-n-1}$ generate $S^\perp$. 
Similarly, the basis vectors in the tangent planes ${\bf e}_{\alpha}^a$ can be chosen so that the last $n$ vectors ${\bf e}_{N-n}^a, \cdots, {\bf e}^a_{N-1}$, together with the normal direction ${\bm \sigma}^{a}$, are a basis for $S$, while the remaining $ {\bf e}_{\alpha}^a$ with $\alpha <N-n$ generate $S^\perp$. In particular, these can be chosen equal for any $a$, as ${\bf e}^a_{\alpha}=\delta_{\alpha, i}\, {\bf x}_i$ for $i <N-n$. With this choice, the covariances between the first $N-n-1$ components of the gradients do not depend on the corresponding directions ${\bf e}^a_{\alpha}$, and depend trivially on the overlaps $q_{ab}$. The covariances between the last components are instead more complicated functions of $q_{ab}$, which depend explicitly on the choice of the basis in $S$. Optimal choices for the basis can be made, to simplify the calculations; we discuss an example in Appendix \ref{app:Hessian}. Regardless of these choices,  Eq. \eqref{eq:Full} can be rewritten in terms of the overlaps alone, as:
\begin{equation}\label{eq:Full3}
\langle{\mathcal{N}^n_{N}(\epsilon, \overline{q})}\rangle \hspace{-0.1 cm}=\hspace{-0.1 cm}  \int \hspace{-0.15 cm}\prod_{a < b=1}^n \hspace{-0.1 cm} dQ_{ab} \, V\tonde{\hat{Q}, \overline{q}}
 \mathcal{E}\tonde{\epsilon, \overline{q}; \hat{Q}} p_{\hat{Q}}({\bf 0}, \epsilon),
\end{equation}
where $\mathcal{E}$ and $p_{\hat{Q}}$ are the expectation value and the joint distribution in Eq.\eqref{eq:Full}, now expressed as a function of the overlap matrix $\hat{Q}$ with components 
\begin{equation}
 Q_{ab}=\delta_{ab}+(1-\delta_{ab})q_{ab},
\end{equation}
while  
\begin{equation}
 V\tonde{\hat{Q}, \overline{q}}\hspace{-0.1 cm}=\hspace{-0.1 cm}\int \hspace{-0.1 cm}\prod_{a=1}^n d{\bm \sigma}^a \delta \tonde{{\bm \sigma}^a\hspace{-0.1 cm} \cdot\hspace{-0.08 cm} {\bf w}_0- \overline{q}}  \prod_{a \leq b} \delta \tonde{Q_{ab}-{\bm \sigma}^{a}\hspace{-0.1 cm}\cdot \hspace{-0.08 cm}{\bm \sigma^{b}}}. \\
\end{equation}
is an entropic contribution. 
  We determine the leading order term in $N$ of each of the three contributions in \eqref{eq:Full3} for $q_{ab} \equiv q$, and subsequently perform the integral with the saddle point method. To simplify the calculation, we choose the bases ${\bf x}_i$ and ${\bf e}_\alpha^a$ so that only one vector has a non-zero overlap with the special direction ${\bf w_0}$: this can be done setting ${\bf x}_N={\bf w_0}$ (hence the name \emph{North Pole}), and choosing ${\bf e}_{N-1}^a$ to be the projection of $ {\bf w_0}$ on the tangent plane of ${\bm \sigma}^a$, ${\bf e}_{N-1}^a=\tonde{{\bf w_0}- \overline{q} {\bm \sigma}^a}/\sqrt{1- \overline{q}^2}$. 
 
 \subsection{The phase space factor: $V\tonde{\hat{Q}, \overline{q}}$}\label{sec:PhaseSpaceFactor}
 \noindent
 The term $V(\hat{Q}, \overline{q})$ is a phase space factor, which accounts for the multiplicity of configurations of replicas satisfying the constraints on the overlap. Its large-$N$ limit can be obtained from the representation:
 \begin{equation*}
  V(\hat{Q}, \overline{q})\hspace{-0.1 cm}=\hspace{-0.1 cm} \int_{-\infty}^\infty \prod_{a,i} d\sigma^a_i \int\hspace{-0.1 cm} \prod_{b \leq c}\frac{d \lambda_{bc}}{2 \pi}\prod_d \frac{d \mu_d}{2 \pi} e^{v( \hat{Q}, \overline{q}; {\bm \mu}, \hat{\Lambda}, \vec{\bm \sigma})},
 \end{equation*}
where $\hat{\Lambda}$ and $\hat{Q}$  are $n \times n$ matrices in replica space with elements $\Lambda_{ab}= (1+\delta_{ab})\lambda_{ab}$ and $Q_{ab}= (1-\delta_{ab})q_{ab}+ \delta_{ab}$, and  
$v( \hat{Q}, \overline{q}; {\bm \mu}, \hat{\Lambda}, \vec{\bm \sigma})=\sum_{a \leq b}i \lambda_{ab}\tonde{{\bm \sigma}^a \cdot {\bm \sigma}^b- q_{ab}}+ \sum_a i \mu_{a}\tonde{{\bm \sigma}^a \cdot {\bf w}_0- \overline{q}} $. Performing the Gaussian integrals over the variables $\sigma^a_i$ and $\mu_a$ we get:
\begin{equation*}
  V(\hat{Q}, \overline{q})= e^{o(N)} (2 \pi)^{\frac{Nn}{2}} \int \prod_{a \leq b}d \lambda_{ab}  \frac{e^{\frac{1}{2} \text{Tr} \quadre{(-i \hat{\Lambda}) \tilde{Q}}}}{\tonde{\text{det}\quadre{-i \hat{\Lambda}}}^\frac{N}{2}},
 \end{equation*}
where $\tilde{Q}_{ab}= Q_{ab}- \overline{q}^2$. After rescaling $-i \hat{\Lambda} \to N \hat{\Lambda'}$, the remaining integral can be computed with a saddle point (which gives $\Lambda'=\tilde{Q}^{-1}$), leading to: 
\begin{equation*}\label{eq:PhaseSpaceFiniten}
\begin{split}
 V(\hat{Q}, \overline{q})&=\text{exp}\grafe{\frac{N}{2} \quadre{n \log \tonde{\frac{2 \pi e}{N}}+ \log \text{det} \tilde{Q} }+ o(N)},
  \end{split}
\end{equation*}
where within the RS ansatz:
\begin{equation*}
 \log \text{det} \tilde{Q}= n \log (1-q)+  \log \tonde{\frac{1- n \overline{q}^2+(n-1)q}{1-q}}.
\end{equation*}
To leading order in $Nn$, we find:
\begin{equation}\label{eq:PhaseSpace}
\begin{split}
  &V(\hat{Q}, \overline{q})=e^{\frac{N n}{2} \quadre{\log \tonde{\frac{2 \pi e (1-q)}{N}} - \frac{\overline{q}^2-q}{1-q}}+ o(N n)}.
  \end{split}
\end{equation}
 This contribution is dominated by $q=\overline{q}^2$, which corresponds to configurations in which the replicas are almost independent with each others, correlated only through the constraint on $\overline{q}$ (indeed, it corresponds to replicas having zero mutual overlap in the portion of phase space that is orthogonal to the special direction ${\bf w_0}$). These configurations are the most numerous, and reproduce the phase space factor obtained in the annealed calculation (when $n=1$), since in that case:
\begin{equation*}\label{eq:PhaseSpaceAnnealed}
 \int d {\bm \sigma} \delta \tonde{{\bm \sigma} \cdot {\bf w_0}- \overline{q}}= e^{\frac{N}{2} \quadre{\log \tonde{\frac{2 \pi e}{N}}+ \log (1- \overline{q}^2)}+ o(N)}.
\end{equation*}
However, they are disfavored by the other terms in \eqref{eq:Full3}, which depend non-trivially on $q$; the competition between these terms leads to a more complicated global saddle point solution $q_{SP}$.

\subsection{Joint density of the gradients and energies and $ p_{\hat{Q}}({\bf 0}, \epsilon)$}\label{sec:JointDensity}
\noindent
We now determine the joint distribution $p_{\hat{Q}}({\bf 0}, \epsilon)$ of the $(N-1)n+ n$ components $(g^a_{\alpha}, h^a)$. This can be obtained from the joint distribution of the gradient components ${\bm \nabla} h^a_i={\bm \nabla} h[{\bm \sigma}^a] \cdot {\bf x}_i$ in the enlarged, $N$-dimensional space, whose covariances read (see Eq. \ref{eq:CovGrad}):
\begin{equation}
 {C}_{ij}^{ab}= p Q_{ab}^{p-1} \delta_{ij}+ p(p-1)Q_{ab}^{p-2} \sigma^b_i \sigma^a_j,
\end{equation}
and averages $\langle{{\bm \nabla} h^a_i}\rangle=- \delta_{i N}\, \sqrt{2 N}\, r f'_k\tonde{\overline{q}}$.  The joint density of the ${\bm \nabla} h^a_i$ is thus:
\begin{equation}\label{eq:GaussianEuclideanGrad}
 p\tonde{\grafe{{\bm \nabla} h^a}_{a=1}^n}= \frac{e^{-\frac{1}{2} \sum_{a,b=1}^n \tonde{
{\bm \nabla} h^a_S}^T \cdot [{\hat{C}}^{-1}]^{ab} \cdot \tonde{{\bm \nabla} h^b_S }}}{(2 \pi)^{\frac{n N}{2}} |\text{det} \;  \hat{C}|^{\frac{1}{2}}} ,
\end{equation}
where
$$ {\bm \nabla} h^a_S ={\bm \nabla} h^a + \sqrt{2 N} r f'_k\tonde{\overline{q}} {\bf x}_N,
$$
and where $[ \hat{C}^{-1}]^{ab}$ is the $ab$ block (in replica space) of the inverse covariance matrix, of dimension $N \times N$. Due to our choice of the reference frame ${\bf x}_i$, each  $\hat{C}^{ab}$ is block-diagonal, $\hat{C}^{ab}= \text{diag}(\hat{A}^{ab}, {\hat{B}}^{ab})$, where $\hat{A}^{ab}$ is an $(N-n-1)\times(N-n-1)$ block with components $\hat{A}_{ij}^{ab}= \delta_{ij}[p \delta_{ab} + p (1-\delta_{ab}) q^{p-1}]$ giving the covariances between the gradients components in $S^\perp$, while ${\hat{B}^{ab}}$ is an $(n+1) \times (n+1)$ block whose elements are the covariances of the gradients components in $S$, which are explicit functions of $q$ and of $\sigma_i^a$. To leading order in $N$ this smaller block can be neglected for the computation of the normalization, and one gets: 
\begin{equation}\label{eq:DeterLow}
\begin{split}
 &|\text{det} \; \hat{C}|=\\
 &\text{exp}\grafe{N n \tonde{\log \quadre{p(1-q^{p-1})} + \frac{q^{p-1}}{1-q^{p-1}}}+ o(Nn)}.\\
 \end{split}
\end{equation}
To get the statistics of the components on the tangent planes, we consider the $N$-dimensional vectors ${\bf \tilde{g}}\quadre{{\bm \sigma}^a}\equiv{\bf \tilde{g}}^a= (g^a_{1}, g^a_{2}, \cdots, g^a_{N-1},\tilde{g}_N^a)$, whose first $N-1$ components are $ g^a_{\alpha}= {\bm \nabla} h[{\bm \sigma}^a] \cdot {\bf e}^a_\alpha$, while the last component $\tilde{g}_N\quadre{{\bm  \sigma}^a} = \tilde{g}_N^a ={\bm \nabla} h[{\bm \sigma}^a] \cdot {\bm \sigma}^a$ equals to:
\begin{equation}\label{eq:RadialDirection}
\tilde{g}_N^a = p\; h[{\bm \sigma}^a]+  \sqrt{2 N} r \quadre{p f_k(\overline{q})- f'_k(\overline{q})\overline{q}},
\end{equation}
and it is thus related to the value of the functional at the point ${\bm \sigma}^a$. The vectors ${\bm \nabla} h[{\bm \sigma}^a]$ and ${\bf \tilde{g}}^a$ are related by a unitary rotation: the joint density $p_{\hat{Q}}({\bf 0},\epsilon)$ of $({\bf g}^a, h[{\bm \sigma}^a])$ evaluated at $({\bf 0}, \sqrt{2 N}\, \epsilon)$ is easily obtained from \eqref{eq:GaussianEuclideanGrad} with a change of variables.  \\
To determine \eqref{eq:GaussianEuclideanGrad}, we introduce the $N n$-dimensional vectors $ {\bm \xi}_1= \tonde{{\bm \sigma}^1, \cdots, {\bm \sigma}^n}$ and ${\bm \xi}_2= \tonde{{\bf w_0}, \cdots,{\bf w_0}}$ (we remind that we chose ${\bf x}_N = {\bf w_0}$), so that:
\begin{equation*}\label{eq:QuadExpanded0}
 p_{\hat{Q}}({\bf 0},\epsilon)= \frac{ e^{-N Q_{p,k,r}^{(n)}(\epsilon,  \overline{q},q)+ o(N)}}{(2 \pi)^{\frac{n N}{2}} |\text{det} \;  \hat{C}|^{\frac{1}{2}}},
\end{equation*}
with
\begin{equation}\label{eq:quad1}
\begin{split}
 &Q_{p,k,r}^{(n)}(\epsilon,  \overline{q},q)=u^2\; {\bm \xi}_1 \hspace{-0.05 cm}\cdot\hspace{-0.05 cm} \hat{C}^{-1} \hspace{-0.1 cm}\cdot \hspace{-0.05 cm}{\bm \xi}_1+[r f'_k(\overline{q})]^2 \; {\bm \xi}_2 \cdot \hat{C}^{-1}\hspace{-0.05 cm}\cdot {\bm \xi}_2\\
 & +r f_k'(\overline{q}) \,u\, \tonde{{\bm \xi}_1 \hspace{-0.05 cm}\cdot\hspace{-0.05 cm}\hat{C}^{-1} \hspace{-0.08 cm}\cdot{\bm \xi}_2+ {\bm \xi}_2\hspace{-0.05 cm}\cdot \hspace{-0.05 cm}\hat{C}^{-1}\hspace{-0.05 cm} \cdot\hspace{-0.05 cm}{\bm \xi}_1},
 \end{split}
 \end{equation}
and 
\begin{equation}\label{eq:u}
 u=u(\epsilon, \overline{q})=p  \epsilon+  r \quadre{p f_k(\overline{q})- f'_k(\overline{q})\overline{q}}.
\end{equation}
 Note that in \eqref{eq:quad1} the matrix $\hat{C}^{-1}$ is contracted with the vectors ${\bm \sigma}^a$, so that the quadratic form depends only on the overlaps $q, \overline{q}$. The exponent \eqref{eq:quad1} can be explicitly computed noticing that the vectors ${\bm \xi}_1$ and $ {\bm \xi}_2$, together with the vector ${\bm \xi}_3= \tonde{\sum_{a \neq 1}{\bm \sigma}^a, \cdots, \sum_{a \neq n}{\bm \sigma}^a}$, form a closed set under the action of the matrix $\hat{C}^{-1}$: the inversion of the correlation matrix can be performed in the restricted subspace spanned by these three vectors, and the matrix elements of $\hat{C}^{-1}$ within this subspace suffice to get \eqref{eq:quad1}. We refer to the Appendix \ref{app:QuadraticForm} for the details of this computation. As a result, we obtain 
\begin{equation*} \label{eq:FullGenn}
\begin{split}
 &Q_{p,k}^{(n)}(\epsilon,  \overline{q},q)=n r^2 \overline{q}^{2 k-2}\left(1-\overline{q}^2\right) \frac{ 1}{p \left(1+(n-1) q^{p-1}\right)}\\
 &+n \left(\frac{r \overline{q}^k}{k}+ \epsilon\right)^2\frac{q^p-q^2-p q^p(1-q)(1+(n-1)q)}{D(q)}\\
 &+2n (n-1) r \overline{q}^k  \left(\frac{r \overline{q}^k}{k}+\epsilon\right) \frac{q^{p+1} (1-q)   }{D(q)}\\
 &-n(n-1)r^2 \overline{q}^{2 k}\frac{(1-q)   q^{p} \quadre{1+(n-1) q^p-p (1-q)} }{p \left(1+(n-1) q^{p-1}\right) D(q)},\\
   \end{split}
\end{equation*}
where $D(q)$ is given in \eqref{eq:D(q)}. 
In the limit of a single replica $n \to 1$, the quadratic form reduces to: 
\begin{equation}\label{eq:Qannealed}
  Q_{p,k}^{(1)}(\epsilon,  \overline{q})=\tonde{\epsilon + r \frac{\overline{q}^k}{k}}^2+ \frac{1}{p} \tonde{r \overline{q}^{k-1} \sqrt{1-\overline{q}^2}}^2,
\end{equation}
which is consistent with \eqref{eq:AvGrad}, as it reflects the factorization of the distribution of the gradients and of the rescaled energy fields: the first term in \eqref{eq:Qannealed} corresponds to the Gaussian weight of the energy functional $h \quadre{{\bm \sigma}^a}$, while the second accounts for the non-zero average of the last component of the vector ${\bf g}^a$ (here we used that ${\bf e}_{N-1}^a=\tonde{{\bf x}_N- \overline{q} {\bm \sigma}^a}/\sqrt{1- \overline{q}^2}$). \\
To leading order in $n$, setting $ Q_{p,k}^{(n)}
   \equiv n\, Q_{p,k} + O( n^2)$, we obtain 
\begin{equation}\label{eq:QuadExpanded}
\begin{split}
 & Q_{p,k}(\epsilon,  \overline{q},q)=
 \frac{r^2 \overline{q}^{2k}}{p} \frac{(1-q^p) q^2 }{(q^p-q^2)(1-q^p)-p q^p(1-q)^2}\\
 &+\tonde{\frac{r \overline{q}^k}{k}+ \epsilon}^2\frac{q^p-q^2-p q^p(1-q)^2}{(q^p-q^2)(1-q^p)-p q^p(1-q)^2}\\
  &-2 r \overline{q}^k \tonde{\frac{r \overline{q}^k}{k}+ \epsilon}\frac{q^{p+1}(1-q)}{(q^p-q^2)(1-q^p)-p q^p(1-q)^2}\\
  &+\frac{r^2 \overline{q}^{2k-2}}{p (1-q^{p-1})},\\
     \end{split}
\end{equation}
and combining \eqref{eq:GaussianEuclideanGrad}, \eqref{eq:DeterLow} and \eqref{eq:QuadExpanded} we get \begin{equation}\label{eq:FinalGradientsDistribution}
 p_{\hat{Q}}({\bf 0},\epsilon)= \frac{e^{-\frac{N n}{2} \tonde{\frac{ q^{p-1}}{1-q^{p-1}}+ 2 \,Q_{p,k,r}(\epsilon,  \overline{q},q)} + o(Nn)}}{\quadre{2 \pi p (1-q^{p-1})}^{\frac{Nn}{2}}}.
\end{equation}
This term is dominated by an energy dependent value of $q= q(\epsilon, \overline{q})$. As we argue in the following section, to leading order in $N$ the expectation value $\mathcal{E}$ turns out to be independent on $q$, so that \eqref{eq:FinalGradientsDistribution} is the term responsible for shifting~\cite{footnote5} the saddle-point solution away from the value $q= \overline{q}^2$ maximizing the phase space term \eqref{eq:PhaseSpace}. 

\subsection{The expectation value of the product of determinants $\mathcal{E}$}\label{sec:Determinant}
\noindent
The expectation value $\mathcal{E}$ in \eqref{eq:Full3} is over the joint distribution of the Hessian matrices $\mathcal{H}^a$, conditioned on a particular value of the gradients ${\bf g}^a$ and field $h^a$. Using the identities \eqref{eq:relHessians} and \eqref{eq:RadialDirection}
we get that the Hessians can be written as:
\begin{equation}\label{eq:RelHess2}
\begin{split}
\mathcal{H}^a_{\alpha \beta}
 &\equiv  \mathcal{M}_{\alpha \beta}^a+ \theta_{r,k}(\overline{q})\;  \tonde{ {\bf e}^a_{\alpha} \cdot {\bf w}_0}\tonde{ {\bf e}^a_{\beta} \cdot {\bf w}_0}
 -\tilde{g}_N^a \, \delta_{\alpha, \beta}.
 \end{split}
\end{equation}
Here $
 \theta_{r,k}(\overline{q})= -\sqrt{2 N}\, r f''_k (\overline{q})=- \sqrt{2 N}\, r f''_k (\overline{q})$
is a deterministic term, while $\mathcal{M}^a_{\alpha \beta}\equiv 
  {\bf e}_\alpha^a \cdot \nabla^2 h_{ps}^a\cdot  {\bf e}_\beta^a$, and $\tilde{g}^a_N$ are random variables.  When conditioning to $h^a = \sqrt{2 N} \epsilon$, the random variable $\tilde{g}^a_N$ becomes a deterministic function equal to $\sqrt{2 N} u(\epsilon, \overline{q})$, see \eqref{eq:u}, so that the conditional law of $\mathcal{H}^a$ can be easily obtained from the one of the matrices $\mathcal{M}^a$. In the following, we discuss the conditional law of $\mathcal{M}^a$. We denote with $\tilde{\mathcal{M}}^a$ the random matrices obeying this law, and similarly for ${\tilde{\mathcal{H}}}^a$.\\
As we show below, the exponential scaling of $\mathcal{E}$ is determined by the leading order term (in $N$) of the density of states of the conditioned matrices $\tilde{\mathcal{H}}^a/\sqrt{N}$. This is simply the density of states of $\tilde{\mathcal{M}}^a/\sqrt{N}$, shifted by the constant term $\sqrt{2} u(\epsilon, \overline{q})$: indeed, the term proportional to $\theta_{r,k}(\overline{q})$ is a rank-1 perturbation that modifies the density of states only to lower order in $N$, and does not affect the result for $\mathcal{E}$ to exponential accuracy in $N$. Notice however that despite this term is irrelevant in the computation of the number of stationary points, it has to be taken into account when discussing their stability, see Sec. \ref{sec:Stability} and Appendix \ref{app:ThirdOrderEquations}. 
  
\subsubsection{Conditional law of the Hessians}\label{sec:ConditionalLaw}
\noindent
Consider first a single matrix $\mathcal{M}^a$: before conditioning to the values of the gradients and energy functionals, the distribution of each $\mathcal{M}^a$ is the one of a GOE matrix, with independent entries with variance $\langle [\mathcal{M}^a_{\alpha \beta}]^2 \rangle= p(p-1)(1+\delta_{\alpha \beta})$, see Eq.\eqref{eq:HessTang} (this follows from the fact that the vectors ${\bf e}_i^a$ in the tangent plane are orthogonal to ${\bm \sigma}^a$). This distribution is modified by the conditioning, as the entries of $\mathcal{M}^a$ are correlated to the gradients and energies of all the other replicas. To determine this effect, we partition each $(N-1) \times (N-1)$ matrix $\mathcal{M}^a$ into blocks, 
\begin{equation}\label{eq:BareHess}
\mathcal{M}^a=
\left(\begin{array}{c|c}
  \begin{matrix}
   &&&\\
    &&&\\
   &\mathcal{M}_0^a&&\\
   &&&\\
    &&&\\
  \end{matrix}
  & \mathcal{M}^a_{{1}/{2}} \\
      \hline 
         \tonde{\mathcal{M}^a_{1/2}}^T & \mathcal{M}^a_1
\end{array}\right),
       \end{equation}
where the larger block $\mathcal{M}_0^a$ has dimension $(N-n-1) \times (N-n-1)$, and contains the components $\mathcal{M}^a_{\alpha \beta}$ along directions $\alpha$ and $\beta$ that both belong to the subspace $S^\perp$, $\mathcal{M}_{1}^a$ is $n \times n$ and contains the components with both $\alpha$ and $\beta$ belonging to $S$, and $\mathcal{M}^a_{1/2}$ contains the remaining, mixed components. The same partitioning can be done for each gradient vector ${\bf g}^b= \tonde{{\bf g}_0^b, {\bf g}_1^b}$.\\
As we argue in Appendix \ref{app:Hessian}, the block structure \eqref{eq:BareHess} is preserved  when conditioning, meaning that correlations between components in different blocks are not induced. Moreover, from \eqref{eq:CorelationsHessianGrad} it appears that the only components that are affected by the conditioning are the ones in the blocks $\mathcal{M}_{1/2}^a$ and $\mathcal{M}_{1}^a$, which are correlated with the vector ${{\bf g}^b_{0}}$, and with the vector ${{\bf g}^b_{1}}$ and the fields $h^b$, respectively. In particular, the conditioning induces correlations between the components of $\mathcal{M}_{1/2}^a$ that belong to the same row, and between all the components in the smaller block  $\mathcal{M}_{1}^a$. 
Similarly, non-zero averages are induced for the components in the block $\mathcal{M}^a_{1}$. As a result, $\tilde{\mathcal{M}}^a$ can be written as  
\begin{equation}\label{eq:ConditionedHess}
\mathcal{M}^a\;\; \stackrel{\text{conditioning}}{\longrightarrow}  \;\;
\tilde{\mathcal{M}}^a=
\left(\begin{array}{c|c}
  \begin{matrix}
   &&&\\
    &&&\\
   &\mathcal{M}_0^a&&\\
   &&&\\
    &&&\\
  \end{matrix}
  & \tilde{\mathcal{M}}^a_{{1}/{2}} \\
      \hline 
         \tonde{\tilde{\mathcal{M}}^a_{1/2}}^T & \tilde{\mathcal{M}}^a_1
\end{array}\right),
       \end{equation}
where the blocks are independent with respect to each others, the largest one $\mathcal{M}_0^a$ has a GOE statistics, while the others have correlated entries. Such correlated entries, since their number is not $O(N^2)$, do not impact the {\it density} of eigenvalues of $\tilde{\mathcal{M}}^a$ in the large-$N$ limit, which is instead dominated by the larger block, and is thus a semicircle law, see Appendix \ref{app:Hessian}. Since, as we argue in the following subsection, the eigenvalues density is the only object needed to compute the expectation value in \eqref{eq:Full3} to leading order in $N$, it follows that the correlations induced by the conditioning can be neglected to exponential accuracy. 

\subsubsection{Factorization of the expectation value of the determinants}\label{sec:Factorization}
\noindent
As it follows from \eqref{eq:HessTang}, the Hessian matrices associated to different replicas are non-trivially correlated among each others~\cite{footnote6}, both before and after the conditioning. When computing the expectation value in \eqref{eq:Full3}, however, the correlations between the Hessians of different replicas can be neglected, since the expectation value factorizes to leading order in $N$ as we now explain. \\
Indeed, since the determinant is a linear statistics, the expectation value can be expressed~\cite{footnote7}
as a functional integral over the manifold of the eigenvalue densities $\grafe{\rho^a}_{a=1}^n$ associated to the rescaled matrices $\tilde{\mathcal{H}}^a/ \sqrt{N}$, as:
\begin{equation}\label{eq:FunctionalIntegralDensities}
 \mathcal{E}(q, \overline{q})= N^{\frac{Nn}{2}} \int \prod_a \mathcal D \rho^{a} \frac{e^{\mathcal{F}_N \tonde{\grafe{\rho^{a}}}}}{\mathcal{Z}}  e^{ N \sum_a  \int d\lambda \rho^{a}(\lambda) \log |\lambda|},
\end{equation}
where $\mathcal{Z}= \int \prod_a \mathcal D \rho^{a} \text{exp} \quadre{\mathcal{F}_N \tonde{\grafe{\rho^{a}}}}$ is a normalization. \\
The functional $\mathcal{F}_N\tonde{\grafe{\rho^{a}}}$ couples the eigenvalues densities of the different matrices. The crucial point is that the leading order term in $\mathcal{F}_N \tonde{\grafe{\rho^{a}}}$ scales as $N^2$, as it follows from the fact that the matrices $\tilde{\mathcal{H}}^a$ are shifted GOE matrices deformed by finite-rank perturbations. When computing the functional integral \eqref{eq:FunctionalIntegralDensities} with a saddle point, the saddle point solutions $\rho_{\text{sp}}^a$ are determined just by the minimization of $\mathcal{F}$. As such, they coincide with the marginals of the joint distribution in the space of measures, since
\begin{equation}
 \langle {\rho^{(a)}}\rangle= \frac{1}{\mathcal{Z}}\int \prod_a \mathcal D \rho^{(a)} e^{\mathcal{F}_N \tonde{\grafe{\rho^{(a)}}}} \rho^{(a)} ={\rho}^{(a)}_{\text{sp}}.\\
\end{equation}
This implies that to leading order in $N$, the expectation value of the product of determinants is just the product of expectation values computed with the marginal distribution of the matrices $\tilde{\mathcal{M}}^a$, properly shifted according to \eqref{eq:RelHess2}. In turn, each expectation value can be computed by means of the eigenvalues density of $\tilde{\mathcal{M}}^a$. 

\subsubsection{The GOE computation}\label{sec:GOEcomputation}
\noindent
 Given these observations, and given the symmetry between replicas, the expectation in \eqref{eq:FunctionalIntegralDensities} reduces to
 \begin{equation}
  \mathcal{E}(q)= N^{\frac{Nn}{2}} \text{exp}\grafe{N n \int d \lambda\, \rho_{\text{sp}}(\lambda) \log |\lambda| + o(Nn)},
 \end{equation}
where $\rho_{\text{sp}}(\lambda)$ is the eigenvalues density of the rescaled matrices $\tilde{\mathcal{H}}/\sqrt{N}$, which coincides with the density of $\tilde{\mathcal{M}}/ \sqrt{N}- \sqrt{2} u(\epsilon, \overline{q}) \hat{1}$. Given that the density of $\tilde{\mathcal{M}}/ \sqrt{N}$ is a semicircle law with support $[-2 \sqrt{p(p-1)}, 2 \sqrt{p(p-1)}]$, one finds
\begin{equation}\label{eq:ShiftedSemicircle}
 \rho_{sp}(\lambda)= \frac{\sqrt{4 p (p-1)- \tonde{\lambda + \sqrt{2} u(\epsilon, \overline{q})}^2}}{2 \pi p(p-1)} .
\end{equation}
From here it follows that:
\begin{equation}
\begin{split}
 \int d \lambda \, \rho_{\text{sp}}(\lambda) \log |\lambda|&=
  \frac{1}{2} \log [2 p (p-1)] + I \quadre{\beta (\epsilon, \overline{q})},
 \end{split}
\end{equation}
where $I(y)= I(-y) =\int d\mu {\sqrt{2 -\mu^2}}\log|\mu-y|/ \pi$ is given in \eqref{eq:Inty}, and $\beta(\epsilon, \overline{q})$ in \eqref{eq:beta}.
The contribution of the determinants in \eqref{eq:Full3} thus reads:
\begin{equation}\label{eq:FinalDeterminant}
 \mathcal{E}(\overline{q})= e^{\frac{Nn}{2}  \tonde{\log N + \log [2 p (p-1)] + 2 I\quadre{\beta(\epsilon, \overline{q})}}+ o(Nn)}.
\end{equation}

\subsection{Threshold energy, isolated eigenvalue and complexity of the \emph{stable} stationary points}\label{sec:Stability}
\noindent
The results obtained so far suffice to derive the explicit expression for the quenched complexity, since combining everything we get:
\begin{equation*}\label{eq:FullFinal}
\langle{\mathcal{N}^n_{N}(\epsilon, \overline{q})}\rangle= e^{\frac{N n}{2} \grafe{\log \quadre{2 e (p-1)}+ 2 I \quadre{\beta(\epsilon, \overline{q})} }} \mathcal{I}(\epsilon, \overline{q}),
\end{equation*}
where
\begin{equation*}\label{eq:IntFinal}
\mathcal{I}(\epsilon, \overline{q})= \int dq\; \text{exp}\tonde{-N n \tilde{Q}_{p,k} (\epsilon, \overline{q}, q)+ o(N n)}
\end{equation*}
has to be evaluated, in the limit of large $N$, at the saddle point $q_{\text{SP}}(\epsilon, \overline{q})$ which \emph{maximizes} $\tilde{Q}_{p,k}$, as it is prescribed by the replica method.\\
To conclude this analysis, it is necessary to address the stability of the stationary points counted by the quenched complexity. Indeed, when presenting the results of the saddle point calculation in the following section, we shall focus only on those stationary points which are local minima, meaning that we restrict to values of the parameters $\overline{q}, \epsilon$ for which the typical Hessian of the stationary points is positive-definite. \\
The density of eigenvalues of the Hessian at a typical stationary point is the one of a random matrix obeying the same law as the conditioned ones $\tilde{\mathcal{H}}$, in the limit $n \to 0$. As a matter of fact, the information on the eigenvalues density of the Hessian $\mathcal{H}[{\bm \sigma}]$ is encoded in the resolvent function:
\begin{equation}\label{eq:ResolventRandom}
 R_{{\bm \sigma}}(z) \equiv  \frac{1}{N} \text{Tr} \tonde{z-\frac{\mathcal{H}[{\bm \sigma}]}{\sqrt{N}}}^{-1},
\end{equation}
where $N$ is here the dimension of the matrix and ${\bm \sigma}$ is assumed to be a stationary point. The quantity \eqref{eq:ResolventRandom} is a fluctuating variable even for fixed realization of the random field, as it changes from stationary point to stationary point. To capture its typical behavior, we first average it over all stationary points at given $\overline{q}, \epsilon$ at fixed realization of the field, and subsequently average of the random field itself. This leads to
\begin{equation}\label{eq:ResolventAveraged}
 R(z) \equiv \left\langle \frac{1}{\mathcal{N}(\epsilon, \overline{q})}\int  d{\bm \sigma} R_{{\bm \sigma}}(z) \, \delta ({\bf g} [{\bm \sigma}])\, \chi(\overline{q}, \epsilon) \right\rangle,
\end{equation}
where $\chi(\overline{q}, \epsilon)=\delta({\bm \sigma} \cdot {\bf w_0}- \overline{q}) \delta(h [{\bm \sigma}]- \sqrt{2N}\epsilon)$ enforces the constraints on the overlap with the signal and on the energy density. The above average can be performed by means of the replica trick, using 
the identity $x^{-1}= \lim_{n \to 0} x^{n-1}$. Exploiting the replicated Kac-Rice formula, we get
\begin{equation}\label{eq:ResolventReplicas}
  R(z)= \lim_{n \to 0} \int \prod_{a=1}^n d{\bm \sigma}^a \, \delta \tonde{{\bm \sigma}^a \cdot {\bf w}_0- \overline{q}} \mathcal{F}_{\vec{\bm \sigma}}(\epsilon, z)\, p_{\vec{\bm \sigma}}({\bf 0}, \epsilon),
 \end{equation}
 where now
 \begin{equation}
  \mathcal{F}_{\vec{\bm \sigma}}(\epsilon, z)=  \Big\langle \tonde{\prod_{a=1}^n \left| \text{det}\; \mathcal{H}^a\right|} R_{{\bm \sigma}^1}(z)  \Big|  
  \grafe{
    \begin{subarray}{l} h^a = \sqrt{2 N}\epsilon,\\
   {\bm g}^a={\bf 0} \; \forall  a\end{subarray}}\Big\rangle.
 \end{equation}
Proceeding as before, and using that the resolvent of the Hessian at a stationary point ${\bm \sigma}^1$ is a function only of the eigenvalue density $\rho^1(\lambda)$, we find that \eqref{eq:ResolventReplicas} can be evaluated with the same saddle-point calculation discussed above, and:
\begin{equation}
 R(z)=\lim_{n \to 0} \int d \lambda \frac{\rho_{sp}(\lambda)}{z-\lambda}.
\end{equation}
Here, as before, $\rho_{sp}(\lambda)$ denotes the eigenvalues density of a matrix distributed as $\tilde{\mathcal{H}}/ \sqrt{N}$, which depends on the parameters $\overline{q}, \epsilon$ as well as on the mutual overlap between replicas, evaluated at its saddle point value $q_{SP}(\epsilon, \overline{q})$. Therefore, to discuss the stability one needs to characterize in detail this density, in the limit $n \to 0$. Note that this computation is a quenched one, as it accounts for the fluctuations between the various stationary points at fixed $\overline{q}, \epsilon$. The annealed approximation would correspond to averaging separately the numerator and the denominator in \eqref{eq:ResolventAveraged} over the random field. This does not account for the correlations between the stationary points, and it is reproduced setting $n=1$ in the above expression (instead of taking $n \to 0$).\\
 To leading order in $N$, the density of states of $\tilde{\mathcal{H}}/\sqrt{N}$ is dominated by the large GOE block, and it is therefore the shifted semicircle in Eq. \eqref{eq:ShiftedSemicircle}: stationary points for which part of the support of the semicircle lies in the negative semi-axis have an extensive number $\sim O(N)$ of unstable directions in phase space. Since the location of the support of \eqref{eq:ShiftedSemicircle} depends on the energy density $\epsilon$, a \emph{threshold} energy can be defined, as the energy at which the support touches zero:
\begin{equation}\label{eq:threshold}
 \epsilon_{\text{th}}\tonde{\overline{q}, r}=- \tonde{ \sqrt{\frac{2 (p-1)}{p}} + r \tonde{\frac{p}{k}-1} \frac{ \overline{q}^k}{p}}.
\end{equation}
For fixed value of $\overline{q}$ and of the parameters $p,k,r$, stationary points having energy larger than \eqref{eq:threshold} have an Hessian with extensively many negative eigenvalues, and therefore can not be considered as trapping minima of the energy landscape. \\
The semicircle law accounts for the continuous part of the spectrum of the Hessian, and does not give information on possible isolated eigenvalues of  $\tilde{\mathcal{H}}/\sqrt{N}$, that may not belong to the support of \eqref{eq:ShiftedSemicircle}. Such eigenvalues correspond to isolated poles of the resolvent $R(z)$; they contribute to the density of states with subleading terms of order $N^{-1}$, and are thus irrelevant when computing \eqref{eq:FinalDeterminant}. However, if an isolated eigenvalue exists, it can become smaller than zero at values of $\epsilon <\epsilon_{\text{th}}(\overline{q})$, leading to the instability of the point ${\bm \sigma}$ along some direction in phase space. \\
For the matrices $\tilde{\mathcal{H}}$, isolated eigenvalues can be generated
by the subset of entries that do not belong to the large GOE block (as they are distributed with a different average and variance, induced by the conditioning to the gradients), as well as by the rank-1 perturbation proportional to $\theta_{r,k}(\overline{q})$ in~\eqref{eq:RelHess2}. As a matter of fact, for large random matrices perturbed by low-rank operators, it is known that when the perturbation exceeds a critical value, the extreme eigenvalues detach from the boundary of the support of the density of states of the unperturbed matrix. This transition is akin to the one proved in \cite{BPP} for the Wishart ensamble and known as the \emph{BBP transition}, and it has been shown to occur under quite general conditions \cite{BenaychGeorges, Edwards}.\\
To inspect whether such a transition occurs in the case under consideration, we need to characterize in more detail the law of the matrices $\tilde{\mathcal{H}}/\sqrt{N}$. This requires to explicitly compute the averages and covariances of all the entries of $\tilde{\mathcal{H}}/\sqrt{N}$ with respect to some fixed basis in the subspace $S$, see the details in Appendix~\ref{app:Hessian}.
This in turn allows to derive general equation for the isolated poles of $R(z)$ in the limit $N \to \infty$, see Appendix~\ref{app:ThirdOrderEquations}.\\
As a result, we find that the isolated eigenvalues, if any, are given in terms of the roots of a third order equation having coefficients that are functions of the parameters $\overline{q}, \epsilon$ and of the saddle point value $q_{SP}(\overline{q}, \epsilon)$. Imposing the minimal eigenvalue to be zero gives the boundary of stability of the stationary points: for fixed $r$ and $\overline{q}$, it defines a critical \emph{stability} energy $\epsilon_{\text{st}}(\overline{q}, r)$ such that the typical stationary points are local minima for $\epsilon< \epsilon_{\text{st}}(\overline{q}, r)$, are saddles of finite index for $\epsilon_{\text{st}}(\overline{q}, r)<\epsilon< \epsilon_{\text{th}}(\overline{q}, r)$, and are saddles of extensive index otherwise. In particular, for the values of parameters that we inspect, we find that in the regime $\epsilon_{\text{st}}(\overline{q}, r)<\epsilon< \epsilon_{\text{th}}(\overline{q}, r)$ there is one single isolated eigenvalue that is negative, whose eigenvector has a finite projection on the direction of ${\bf w_0}$; thus, this instability is an instability toward the signal.\\
In summary, the complexity of the \emph{stable} stationary points is therefore given by \eqref{eq:FinalComplexity}, endowed with the condition of the energy being smaller than the \emph{threshold} energy \eqref{eq:threshold} or, when the isolated eigenvalue exists, of the \emph{stability} energy $\epsilon_{\text{st}}(\overline{q}, r)$.

\subsection{Mapping between complexity at different $k$}\label{sec:mapping}
\noindent
Before presenting the results, we derive the mapping \eqref{eq:ExplicitMapping} relating the complexity for different values of $k$. \\
Suppose that ${\bm \sigma}$ is a stationary point of the functional \eqref{eq:EnegyDensity} for a fixed $k$ and for a given value of $r= r_k$ (we now make explicit the dependence on $k$ and $r$ writing $h_{k,r} \quadre{{\bm \sigma}}$), with overlap $\overline{q}$ and with energy density $\epsilon= \epsilon_k$. Then, the point ${\bm \sigma}$ is also a stationary point of the functional \eqref{eq:EnegyDensity} with $k=1$, provided that $r=r_1^{\text{eff}}$ is chosen so that:
\begin{equation}\label{eq:rEff}
 r_1^{\text{eff}}\tonde{r_k, \overline{q}}= \frac{ f'_k(\overline{q})}{f'_1(\overline{q})} r_k.
\end{equation}
In this case, ${\bm \sigma}$ has overlap $\overline{q}$ with ${\bf w_0}$, and has energy density:
\begin{equation}\label{eq:enEff}
 \epsilon_{1}^{\text{eff}}\tonde{r_k, \epsilon_k, \overline{q}}= \epsilon_k+ r_k\tonde{f_k(\overline{q}) - \frac{ f'_k(\overline{q})}{f'_1(\overline{q})}f_1(\overline{q})}.
\end{equation}
Indeed, for ${\bm \sigma}$ to be a stationary point at a fixed $k$, it must hold ${\bm \nabla}h_{k, r_k} \cdot {\bf e}_\alpha\quadre{{\bm \sigma}}=0$, which implies:
 $${\bm \nabla}h_{\text{ps}} \quadre{{\bm \sigma}} \cdot {\bf e}_\alpha\quadre{{\bm \sigma}} =
 \begin{cases}
  0  &\text{  if  } \alpha <N-1\\
  r_k f'_k(\overline{q}) \sqrt{1- \overline{q}^2} &\text{  if  } \alpha =N-1,
  \end{cases}
  $$
where we exploited our choice of bases ${\bf x}_N= {\bf w_0}$ and ${\bf e}_{N-1}\quadre{{\bm \sigma}}=\tonde{{\bf x}_N- \overline{q} {\bm \sigma}}/\sqrt{1- \overline{q}^2}$. Moreover, $h_{\text{ps}} \quadre{{\bm \sigma}}= \epsilon_k+ r_k f_k(\overline{q})$. This in turn implies that $ {\bm \nabla}h_{1,r_1} \quadre{{\bm \sigma}} \cdot {\bf e}_\alpha\quadre{{\bm \sigma}}=0$ for $\alpha< N-1$, while $ {\bm \nabla}h_{1,r_1} \quadre{{\bm \sigma}} \cdot {\bf e}_{N-1}\quadre{{\bm \sigma}}=(r_k f'_k(\overline{q})- r_1 f'_1(\overline{q})) \sqrt{1- \overline{q}^2}$. Thus, this is a stationary point for $k=1$ if $r_1$ is chosen as \eqref{eq:rEff}. In this case, it is easy to check that its energy density equals \eqref{eq:enEff}. It follows from this that the knowledge of the curves \eqref{eq:FinalComplexity} for $k=1$ is sufficient to reconstruct the curves at any larger $k$, via the mapping \eqref{eq:ExplicitMapping}. 
We however remind that the analysis of the instability of the stationary point induced by the isolated eigenvalue is strongly dependent on $k$, and thus has to be performed separately for any case.

\subsection{The results of the Kac-Rice calculation}\label{sec:Results}
\noindent
We are now ready to discuss concrete results. 
In the following we report the curves resulting from the computation of \eqref{eq:FinalComplexity}, focusing on the cases $p=3$ and $p=4$. \\
For each of the values of $k$ that we consider, we find the following general features: as long as $r<r_c$ (and, in most cases, also for $r>r_c$), there are values of $\overline{q}, \epsilon$ for which the quenched complexity is positive and the typical Hessian of the stationary points is positive definite, indicating the presence of exponentially many local minima of the energy functional. In particular, at fixed latitude $\overline{q}$ this occurs over a finite range of energies $\epsilon^*(\overline{q}) \leq \epsilon \leq \epsilon_{\text{th}}(\overline{q})$, with $\epsilon_{\text{th}}(\overline{q})$ replaced by $\epsilon_{\text{st}}(\overline{q})$ whenever the isolated eigenvalue exists. We find that $\Sigma_{p,k}( \epsilon, \overline{q})$ is monotone increasing in this energy range, implying that the most numerous stable stationary points at a given $\overline{q}$ are the ones at higher energy. At the other extreme of the support $\epsilon^*(\overline{q})$, the quenched complexity vanishes, $\Sigma_{p,k}(\epsilon^*(\overline{q}), \overline{q})=0$.\\
We denote with $\epsilon^*(r)$ the absolute minimum of the energies over all $\overline{q}$, and with $q^*(r)$ the corresponding latitude; these values coincide with the ones found solving the RSB equations in Sec.~\ref{sec:ReplicaEquations}. We use the notation $\overline{q}_{\text{num}}(r)$ for the latitude where the largest number of stationary points is found, for any fixed $r$.\\
At the transition point $r_c$ and at the latitude $\overline{q}_c$ given in \eqref{eq:FinalComplexityAnnealed}, the support of the positive part of the complexity shrinks to a single point $\epsilon^*= \epsilon_{\text{th}}=\epsilon_c$, where $\Sigma_{p,k}(\epsilon_c, \overline{q}_c)=0$.  Moreover, the whole complexity curve at this latitude coincides with the annealed one, Eq.~\eqref{eq:FinalComplexityAnnealed}. The same remains true for larger $r$: the annealed complexity is exactly zero at values of $\overline{q}^*, \epsilon^*$ which coincide with the solution of the RS limit of the saddle point equations in Sec.~\ref{sec:ReplicaEquations}, and which give the latitude and energy of an isolated minimum of the energy landscape. For $k>1$ and for some values of $r$, beyond this isolated minimum there is a residual band containing exponentially many local minima, at smaller overlap $\overline{q}$ with the signal. \\
In the following, we present in more detail the results for each of the cases presented qualitatively in Sec.~\ref{sec:SummaryResults}.
\subsubsection{Case I}
\begin{figure}[!htbp]
\includegraphics[width=\linewidth]{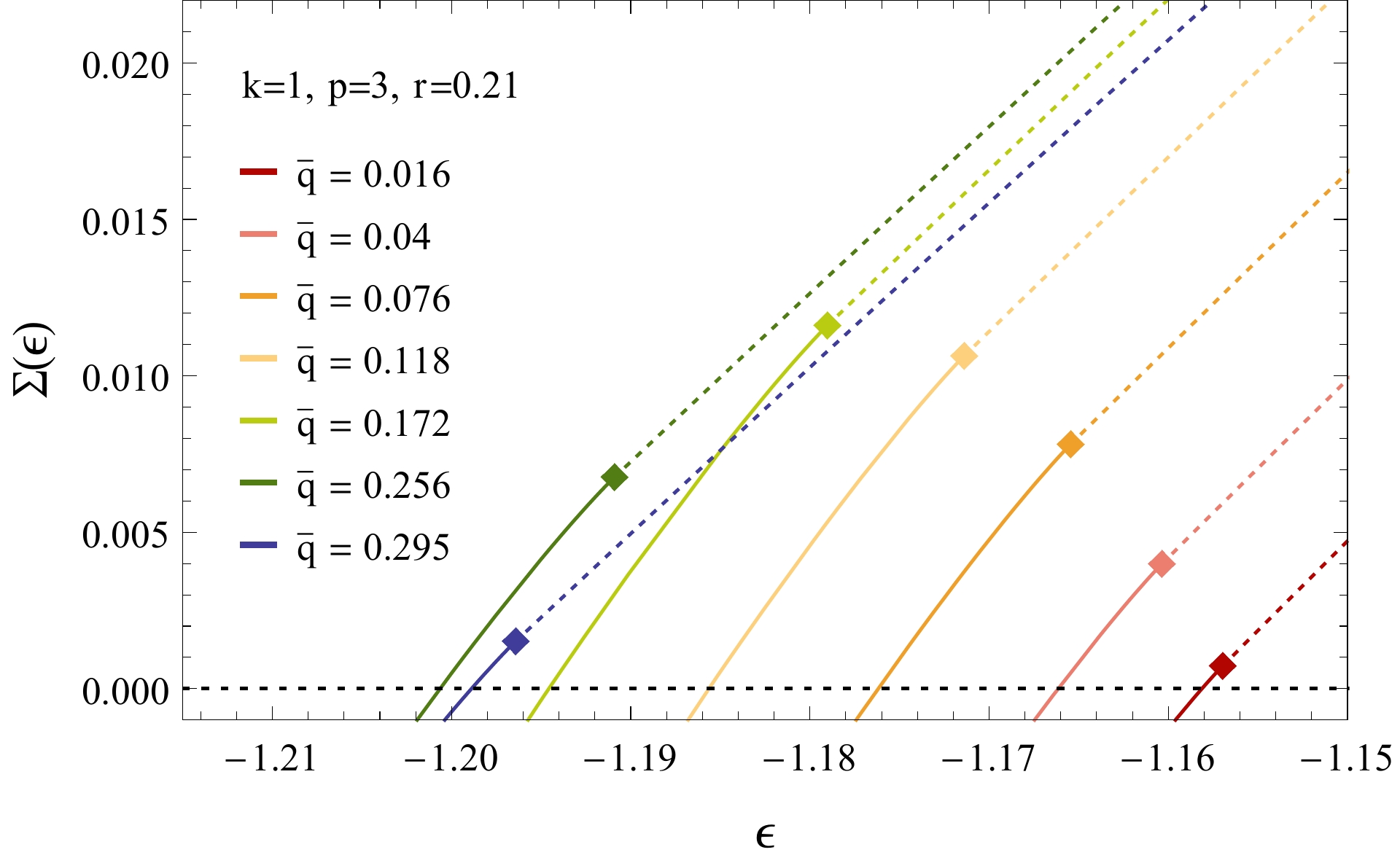}
    \caption{Complexity curves $\Sigma_{3,1}(\epsilon, \overline{q})$ as a function of the energy density $\epsilon$, for $r \approx 0.21$ and different values of $\overline{q}$. The squares denote the \emph{threshold} energy $\epsilon_{\text{th}}(\overline{q})$: the points having this energy are the most numerous \emph{stable} stationary points at the latitude $\overline{q}$. The most numerous stable states are at $\overline{q}_{\text{num}} \approx 0.16$, while the ones with smallest energy are at $\overline{q}^* \approx 0.26$. The minimal and maximal latitude are $\overline{q}_m \approx 0.01$ and $\overline{q}_M \approx 0.304$. }\label{fig:Complexitiesk1p3}
 \end{figure}

\begin{figure}[!htbp]
\subfloat[][]{\includegraphics[width=\linewidth]{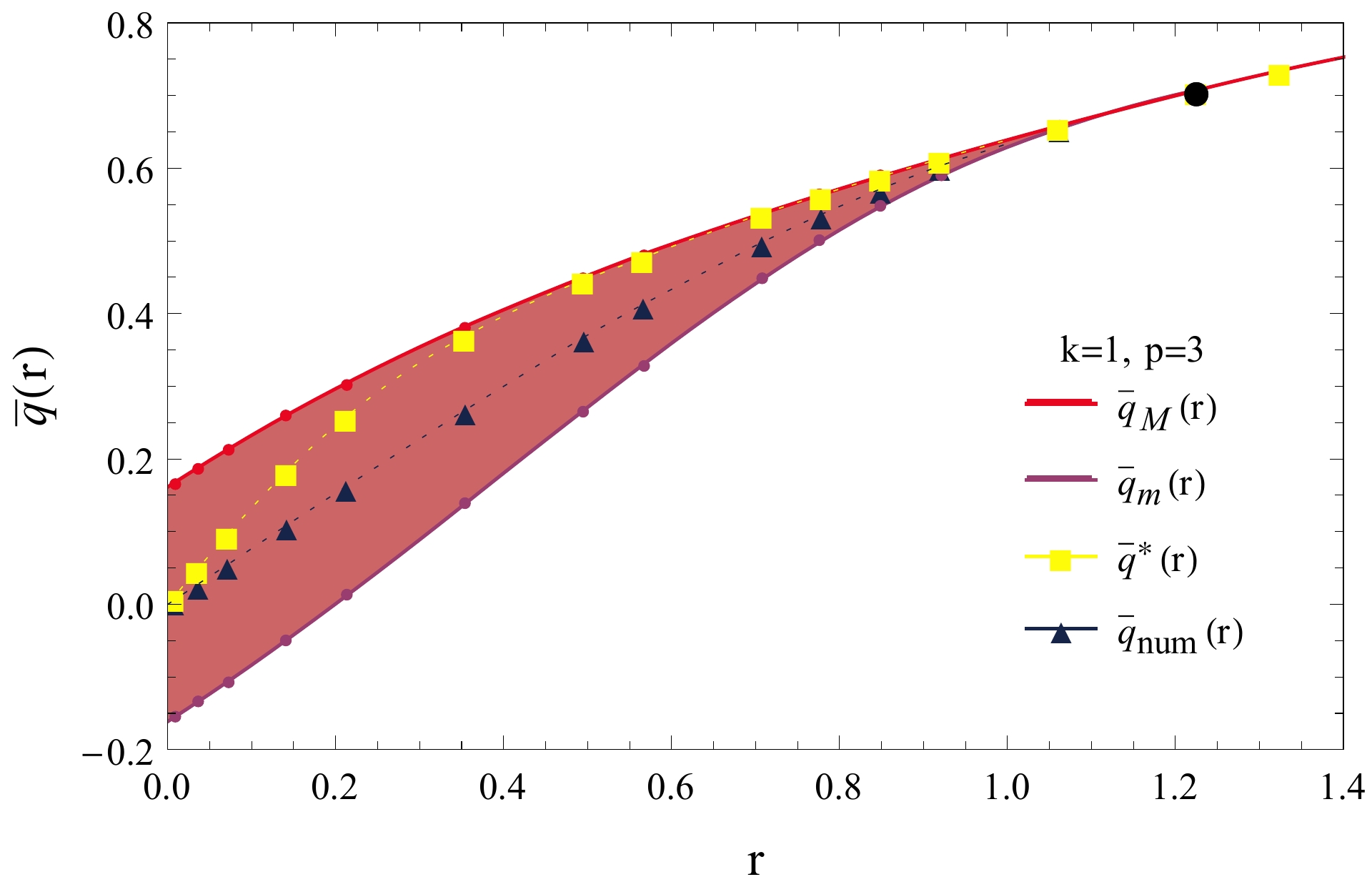}\label{fig:Bandek1p3}}\\
\subfloat[][]{ \includegraphics[width=\linewidth]
 {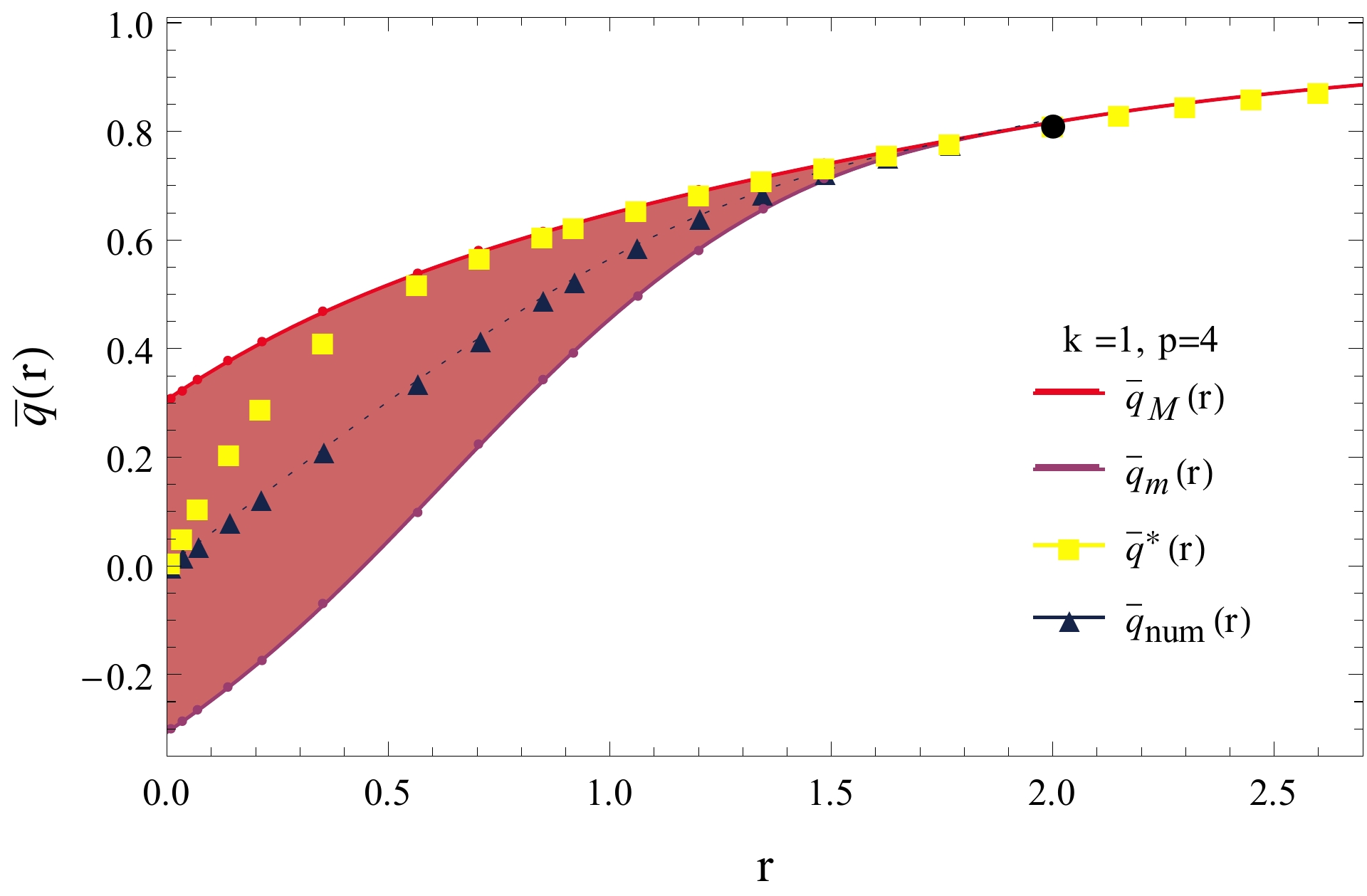}\label{fig:Bandek1p4}} \caption{The red strips denotes the latitudes where exponentially many stationary points are found, with energy smaller than the threshold energy. The points along the boundary lines are obtained numerically, while the continuous curves are interpolations. The yellow squares are the latitudes of the deepest minima, and the blue triangle the ones of the most numerous states. The bands collapse at $r_c=\sqrt{3/2} \approx 1.22$ for $p=3$, and $r_c=2$ for $p=4$, marked by the black points in the figures. The corresponding value of $\overline{q}$ is $\overline{q}_c \approx 0.71$ for $p=3$, and  $\overline{q}_c \approx 0.82$ for $p=4$. }\label{fig:Bandek1}
\end{figure}

\noindent
Instances of the complexity curves $\Sigma_{p,k}( \epsilon, \overline{q})$ in the case $k=1$, $f_1(x)=x$ are given in Fig.~\ref{fig:Complexitiesk1p3}, for $p=3$ and fixed $r<r_c$. The curves are obtained solving numerically the saddle point equations for $q$ for each value of the parameters $\overline{q}, \epsilon$.
For $k=1$ we find that there is no isolated eigenvalue exiting the bulk of the semicircle: thus, for each $\overline{q}$ the maximal energy where stable stationary points are found is $\epsilon_{\text{th}}(\overline{q})$, which is marked with the squares in Fig.~\ref{fig:Complexitiesk1p3}.\\
The curves show the following trend: below a minimum value $\overline{q}_m$, the complexity is positive only for the states which have energy above the \emph{threshold}, and are therefore unstable. At $\overline{q}_m$, the equality $\epsilon^*(\overline{q}_m,r)= \epsilon_{\text{th}}(\overline{q}_m,r)$ holds, meaning that at this latitude there are only marginally stable (and unstable) stationary points. For larger latitudes, as $\overline{q}$ increases the energy interval in which the complexity is positive (and the points are stable) gets wider and moves toward smaller energies, until the maximal width is reached at $\overline{q}_{\text{num}}$. At larger $\overline{q}$, the energy interval start shrinking, and the minimal energies  $\epsilon^*(\overline{q})$ decrease until the absolute minimum is reached at $\overline{q}^*$; for $\overline{q} > \overline{q}^*$ the trend is reversed and $\epsilon^*(\overline{q})$ starts increasing, until it collapses to $\epsilon_{\text{th}}(\overline{q})$ at $\overline{q}_M$. Analogous results are obtained for different values of $r$ below $r_c$, as well as for $p=4$.  \\
In Fig.~\ref{fig:Bandek1}, we plot the bands $\overline{q}_m(r) \leq \overline{q} \leq \overline{q}_M(r)$ containing exponentially many local minima, as a function of $r$. These bands correspond to the red ones plotted pictorially in Fig.~\ref{Fig-k1}. For each of the $\overline{q}$ within the bands, the quenched complexity behaves as in  Fig.~\ref{fig:Complexitiesk1p3}. As $r$ increases, the bands gets wider and subsequently shrink and collapse to $\overline{q}_c$ at $r=r_c(p)$, corresponding to the black points in the figures. Here the minimum becomes unique, and it is marginally stable. This landscape phase transition at $r_c$ is signaled by the fact that the saddle point solution $q_{\text{SP}}$ converges to $1$, meaning that all the replicas coincide, and that the quenched complexity becomes equal to the annealed one. This corresponds to the recovery of the RS symmetry in the replica calculation of Sec.~\ref{sec:ReplicaEquations}. 
\subsubsection{Case II}
\begin{figure*}[!htbp]
\subfloat[][]{\includegraphics[width=.5\linewidth] {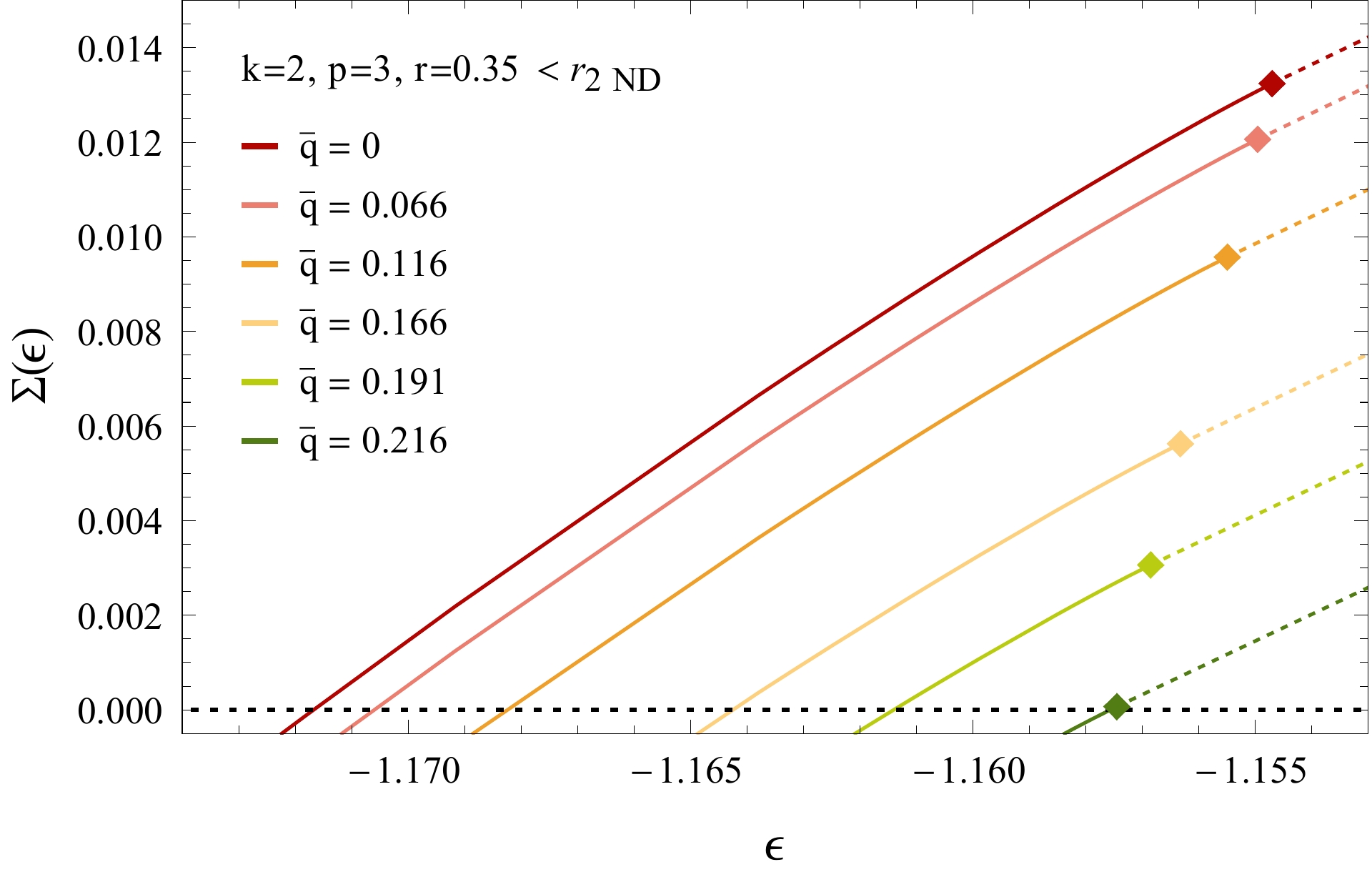}\label{fig:Complexitiesk2p3smallr}}
\subfloat[][]{\includegraphics[width=.5\linewidth] {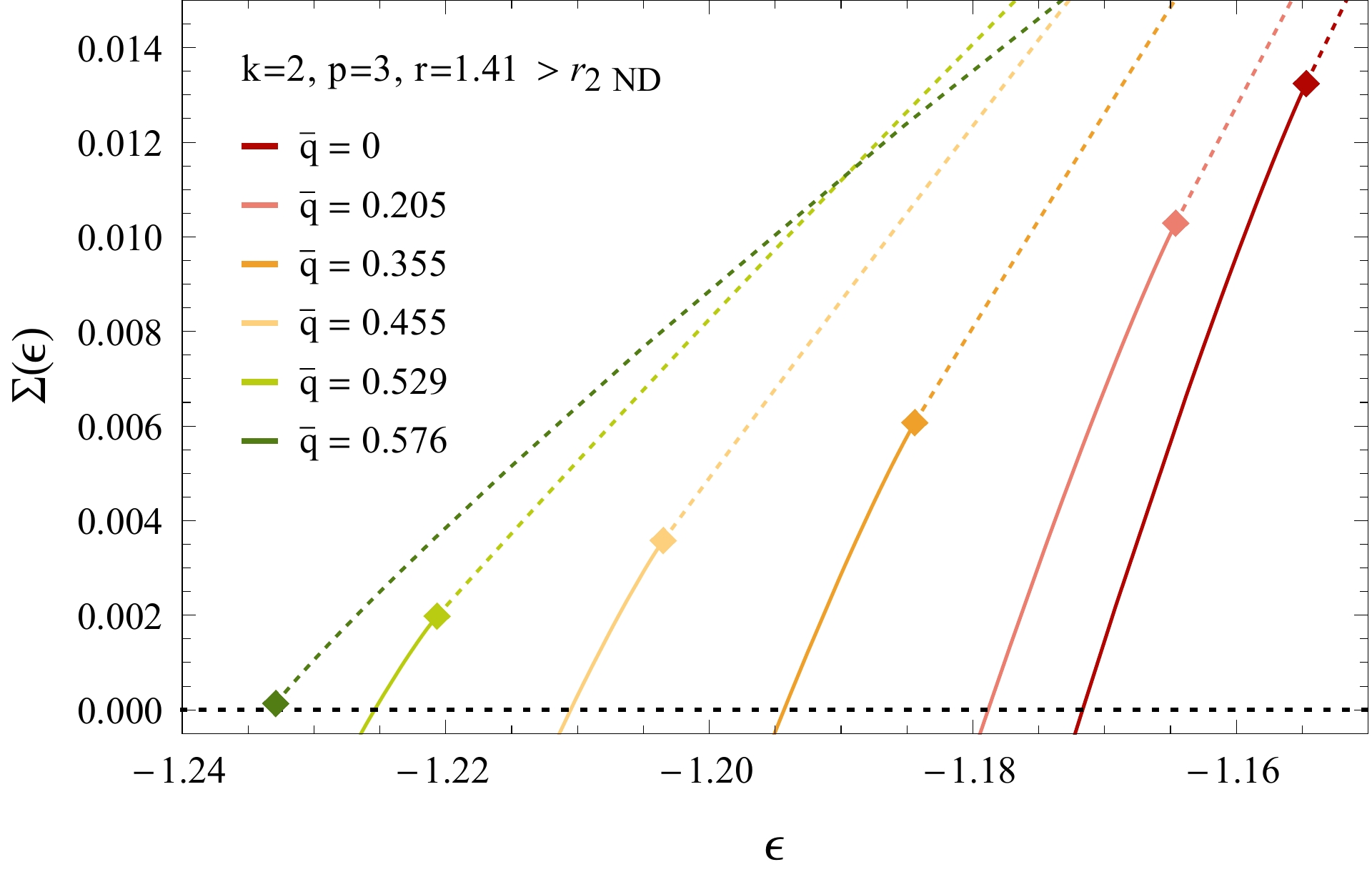}\label{fig:Complexitiesk2p3intermediater}} 
 \caption{Complexity curves $\Sigma_{3,2}(\epsilon, \overline{q})$ as a function of the energy density $\epsilon$, for different values of $\overline{q}$ and $r<r_c$. The squares along the curves mark the \emph{threshold} energy $\epsilon_{\text{th}}(\overline{q})$. (a) For $r \approx 0.35< r_{\text{2ND}}$, the minimal and maximal latitude are $\overline{q}_M =-\overline{q}_m \approx 0.22$, and both the most numerous stable states and the deepest ones are at the equator,  $\overline{q}_{\text{num}}=0=\overline{q}^*$.  (b) For $r_{\text{2ND}} <r= \sqrt{2}<r_c$, the minimal and maximal latitude are $\overline{q}_M= -\overline{q}_m \approx 0.5773$.  The most numerous stable states are at the equator, $\overline{q}_{\text{num}}=0$, while the deepest ones are at $\overline{q}^*\approx 0.5771$. 
 }\label{fig:Complexitiesk2p3}
\end{figure*}

\begin{figure*}[!htbp]
\subfloat[][]{\includegraphics[width=.5\linewidth] {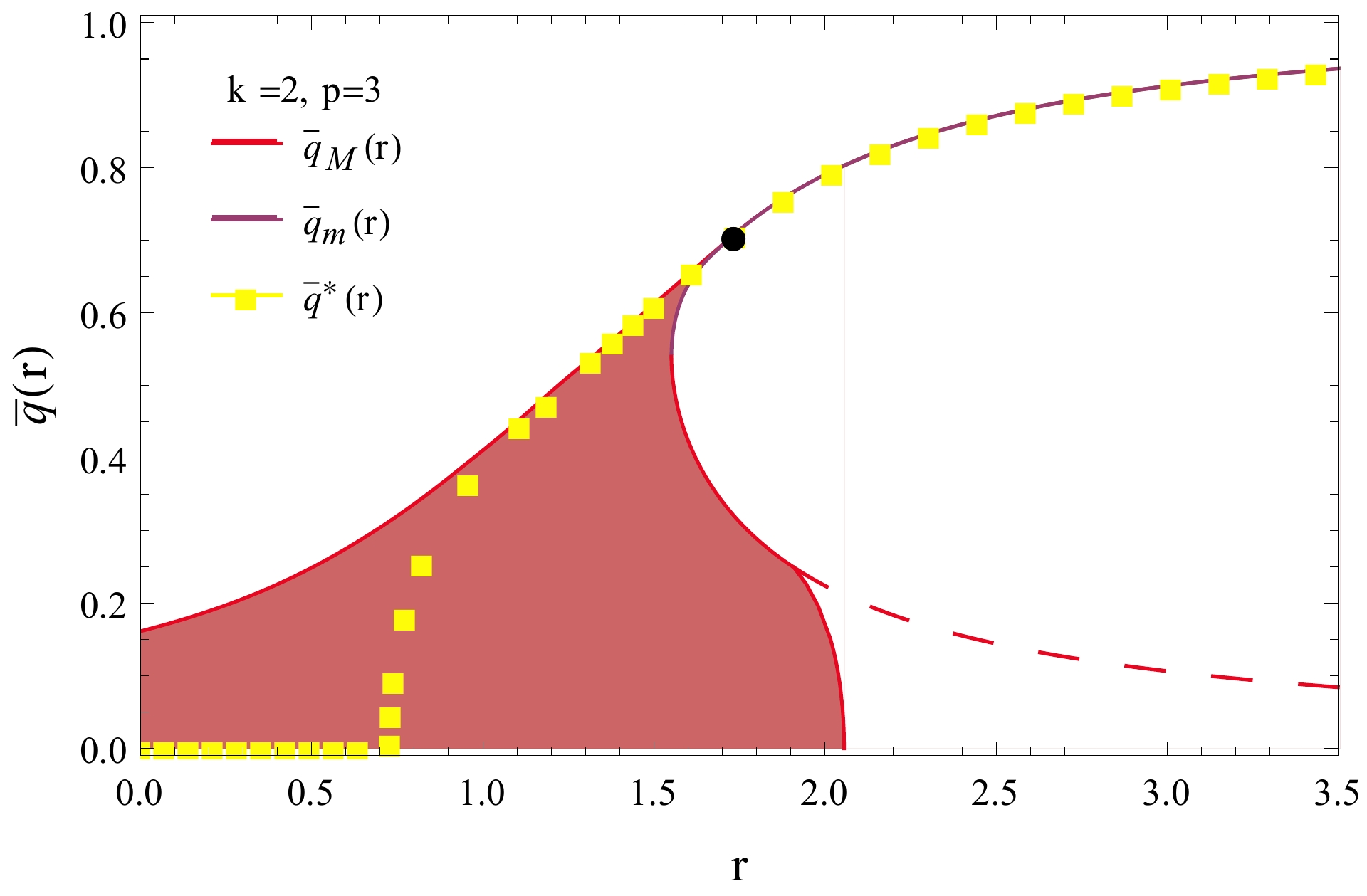}\label{fig:Bandek2p3}}
\subfloat[][]{\includegraphics[width=.5\linewidth] {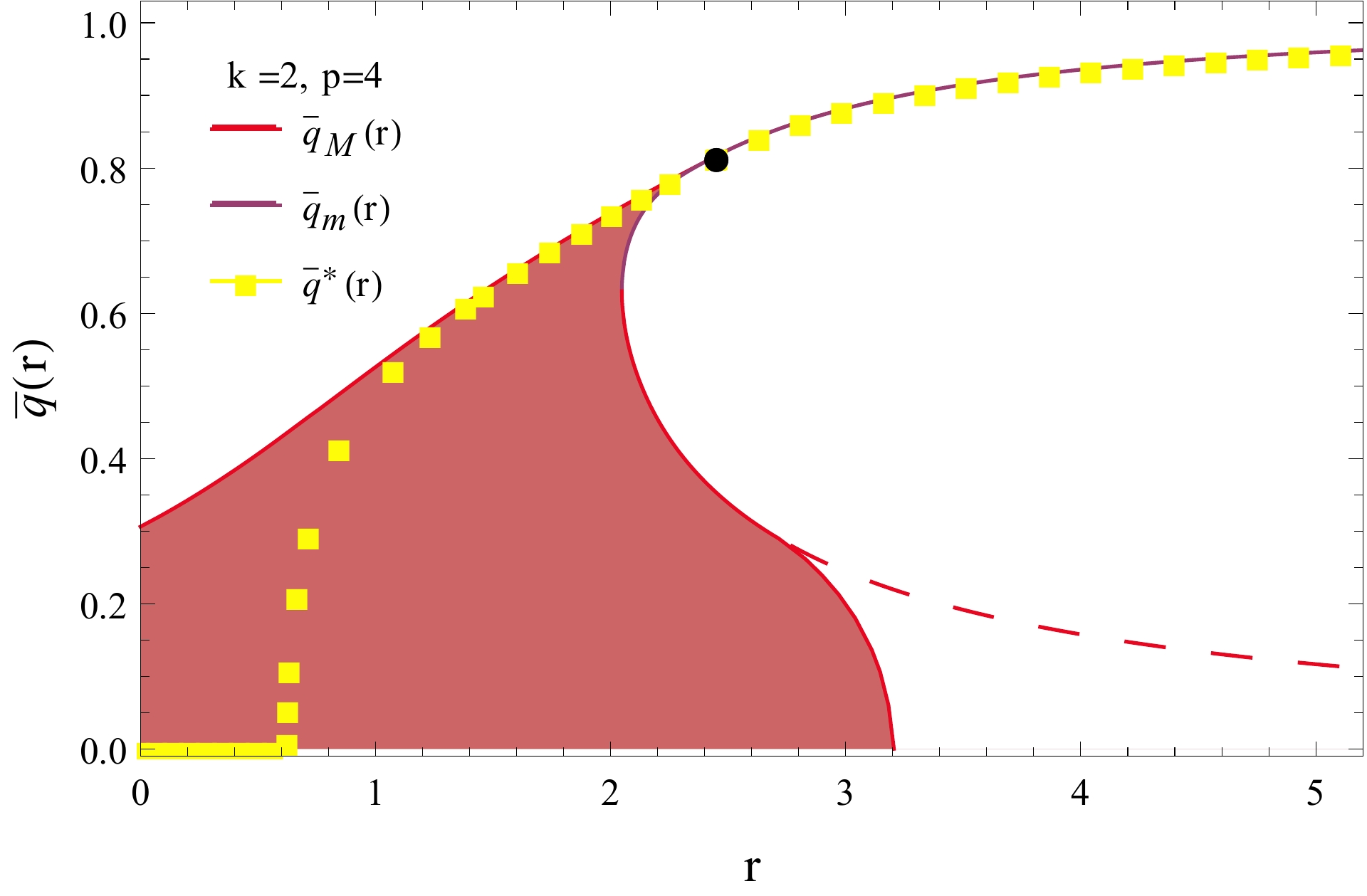}\label{fig:Bandek2p4}}
\caption{The red strip denotes the interval $ \overline{q}_m(r) \leq \overline{q} \leq \overline{q}_M(r)$ containing exponentially many stable stationary points (only positive values of $\overline{q}$ are represented, given the symmetry in $\overline{q} \to -\overline{q}$). The yellow points indicate the latitudes $\overline{q}^*(r)$ of the deepest minima, which detach from $\overline{q}=0$ at $r=r_{\text{2ND}}$ ($r_{\text{2ND}} \approx  0.73$ for $p=3$ and $r_{\text{2ND}} \approx0.62$ for $p=4$). At a higher value of $r$ ($\approx  1.55$ for $p=3$ and  $\approx 2.05$ for $p=4$), the band of states splits into two. The resulting smaller band at high overlap collapses to the RS solution at $r_c \approx 1.73$ for $p=3$ and $r_c \approx 2.45$ for $p=4$ (black point in the figure). The larger band enclosing the equator shrinks for increasing $r$, and it disappears at $r \approx 2.06$ for $p=3$, and $r \approx 3.21$ for $p=4$. The dashed line is the boundary of the band computed without accounting for the isolated eigenvalue: states below this line but outside the colored band have energy smaller than the threshold energy, but are unstable because of the isolated eigenvalue.}\label{fig:Bandek2}
\end{figure*}

\noindent
According to the analysis of Sec.~\ref{sec:ReplicaAnalysis} and of Ref.~\cite{sherrington1}, for $k=2$, $f_2(x)=x^2/2$, the minima of the energy landscape undergo a second order transition at $r= r_{\text{2ND}}< r_c$. The transition marks the boundary between two different behaviors of the complexity curves, see Fig.~\ref{fig:Complexitiesk2p3}: for $r< r_{\text{2ND}}$, the energy interval containing exponentially many states is maximally large at the equator, where both the deepest and the most numerous states lie. For $r_{\text{2ND}}<r<r_c$, instead, the most numerous states remain at the equator and have $\epsilon= \epsilon_{\text{th}}$, but the deepest states move toward a higher overlap $\overline{q}^*>0$ with the signal. At $r_{\text{2ND}}$, the states of minimal energy $\epsilon^*(r)$ detach from the equator, moving toward larger latitudes. The features of the bottom of the landscape (that is, the spectrum of the minimal energies $\epsilon^*(\overline{q}, r)$, the thermodynamic energies $\epsilon^*(r)$ and the value of $r_{2ND}$) can all be obtained from the corresponding $k=1$ curves $\epsilon^*_1(\overline{q}, r_1)$ satisfying $\Sigma_{p,1}(\epsilon_1^*(\overline{q},r_1),   \overline{q}; r_1)=0$, as we discuss in more detail in Appendix \ref{app:Results}.\\
\noindent Consider now the other transition in the energy landscape, which occurs when the strip containing exponentially many stationary points splits into three different bands (in the following, we restrict to positive values of $\overline{q}$: due to the symmetry, the landscape at negative overlap is specular to the one at positive overlap). The strip containing the stationary points for $k=2$ can be identified exploiting again the mapping \eqref{eq:ExplicitMapping},
with the caveat that the stationary points so determined are stable only in the sense of the \emph{threshold}, and the analysis of the sign of the isolated eigenvalue has to be performed separately. We give the details of the mapping in Appendix \ref{app:Results}. The resulting bands are shown in Fig.~\ref{fig:Bandek2}, where one sees that they split at $r \approx 1.55$ for $p=3$, and $r \approx 2.05$ for $p=4$. 
For $r$ larger than this splitting point, the band closer to the \emph{North Pole}, which is the one containing the deepest minima, shrinks until it collapses to a single state at the RS transition, while the band enclosing the equator, which is the one containing the most numerous minima, shrinks to zero only asymptotically (this corresponds to the dashed lines in Fig.~\ref{fig:Bandek2}). This implies that, for any value of $r$, there is a strip of finite width and small overlap $\overline{q}$ with the signal, containing exponentially many stationary point with energy smaller than the \emph{threshold}. To conclude the analysis of the landscape, it is necessary to investigate the possible instability of these points due to the presence of a negative, isolated eigenvalue. 
For $k=2$, the isolated eigenvalue exists only for sufficiently large $r$, and it renders unstable, for each $\overline{q}$ for which it exists, the stationary points at higher energy $\epsilon_{\text{st}}(\overline{q},r)< \epsilon < \epsilon_{\text{th}}(\overline{q},r)$. We refer to the Appendix~\ref{app:Results} for a more detailed analysis of this instability, and report here only its main consequences. First, if this instability is accounted for, we find that for $r>r_c$ the most numerous non-unstable points are still at $\overline{q}=0$, but are no longer marginally stable. Rather, they have an energy $\epsilon_{\text{st}}(0,r)$ smaller than the \emph{threshold} energy, and have one flat direction in their Hessian, corresponding to the isolated eigenvalue being zero. For general $\overline{q}$, as $r$ increases $\epsilon_{\text{st}}(\overline{q},r)$ decreases, until it becomes smaller than the lower bound $\epsilon^*(\overline{q},r)$, implying that all the points at the given latitude are unstable because of a single negative eigenvalue (See Fig. \ref{fig:Isolatedk2p3} in Appendix \ref{app:Results}). This happens first for the larger values of $\overline{q}$ belonging to the band: thus, the  band of those stationary points gets narrower around the equator, from above. At a finite value of $r$ ($r \approx 2.06$ for $p=3$ and $r \approx 3.21$ for $p=4$), also the last stationary points at the equator become unstable (this value of $r$ can be computed within the annealed approximation, see the comments at the end of Appendix \ref{app:ThirdOrderEquations}). For larger $r$, there is a unique stable minimum, that is the minimum of the annealed complexity. 
\subsubsection{Case III}
\begin{figure*}[!htbp]
\subfloat[][]{\includegraphics[width=.5\linewidth] {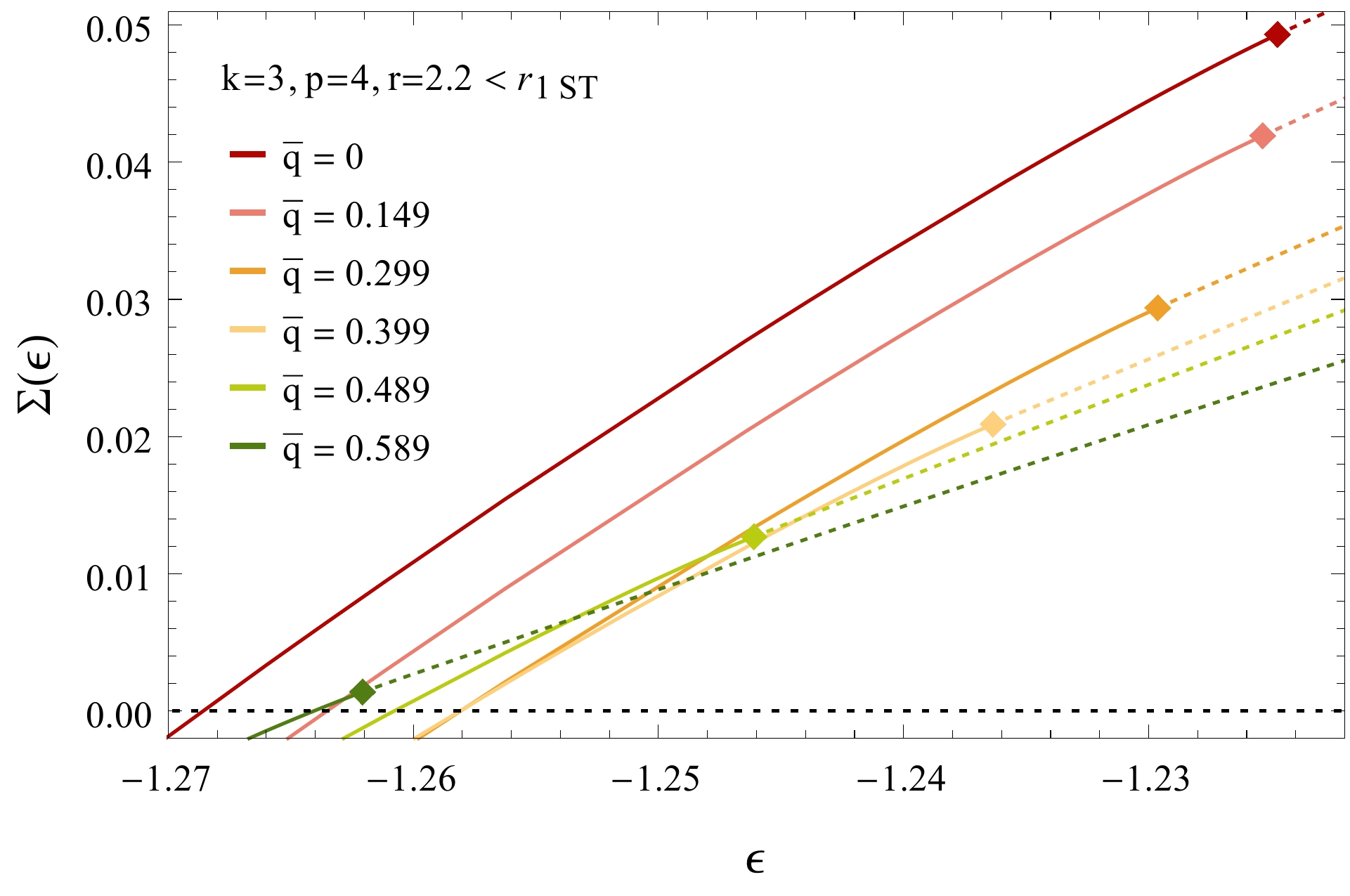}\label{fig:Complexitiesk3p4smallr}}
\subfloat[][]{\includegraphics[width=.5\linewidth] {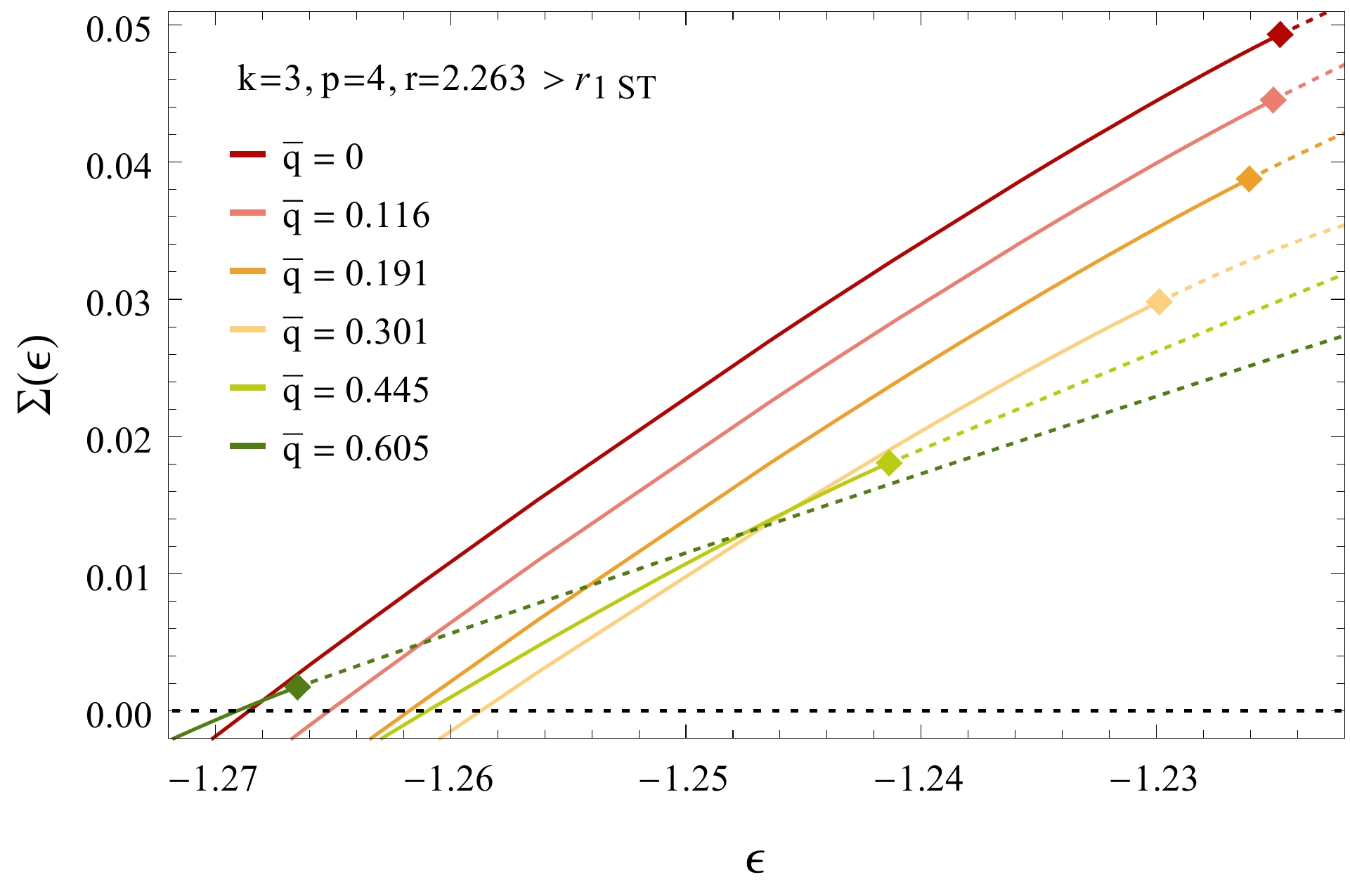}\label{fig:Complexitiesk3p4intermediater}} 
 \caption{
 Complexity curves $\Sigma_{4,3}(\epsilon, \overline{q})$ as a function of the energy density $\epsilon$, for positive values of $\overline{q}$ and $r<r_c$. The squares along the curves mark the \emph{threshold} energy $\epsilon_{\text{th}}(\overline{q})$. (a) For $r \approx 2.2< r_{\text{1ST}}$, both the most numerous stable states and the deepest ones are at the equator,  $\overline{q}_{\text{num}}=0=\overline{q}^*$. The minimal energy $\epsilon^*(\overline{q})$ at which the complexity crosses zero is non-monotonic, and has a second minimum at $\overline{q}^*_2=0.584$. The maximal latitude is $\overline{q}_M \approx 0.597$. (b) For $r \approx 2.263> r_{\text{1ST}}$, the most numerous stable states are at the equator, $\overline{q}_{\text{num}}=0$, while the deepest ones are at $\overline{q}^*\approx 0.609$. The maximal latitude is $\overline{q}_M\approx 0.617$.  
 }\label{fig:Complexitiesk3p4}
\end{figure*}

\begin{figure*}[!htbp]
\subfloat[][]{\includegraphics[width=.5\linewidth] {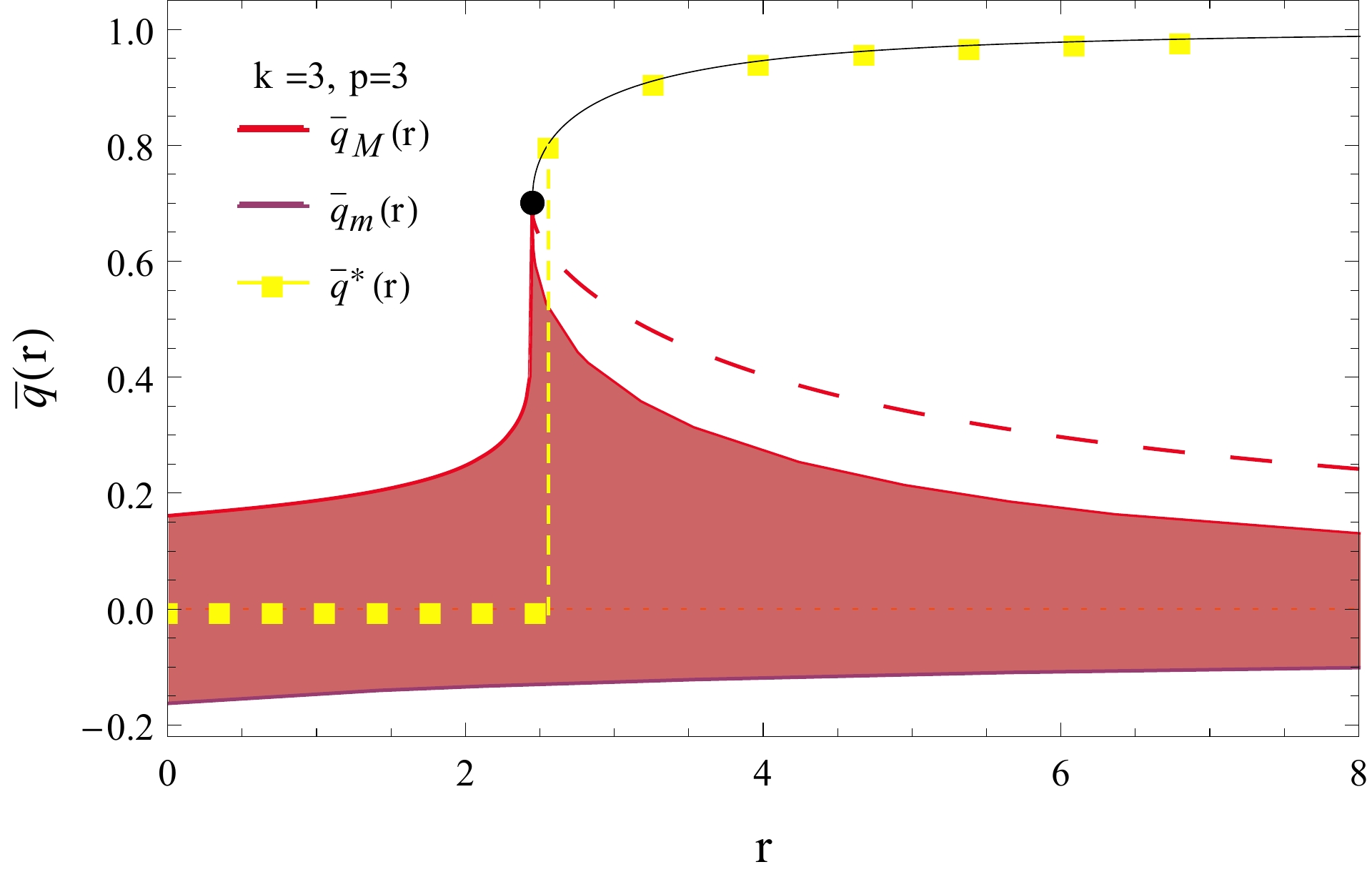}\label{fig:Bandek3p3}}
\subfloat[][]{\includegraphics[width=.49\linewidth] {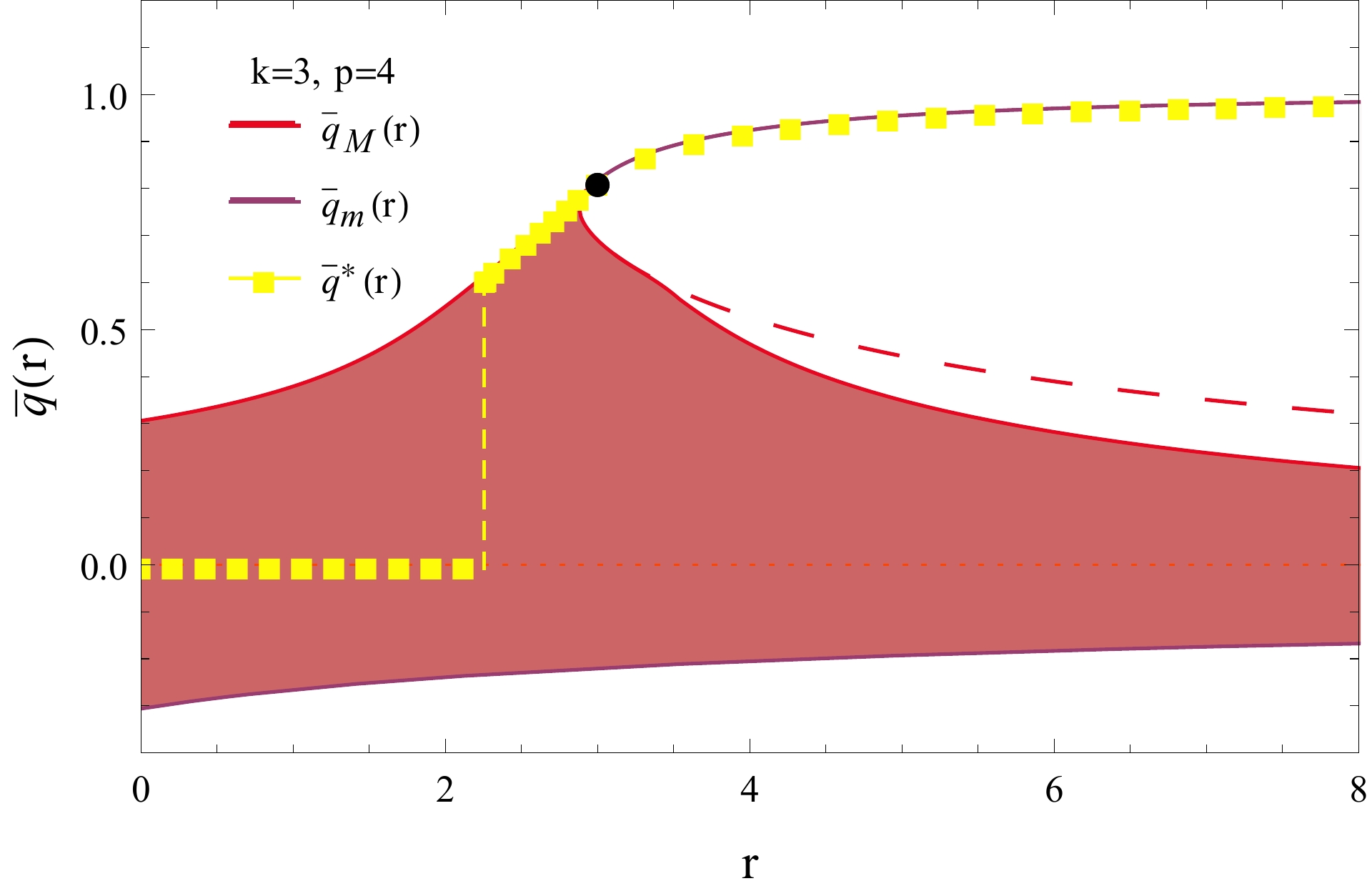}\label{fig:Bandek3p4}}
\caption{
The red strips denote the interval $\overline{q}_m(r) \leq \overline{q} \leq \overline{q}_M(r)$ containing exponentially many \emph{stable} stationary points. The dashed lines are the boundary of the band computed without accounting for the isolated eigenvalue (i.e., states below this line that do not belong to the red strip have energy smaller than the threshold energy, but are unstable due to the isolated eigenvalue). The yellow squares indicate the latitudes $\overline{q}^*(r)$ of the deepest minima. The most numerous minima are always at the equator, and are marginally stable. 
(a) For $p=3=k$, the RS transition occurs at $r_c \approx 2.449$ (black point in the figure). For $r \geq r_c$, the isolated eigenvalue renders unstable some stationary points with energy below the \emph{threshold}. At $r_{\text{1ST}}= 2.56$ the deepest minimum in the landscape becomes the isolated minimum at high overlap with the signal. (b) For $k=3, p=4$, at $r \approx 2.89$ the band  splits into two, and at $r_c= 3$ (black point in the figure) the strip at larger overlap undergoes the RS transition. The deepest minima detach from $\overline{q}=0$ at $r_{\text{1ST}} \approx  2.26$.}
\label{fig:Bandek3}
\end{figure*}

\noindent
In this case, the transition at the bottom of the energy landscape is of first order. What distinguishes the two options presented in Sec.~\ref{sec:SummaryResults} is whether this thermodynamic transition occurs before or after the band of stationary points separates into two distinct strips, and the strip at larger overlap $\overline{q}$ undergoes the RS transition. \\
The second case (\emph{Option B}) is realized, for instance, for $k=3$, $p=4$. In this case, the curves $\Sigma_{p,k}(\epsilon, \overline{q})$ behave in the following way: for small $r$, they are monotone decreasing for increasing $\overline{q}$ (they look like their $k=2$ counterpart in Fig.~\ref{fig:Complexitiesk2p3}~(a)), so that both the deepest and the most numerous states are at the equator. At a spinodal point $r_{1\text{SP}} \approx 2.01$, a local minimum in $\epsilon^*(\overline{q})$ appears at a latitude $\overline{q}^*_2(r)>0$, so that for $r>r_{1\text{SP}}$ the curves are no longer monotone, see Fig.~\ref{fig:Complexitiesk3p4}~(a). The absolute minima remain however at the equator, $\overline{q}^*=0$. The latitude  $\overline{q}^*_2$ of the second minimum increases with $r$, and its energy decreases; at the first order transition $r_{\text{1ST}}$, its energy become smaller than the energy of the minima at the equator (that is the ground states of the unperturbed $p$-spin model), and $\overline{q}^*$ jumps discontinuously from zero to a finite value $\overline{q}^*_2(r_{\text{1ST}})$, see Fig.~\ref{fig:Complexitiesk3p4}~(b). \\
The value of $r_{1\text{SP}}$, the latitudes of the second minima $\overline{q}^*_2(r)$ and the corresponding energies can be obtained via the mapping from the curves at $k=1$, as we discuss in Appendix~\ref{app:Results}. The bands of latitudes corresponding to positive complexities below the \emph{threshold} energy can also be obtained from $k=1$, in a way analogous to the one discussed in the Appendix for $k=2$. A major difference with respect to \emph{Case II} concerns the effect of the isolated eigenvalue, since for $k \geq 3$ the states at the equator are not destabilized by it (see the details in Appendix~\ref{app:Results}). Thus, in this case the threshold states of the unperturbed $p$-spin model are the most numerous stable minima, for any~$r$. This case is summarized in Fig.~\ref{fig:Bandek3}~(b).\\
\noindent Finally, we consider the case $k=3$, $p=3$, which realizes \emph{Option A} of Sec.~\ref{sec:SummaryResults}. In this case we find that $r_{1\text{SP}}= r_c$. The curves $\Sigma_{p,k}(\epsilon, \overline{q})$ behave similarly to the ones in Fig.~\ref{fig:Complexitiesk2p3}~(a) for any $r <r_c \approx 2.45$. As $r$ approaches $r_c$ from below, the band of stationary points rapidly grows, and at $r_c$ it reaches its maximal width, incorporating $\overline{q}_c$ (i.e., $\overline{q}_M(r_c)= \overline{q}_c$). Exactly at this latitude $\overline{q}_c$, the saddle point $q_{\text{SP}}=1$ reaches one, and the quenched complexity becomes equal to the annealed one, having positive support for a single value of the energy density $\epsilon_c$. The curve of minimal energies $\epsilon^*(\overline{q})$ has a minimum at $\overline{q}=0$, and it is flat at $\overline{q}_c$, where it intersects the \emph{threshold} energy $\epsilon_{\text{th}}$ (which for $p=k$ is independent of $\overline{q}$ and $r$, and equals to the threshold of the unperturbed $p$-spin model). Therefore, the second minimum of $\epsilon^*(\overline{q})$ appears exactly at $r_c$, and at this point it coincides with the RS solution. At larger values of $r$, the minimum of the annealed complexity is isolated (it departs from the band containing all the other minima), and becomes energetically favorable at $r_{\text{1ST}} \approx 2.56$. The band at small overlap shrinks asymptotically around the equator. \\
Thus, in this case the band of minima is connected up to $r_c$, and it splits exactly at the critical point, see  Fig.~\ref{fig:Bandek3}~(b). The analysis of the isolated eigenvalue shows that for large enough $r$, the eigenvalue renders unstable the points at higher overlap in the strip enclosing the equator, but it does not affect the most numerous, marginally stable states at the equator, nor the minimum of the annealed complexity, which is stable for any $r>r_c$. 

\section{Comparison between Kac-Rice and replica method}\label{sec:Monasson}
\noindent
As pointed out in the previous section, the information on the thermodynamics provided by the replica calculation is fully recovered from the Kac-Rice results, by analyzing the spectrum of minima $\epsilon^*(\overline{q},r)$ satisfying $\Sigma_{p,k}(\epsilon^*(\overline{q},r), \overline{q};r)=0$. As first pointed out in \cite{Monasson}, the thermodynamical replica method 
can also be used to obtain information on the number of critical points. 
In this section, by comparing the predictions of the two calculation schemes 
concerning the configurational entropy, i.e. the complexity of the most numerous stationary points $\Sigma_c(\epsilon)$,
we show that the thermodynamical replica method is not able to reproduce the full Kac-Rice results and leads to partially incorrect predictions. This is an important point since, although the method could be probably amended, its present form which is often used for the purpose of computing the configurational entropy
fails for the models we consider. We highlight below the two different reasons for failure.\\
The replica formalism allows to sample local minima at energy higher than the equilibrium one by not imposing the saddle point condition on $\beta m$ (the third among Eqs.~\ref{spS}), and using $m$ as a parameter, which plays the role of an effective inverse temperature.
By lowering $m$ ($\beta m$ in the $T=0$ case) the remaining saddle point equations describe the macroscopic features of the most numerous local minima at higher energy.
In particular, the expression contained in the disregarded saddle point equation \eqref{spS} gives the corresponding intensive log multiplicity of these minima, i.e. their configurational entropy $S_c$:\\
\begin{equation}\label{Sc}
\begin{split}
S_c&= -\frac{\beta^2 m^2(1 - q_0^p)}{4} -\frac{1}{2}
  \log\left[\frac{\beta(1-q_1)}{\beta(1-q_1) + \beta m(1 - q_0)}\right] \\
   & +\frac{\beta^2 m^2(1 - q_0)(q_0 - \overline q^2)}{2(\beta(1-q_1) + \beta m(1 - q_0))^2}\\
      &-\frac{\beta m(1- q_0)}{2(\beta(1-q_1)+ \beta m (1-q_0))}\ .
   \end{split}
\end{equation}
The stability of the metastable minima, whose multiplicity is accounted for by \eqref{Sc}, is
checked by analyzing the stability with respect to fluctuations in the overlap matrix $Q_{ab}$, see Appendix \ref{app:StabilityMonasson}.\\
The entropy $S_c$ can then be compared with the Kac-Rice complexity of the most numerous stationary points. The latter is obtained, for each energy density $\epsilon$, as the maximum of the curves $\Sigma_{p,k}(\epsilon, \overline{q})$ over those latitudes $\overline{q}$ that correspond to stationary points that are \emph{stable} at the energy~$\epsilon$.

\begin{figure}[!htbp] 
\includegraphics[width=\linewidth]{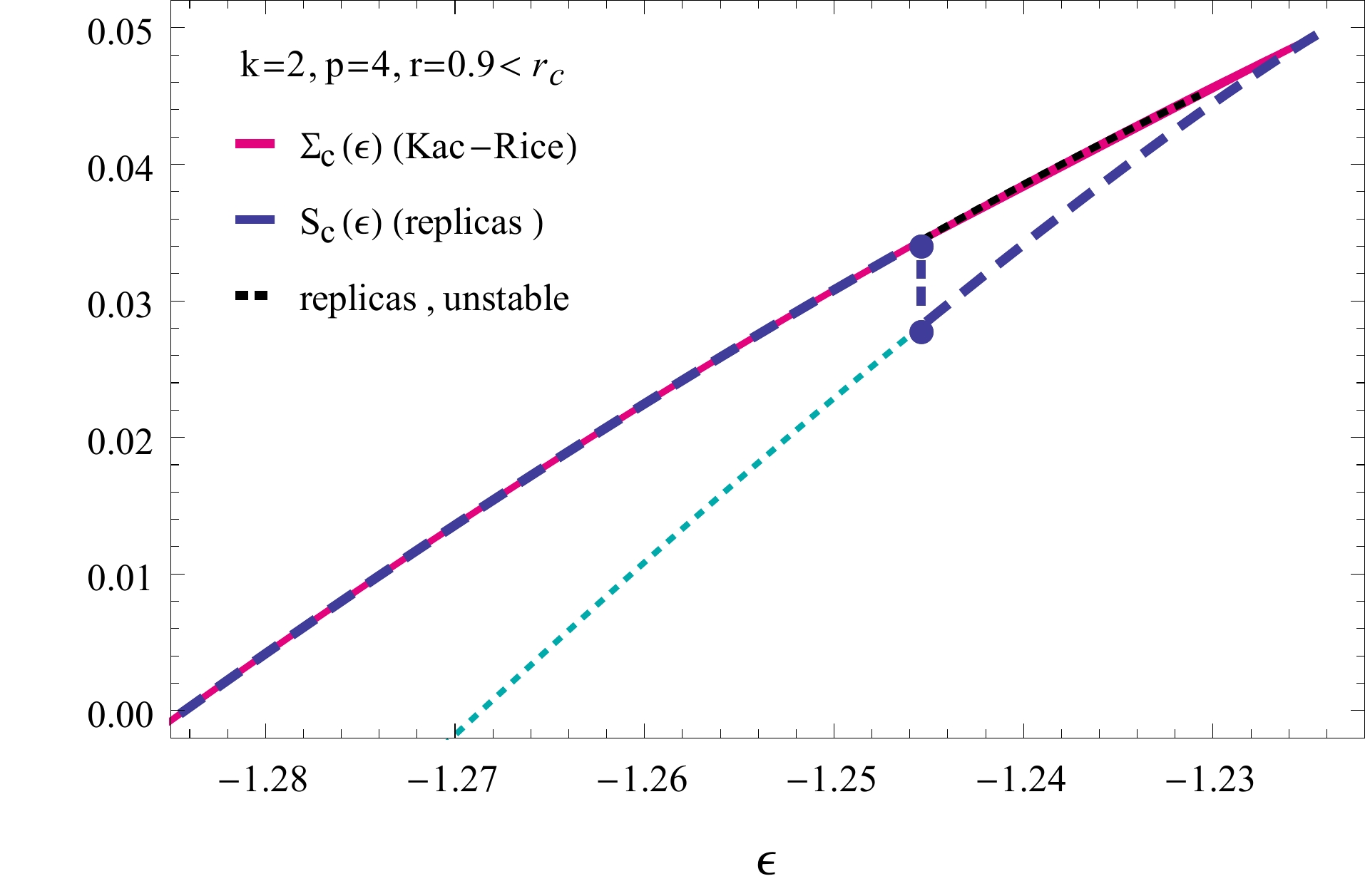} 
\caption{Comparison between the configurational entropies $S_c(\epsilon)$ and $\Sigma_c(\epsilon)$ obtained with the replica and Kac-Rice calculation, respectively. The two curves coincide below $\epsilon \approx -1.245$ (blue point), which is the energy at which the replica solution with $\overline{q} \neq 0$ becomes unstable. For higher energies, the curve computed with replicas is contributed by the states at $\overline{q}=0$ (which are a solution of the replica equations for every energy density), while the Kac-Rice curve is contributed  by marginally stable states whose $\overline{q}$  does not satisfy the stationarity condition of the replica action.}\label{fig:Monasson} 
\end{figure}

\begin{figure}[!htbp] 
\includegraphics[width=\linewidth]{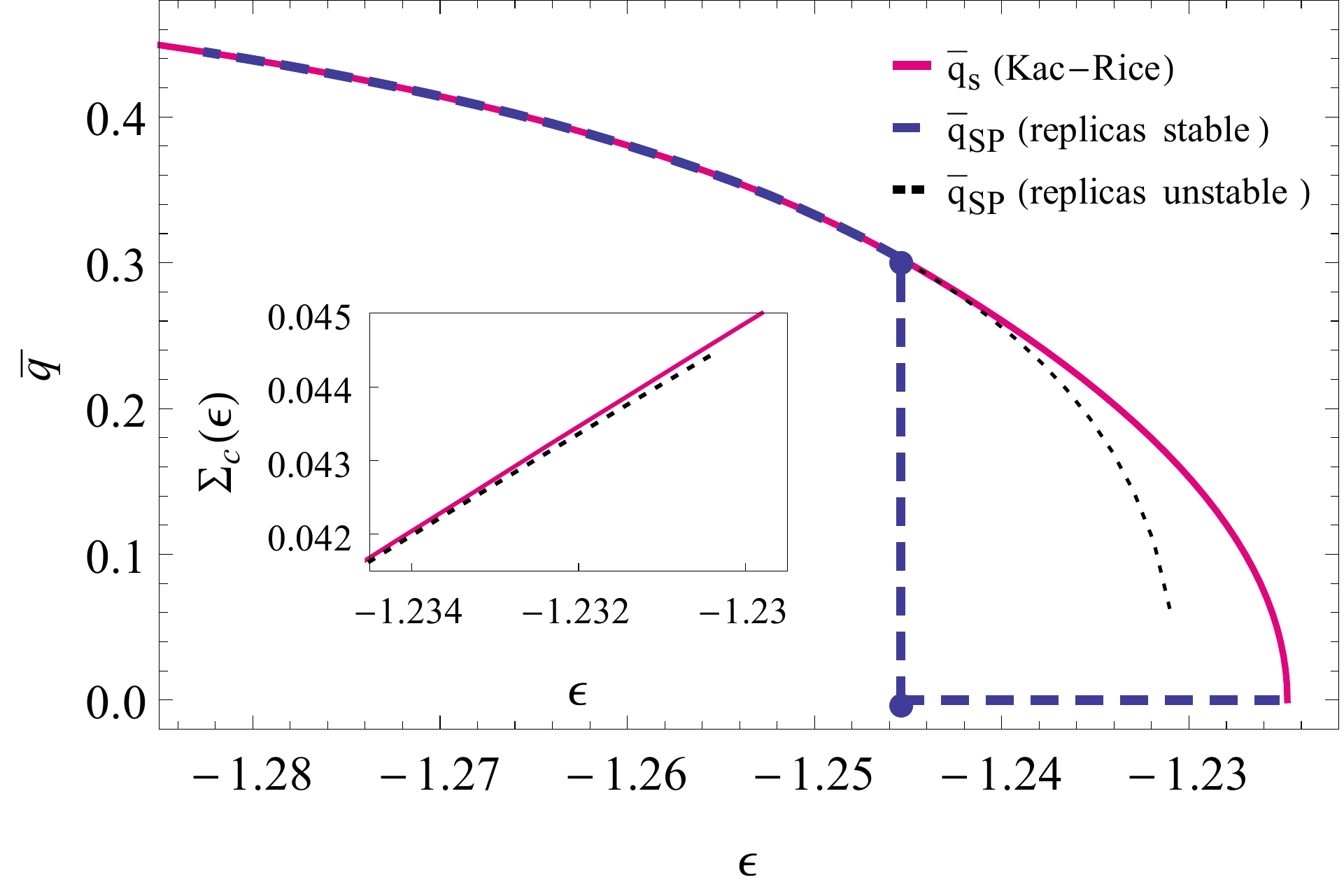} 
\caption{Comparison between the  $\overline{q}_s$ that maximize the Kac-Rice complexities at fixed energy density, and the $\overline{q}_{SP}$ that solve the saddle-point equations for the replica action at fixed $\beta m$ (for $k=2, p=4$ and $r=0.9$). Above $\epsilon \approx -1.245$, the solution of the replicas saddle-point equations with $\overline{q}\neq 0$  corresponds to unstable states (black dotted curve). \emph{Inset.} Zoomed comparison between $\Sigma_c(\epsilon)$ (red) and the unstable part of $S_c(\epsilon)$ resulting from the replica calculation (dashed black), corresponding to a negative value of the replicon eigenvalue.}\label{fig:Monasson2} 
\end{figure}
\begin{figure}[!htbp] 
\includegraphics[width=\linewidth]{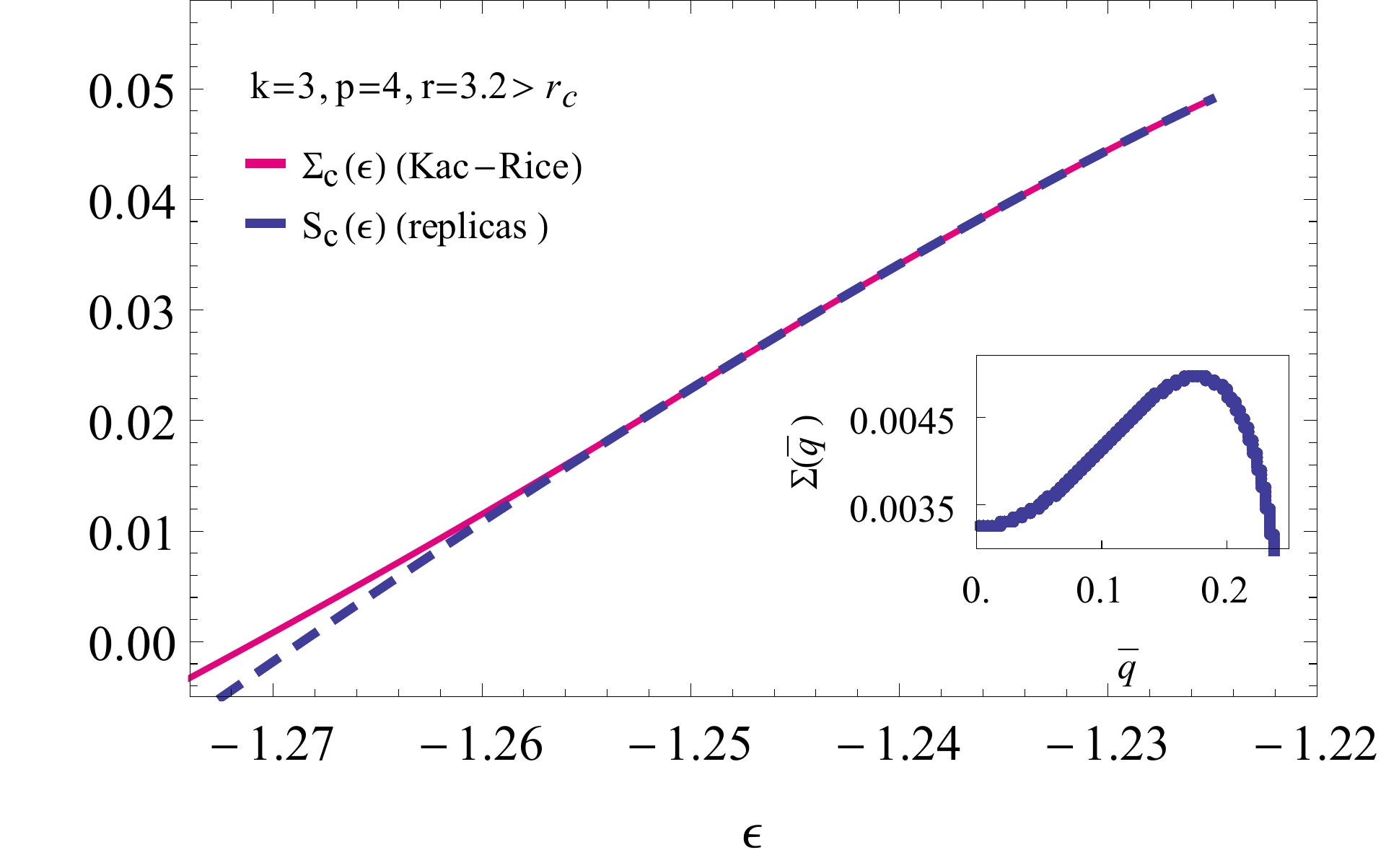} 
\caption{Comparison between the complexity of the most numerous states at fixed energy computed with the replica and Kac-Rice calculation, respectively. The two curves coincide above $\epsilon \approx -1.255$, where the curves are contributed by points at $\overline{q}=0$. At lower energy densities, the replica equations reproduce the complexity of the points at $\overline{q}=0$, while from the Kac-Rice calculation it emerges that there are more numerous stable points at higher overlap $\overline{q}>0$, among which the most numerous ones have an Hessian with a single zero mode. Their complexity is given by the red curve. \emph{Inset. } Complexity as a function of the latitude $\overline{q}$ for a fixed value of energy $\epsilon=-1.266 $. The curve has a maximum at $\overline{q}\approx 0.18$, where the isolated eigenvalue of the Hessian is exactly zero. The points at larger latitudes are saddles of index one.}\label{fig:Monasson3} 
\end{figure}

 \noindent We find that there are regimes in which the two calculations are not equivalent, with the replica calculation failing to identify part of the complexity curve resulting from Kac-Rice. As an illustrative case, we consider the parameters $k=2$, $p=4$.\\ 
For $r<r_c$, the stability of the stationary points at fixed latitude is determined only by the bulk of the eigenvalues density of the Hessian, since no isolated eigenvalue is present. Therefore, the constrained maximization of the Kac-Rice complexities reads:
\begin{equation}\label{eq:EqMaxComp}
 \Sigma_c(\epsilon)= \max_{\overline{q}:\, \epsilon \leq \epsilon_{\text{th}}(\overline{q})} \Sigma_{p,k}(\epsilon, \overline{q}).
\end{equation}
As long as $r< r_{2\text{ND}}$, at fixed $\epsilon$ the curves $\Sigma(\epsilon, \overline{q})$ are monotone decreasing in $\overline{q}$, with a maximum at $\overline{q}=0$. In this case, $\Sigma_c(\epsilon)$ coincides with the complexity of the stationary points at the equator, and the quantity \eqref{Sc} reproduces it. \\
For $r_{2\text{ND}}< r<r_c$, the two complexity curves coincide only in the lowest part of the energy domain (see Fig.~\ref{fig:Monasson} for a comparison between the curves obtained with the two methods for $r=0.9$). 
More precisely, the curves coincide for the energies $\epsilon$ for which the maximum in \eqref{eq:EqMaxComp} is attained inside the interval, at a $\overline{q}_s$ satisfying $\epsilon < \epsilon_{\text{th}}(\overline{q}_s)$. This means that the most numerous states at these energies have Hessian gapped away from zero, and are at latitudes satisfying $\partial \Sigma_{p,k}(\epsilon, \overline{q}_s) / \partial \overline{q}=0$. In this case, $\overline{q}_s$ coincides with the value of $\overline{q}$ selected by the saddle point equations of the replica calculation (see Sec. \ref{sec:ReplicaEquations}), and we recover $S_c(\epsilon)= \Sigma_c(\epsilon)$. \\
 In the second part of the curve, instead, the maximum is attained at the boundary of the interval, at latitudes $\overline{q}_s$ such that $\epsilon= \epsilon_{\text{th}}(\overline{q}_s)$. This part of the curve $\Sigma_c(\epsilon)$ is thus contributed by points that are marginally stable, and which do not fulfill the stationarity condition $\partial \Sigma_{p,k}(\epsilon, \overline{q}) / \partial \overline{q}=0$. This piece of curve is not recovered by the replica scheme: rather, in this energy regime the replica solution corresponding to saddle point values $\overline{q} \neq 0$ results in a different entropy curve (the black-dashed line in the Inset of Fig. \ref{fig:Monasson2}) that has to be disregarded, being unstable with respect to the replicon criterion recalled in Appendix \ref{app:StabilityMonasson}. On the other hand, the dashed blue line in the figure corresponds to $\overline{q}=0$, which is always a solution of the saddle point equations in the replica calculation, that coincides with the Kac-Rice curve only for the highest energies. Thus, the replica result is inconsistent with the Kac-Rice one at intermediate energy densities.
 As $r$ increases toward $r_c$, the interval of energies in which the two curves coincide shrinks, so that the replica calculation allows to recover only a very small portion of the configurational entropy obtained via Kac-Rice, the part of curve contributed by strictly stable points. The situation outlined above highlights the first way in which the replica method 
 can fail: the correct result is recovered {\it only when} the largest contribution to the configurational entropy at fixed energy 
 is given by a $\overline{q}$ such that $\partial \Sigma_{p,k}(\epsilon, \overline{q}) / \partial \overline{q}=0$. 
 The reason is that the configurations taken into account by the replica method if $\partial \Sigma_{p,k}(\epsilon, \overline{q}) / \partial \overline{q}\ne 0$ do not correspond to true minima since they have a non-zero gradient in the north-pole direction. \\
 Let's now focus on the other way in which the usual replica method to compute the configurational entropy can fail. For $r>r_c$ (but smaller than the value of $r$ at which the landscape becomes completely convex), for the smaller energies $\epsilon$ the curves $\Sigma(\epsilon, \overline{q})$ are monotonic in $\overline{q}$, with a local minimum at $\overline{q}=0$ and a global maximum at a latitude $\overline{q}>0$ (see the Inset in Fig. \ref{fig:Monasson3}), while at larger energies the minimum at the equator becomes the maximum of the curve. Therefore, at small energies $\Sigma_c(\epsilon)$ is contributed by the points at the latitudes $\overline{q}>0$ of the maximum.
 The replica calculation reproduces nevertheless only the complexity at $\overline{q}=0$, even when there is a full spectrum of more numerous (stable) points at higher overlap with the signal (see Fig. \ref{fig:Monasson3} for the case $r=3$). The reason for this is that precisely at the latitude where the complexity has a maximum, the isolated eigenvalue of the Hessian is exactly equal to zero. Thus, the lower-energy part of the curve $\Sigma_c(\epsilon)$ obtained from Kac-Rice is contributed by stationary points that have the Hessian with a single zero mode, which are known to be 
 not captured by the standard replica calculation \cite{Annibale,CLR1,CLR2}. 
The physical reason is that these stationary points 
do not correspond to the zero-temperature limit of stable states. In fact, as shown in \cite{Aspelmeier}, they correspond to minima characterized by finite barriers. This situation has been already found 
in computation of  the multiplicity of TAP solutions in mixed models \cite{Annibale} and in models exhibiting a full
replica symmetry breaking phase \cite{CLR1, CLR2}.\\

\section{On the spiked tensor case}\label{sec:Spiked}
\noindent
As we have previously remarked, the Hamiltonian $H_{p,k}(r)$ in the case $k=p$ is related to spiked tensor model~\cite{montanari}, i.e., to the inference problem of detecting a low-rank, additive perturbation of a symmetric Gaussian tensor, which has attracted a lot of attention recently. \\
In this section we specifically present our analysis on this system, focusing in the case $k=p \geq 3$ for concrete results. Some of these observations are already stated in~\cite{montanari,krzakala,bandera,Chen}. 
We also discuss which properties the annealed computation \cite{MontanariBenArous} cannot capture. \\
The inference task in the spiked tensor problem consists in reconstructing the unknown vector ${\bf v}_0$ from the observation of a random $p$-tensor with components
\begin{equation}\label{eq:Spiked}
 W_{i_1, \cdots, i_p}= p!J_{i_1, \dots, i_p}- \frac{r}{p} ({\bf v}_0)_{i_1} \cdots ({\bf v}_0)_{i_p},
\end{equation}
where the random couplings $J_{i_1, \dots, i_p}$, symmetric with respect to a permutation of the indices, 
correspond to the noise and ${\bf v}_0$ (the signal) is generated at random from a spherical prior distribution. In particular, one is interested in identifying the strong detection threshold~\cite{bandera}, i.e. the critical signal-to-noise ratio below which the spiked model is statistically indistinguishable from the un-spiked one with $r=0$, and the detection threshold, above which an estimator $\hat{\bf s}$ of the signal having a finite overlap with ${\bf v}_0$ in the limit $N \to \infty$ exists (for a precise definition of statistical indistinguishability see \cite{bandera}) . In the matrix case $p=2$, the two thresholds coincide~\cite{MontanariMatricesPCA, OnatskiMatrixPCA, BandeiraMatrixPCA}. They are given by the signal-to-noise ratio at which the smallest eigenvalue of the matrix pops out from the semi-circle; the corresponding  eigenvector is correlated with the signal. In the tensor case, rigorous bounds on both thresholds are given in \cite{montanari, MontanariMatricesPCA, bandera}, while the sharp threshold given by the Minimal Mean Squared Error estimator is determined in \cite{krzakala}. \\
The connection with the analysis presented above emerges when considering the maximum-likelihood estimator $\hat{{\bf s}}_{\text{ML}}$ of ${\bf v}_0$ ~\cite{montanari}. It is immediate to see that this corresponds to the vector that maximizes the injective norm of the tensor:  
\begin{equation}
 \hat{{\bf s}}_{\text{ML}}=\underset{{\bf s}: ||{\bf s}||^2=N}
{\text{argmax}}\langle {\bf W}, 
{\bf s}^{\otimes p}\rangle,
\end{equation}
where $\langle \cdot, \cdot \rangle$ denotes the tensor product. This coincides, up to a global sign flip of the energy functional $H_{p,k}(r)$, with its absolute minimum; therefore, a reliable estimate of the signal by means of $\hat{{\bf s}}_{\text{ML}}$ is possible whenever the global minimum of the landscape acquires a non-zero overlap with the special direction of the signal, i.e., whenever $r \geq r_{1\text{ST}}$. The thermodynamic transition thus gives (in general) and upper bound to the detection threshold. On the other hand, the performance of algorithms \cite{montanari} aiming at reconstructing the signal is expected to depend on the full structure of metastable states, encoded in the complexity. \\
The analysis presented in the previous sections and in Refs.~\cite{montanari,krzakala,bandera,Chen} lead to the following picture for the spiked-tensor model ($p=k \geq 3$):
\begin{itemize}
 \item[(i)] The spinodal point $r_{1\text{SP}}$, where a high-overlap metastable minimum appears in the curve $\epsilon^*(\overline{q})$ (or, equivalently, where a second solution appears for the replicas equations) is exactly equal to the point $r_c(p)$ where the trivialization of the portion of the landscape at high overlap with the signal occurs. Moreover, there is no splitting of the band of minima for $r<r_c$.
 This implies that the portion of landscape close to the high-overlap minimum is not rugged for all $r$s such that the high-overlap minimum exists, and no intermediate phase with metastability at high-overlap with the signal is present.
 \item[(ii)] The transition points $r_c$ and $r_{1\text{ST}}$ are related, respectively, to the dynamical and statical transition temperatures ($\beta_d$ and $\beta_s$) of the pure spherical $p$-spin model; more precisely:
 \begin{equation}\label{eq:TransitionTemperatures}
r_c= \frac{p}{2} \beta_d, \quad r_{1\text{ST}} = \frac{p}{2} \beta_s.
 \end{equation}
 \item[(iii)] For $r_{\text{1SP}}\leq r \leq r_{\text{1ST}}$ and for most energy densities $\epsilon$, the complexity is non-monotonic in the overlap $\overline{q}$; however, for any $\epsilon$ the most numerous minima are found to be orthogonal to the signal, at $\overline{q}=0$. 
\end{itemize}
The equality $r_{1\text{SP}}= r_c$ is shown in Appendix \ref{app:Results}. It implies that for $p=k$ the thermodynamic transition always occurs when the high-overlap part of the energy landscape is convex.
This allows for the annealed Kac-Rice computation \cite{MontanariBenArous} to correctly capture the transition value $r_c$ although quenched and annealed complexity do not coincide for $r<r_c$. For other models,
such as $p=3, k=4$ for which the landscape is rugged close to global high-overlap minimum at $r_{1\text{ST}}$,
this is no longer the case and the quenched computation is needed to also correctly describe the transition.  \\
The first identity in \eqref{eq:TransitionTemperatures} can be read explicitly from Eq.~\ref{eq:rCritico}. The second identity is naturally true for Bayes optimal estimates \cite{krzakala}, as it holds in general along the Nishimori line (which corresponds to the line, in the $(r, \beta)$ phase diagram, where $\beta= 2 r/p$). The fact that the same detection threshold is found with the maximum likelihood estimator follows from the properties of the thermodynamic phase diagram~\cite{sherrington1}, where the first-order transition line appears to be  independent of temperature, thus implying that the same $r_{1\text{ST}}$ found on the Nishimori line is recovered at $T=0$ (note however that this is a peculiarity of the spherical case and does not hold in general \cite{krzakala}). The property (iii) implies that the quenched complexity $\Sigma_c(\epsilon)$  
is identical for the spiked and the un-spiked model for $r_{\text{SP}}\le r \le r_{1\text{ST}}$, which is 
consistent with the strong detection threshold being at $r_{1\text{ST}}$. \\
Given the structure of the energy landscape, we expect that for all $r$ not diverging with $N$, physical dynamics starting from random initial conditions behaves as in the un-spiked model, i.e., the system remains stuck in the vicinity of the most numerous, marginally stable states that lie at the equator $\overline{q}=0$, thus being unable to recover any information on the signal. The approximate message passing algorithm is known to fail as well \cite{krzakala}. On the other hand, for $r> r_{1\text{ST}}$ the dynamics with a warm start should converge to the global minimum of the energy landscapes over time scales of $O(1)$, due to the smoothness of the landscape in its vicinity \cite{footnote8}. Polynomial-time algorithms are instead known to succeed for $r$ scaling as $N^{(p-2)/4}$, see  \cite{montanari} and references therein. \\
Finally, it is proven in \cite{Chen} that for the Ising spiked tensor defined on the hypercube ($s_i= \pm 1$), the strong detection and detection threshold coincide, being both equal to the threshold given by the minimal mean square error estimator \cite{krzakala}. The proof relies on the bound (for large $N$) of the fluctuations of the free energy of the Ising $p$-spin model around its average value, in the high-$T$ phase. A similar bound should hold for the spherical case, since the variance of the intensive free-energy is found to be of order $1/N$ by the replica method (the variance can be directly obtained using the RS approximation to compute the $O(n^2)$ term of the replicated free energy \cite{crisantisommers}).  In consequence, we expect that this argument can be extended to the spherical case, thus implying that both thresholds are given by the maximum-likelihood estimator.

 \section{Discussion and conclusion}\label{sec:Conclusion}
 \noindent
We have analyzed the evolution of an archetypical model of high-dimensional landscapes generated by  
an energy function in which random fluctuations compete with a deterministic contribution favoring a single minimum.
For entropic reasons the overall majority of the minima 
created by the randomness lie in a region different from the one favored by the deterministic contribution. 
By increasing the strength of the deterministic contribution, and depending on the form of the latter, different behaviors and geometric phase transitions, that we have classified and thoroughly analyzed, can take place. As discussed in the introduction, our results provide guidelines for current problems in several different fields, and a full analysis of the energy landscape of the spiked-tensor model which recently attracted a lot of attention \cite{montanari,krzakala,bandera, Chen, MontanariBenArous}. In particular, our analysis is useful to understand how the dynamics governed by gradient descent (and stochastic versions of it) proceed 
in such landscapes. The region of bad and numerous local minima that we called the equator is a trap for the dynamics. 
Only in case of a sufficiently warm start, i.e. if the initial condition of the dynamics has a finite overlap with the special direction ${\bf v_0}$ selected by the deterministic contribution, the
system can end up close to ${\bf v_0}$, although not necessarily in the global minimum since many additional good local minima can be present.  \\
The other main contribution of our work is methodological. We have developed a framework based on the Kac-Rice method that allows to compute the quenched complexity, opening the way to full analysis of random landscapes in many different contexts. We have shown that it is superior to previous frameworks used in the literature. Indeed, the usual replica method fails in some cases, as demonstrated in this work, whereas the super-symmetry one is in comparison quite obscure.
Instead, the Kac-Rice formalism we developed is free of ambiguities, straightforward although complex, and likely to be transformed in a rigorous formalism in a not too distant future. \\
  { \it Acknowledgements. } We thank A. Bandeira, S.Sarao, P. Urbani and L. Zdeborova for useful discussions. This work was partially supported by the grant from the Simons Foundation ($\sharp$454935, Giulio Biroli).
 


\section{Appendices}

\subsection{Details on the replica calculation}\label{app:Replicas}
\noindent
In this Appendix we provide some additional details on the replica analysis presented in Sec. \ref{sec:ReplicaAnalysis}.\\
At finite $\beta$, the replicated action $S$ evaluated within the $1$RSB \emph{ansatz} for the overlap matrix $Q_{ab}$ reads:
\begin{eqnarray}
S_{\rm 1RSB}=\frac{\beta^2}{4}[1-q_1^p+m(q_1^p-q_0^p)]+ \nonumber \\
	r\beta f_k(\overline{q})+\frac{1}{2}\log(1-q_1)+\\
	\frac{1}{2m}\log\left(\frac{1-q_1+m(q_1-q_0)}{1-q_1}\right)+ \nonumber \\
	\frac{1}{2}\frac{q_0-\overline{q}^2}{1-q_1+m(q_1-q_0)} \ \nonumber.
\end{eqnarray}
The saddle point equations for the four parameters $q_1, q_0, m$ and $\overline{q}$ equal to:
\begin{eqnarray*}
\frac{\beta^2p}{2}q_1^{p-1}=\frac{1}{m}\left(\frac{1}{1-q_1}-\frac{1}{1-q_1+m(q_1-q_0)}\right)+\\
	\frac{q_0-\overline{q}^2}{[1-q_1+m(q_1-q_0)]^2} \nonumber,
\end{eqnarray*}
\begin{eqnarray*}
\frac{\beta^2p}{2}q_0^{p-1}=\frac{q_0-\overline{q}^2}{[1-q_1+m(q_1-q_0)]^2},
\end{eqnarray*}
\begin{eqnarray*}
\frac{\beta^2}{2} (q_1^p-q_0^p)+\frac{1}{m^2}\log\left(\frac{1-q_1}{1-q_1+m(q_1-q_0)}\right)+\\
	\frac{1}{m}\frac{q_1-q_0}{1-q_1+m(q_1-q_0)}-\frac{(q_1-q_0)(q_0-\overline{q}^2)}{[1-q_1+m(q_1-q_0)]^2}=0 \nonumber
\end{eqnarray*}
and
\begin{eqnarray}
\overline{q}=r\beta[1-q_1+m(q_1-q_0)]f'_k(\overline{q}).
\end{eqnarray}
In the zero temperature limit $\beta \to \infty$, 
 the variables of order one are $\beta(1-q_1)$ and $\beta m$. Performing this limit in the above equations one recovers the expressions given in the main text.\\
The expansion of the saddle point equations in $q_1-q_0$ gives rise to four different equations; three of them are useful to get the variables $\overline q$, $q_0=q_1=q$ and $m$ at the continuous transition between 1RSB structure to RS structure of the high overlap phase. The fourth equation fixes the position of the continuous transition line on the phase diagram $T, r$, {\it i.e.} 
for a given $T$ it gives the corresponding $r_c(T)$.
At zero temperature, when expressed in terms of $\beta(1-q), \beta m,$ and $\overline q$, the equations read as follows:
\begin{eqnarray}\label{eq1t}
\overline{q}=rf'_k(\overline{q})\beta(1-q) \ ,
\end{eqnarray}
\begin{eqnarray}\label{eq2t}
\overline q^2=\frac{p-2}{p-1} \ ,
\end{eqnarray}
\begin{eqnarray}\label{eq3t}
\frac{p(p-1)}{2}=\frac{1}{\beta^2(1-q)^2} \ ,
\end{eqnarray}
and 
\begin{eqnarray}\label{eq4t}
\beta m=\frac{p-2}{2}\beta(1-q) \ .
\end{eqnarray}
Their solution gives the generic expression for $r_c$ and $\overline{q}_c$ reported in the main text, Eqs. \ref{eq:rCritico} and \ref{eq:qc}.

\subsection{Computation of the quadratic form Eq. \eqref{eq:quad1}}\label{app:QuadraticForm}
\noindent
In this Appendix we provide some details on the computation of the inverse correlation matrix $\hat{C}^{-1}$ in \eqref{eq:quad1}. \\
As pointed out in the main text, for the purpose of computing the quadratic form \eqref{eq:quad1} it suffices to invert $\hat{C}$ within the subspace spanned by the $Nn$-dimensional vectors ${\bm \xi}_1, {\bm \xi}_2$ and ${\bm \xi}_3$, which is closed under the action of $\hat{C}$.
For convenience, we separate the matrix $\hat{C}$ into its diagonal and off-diagonal parts in replica space, $\hat{C}= p \tonde{\hat{D}+ \hat{O}}$ with
\begin{equation}
\begin{split}
 D^{ab}_{ij}&= \delta_{ab} \tonde{\delta_{ij}+(p-1) \sigma^a_i \sigma^a_j},\\
 O^{ab}_{ij}&= (1-\delta_{ab}) \tonde{\delta_{ij}q^{p-1}+(p-1)q^{p-2} \sigma^b_i \sigma^a_j}.
 \end{split}
\end{equation}
It holds $\hat{C}^{-1}=p^{-1} \hat{D}^{-1} \tonde{\hat{1}+ \hat{O} \hat{D}^{-1}}^{-1}$, with $[\hat{D}^{-1}]^{ab}_{ij}= \delta_{ab} \tonde{\delta_{ij}- (p-1)p^{-1} \sigma^a_i \sigma^a_j}$. The operator $\hat{1}+\hat{O}\hat{D}^{-1}$ acts on the chosen vectors as follows:
\begin{equation}\label{eq:ClosedAction}
 \begin{split}
&\tonde{\hat{1}+ \hat{O} \hat{D}^{-1}}\bm{\xi}_1= \bm{\xi}_1+ q^{p-1} \bm{\xi}_3,\\
&\tonde{\hat{1}+ \hat{O} \hat{D}^{-1}}\bm{\xi}_2=  \tonde{1+(n-1)q^{p-1}} \bm{\xi}_2\\
&+(p-1) (1-q) \overline{q}  q^{p-2} \bm{\xi}_3\\
&\tonde{\hat{1}+ \hat{O} \hat{D}^{-1}}\bm{\xi}_3= (n-1) q^{p-1}  \bm{\xi}_1\\
&+ \tonde{1+ q^{p-2}\quadre{(p-1)(1 -(n-1)  q^2)+(n-2) p q}}\bm{\xi}_3.
 \end{split}
\end{equation}
 Given that $\hat{D}^{-1}{\bm \xi}_1= p^{-1}{\bm \xi}_1$ and $\hat{D}^{-1}{\bm \xi}_2=  {\bm \xi}_2- \overline{q}(p-1)p^{-1}  {\bm \xi}_1$, the quadratic form \eqref{eq:quad1} can be straightforwardly rewritten in terms of matrix elements of the operator $\hat{Y}\equiv \tonde{\hat{1}+ \hat{O} \hat{D}^{-1}}^{-1}$. To invert this operator, we introduce an orthonormal basis of the subspace spanned by the vectors ${\bm \xi}_i$:
 \begin{equation}\label{eq:ChangeBasis}
 \begin{split}
  & {\bf v}_1= \frac{{\bm \xi}_1 }{ \sqrt{n}}, \\
   &{\bf v}_2=\frac{\bm{\xi}_2- \overline{q} \bm{\xi}_1 }{\sqrt{ n(1-\overline{q}^2)}} ,  \\
   &{\bf v}_3= \frac{(n-1)(\overline{q}^2-q){\bm \xi}_1- (n-1)\overline{q}(1-q){\bm \xi}_2+ (1- \overline{q}^2){\bm \xi}_3}{\sqrt{(1-\overline{q}^2) A}},
   \end{split}
 \end{equation}
with $A={n(n-1) (1-q) \left(1-n \overline{q}^2+(n-1) q\right)}$, and we write $Y_{ij}= {\bf v}_i \cdot \hat{Y} \cdot {\bf v}_j$. In this basis, using $f_k(\overline{q})= \overline{q}^k/k$ and  $u(\epsilon)=p  \epsilon+  r (p/k-1)\overline{q}^k$, we obtain for \eqref{eq:quad1}:
\begin{equation}\label{eq:quad2}
\begin{split}
&Q^{(n)}_{p,k,r}(\epsilon, \overline{q},q)=n \left(x+r \frac{\overline{q}^k}{k}\right)^2 Y_{11}+\\
&n {r\overline{q}^{k-1}  \sqrt{1-\overline{q}^2}  }\left(x+r \frac{\overline{q}^k}{k}\right) \tonde{\frac{Y_{12}}{p}+Y_{21}}+\\
&n \frac{ \left(1-\overline{q}^2\right) r^2 \overline{q}^{2 k-2}}{p}Y_{22}.
\end{split}
\end{equation}
Using \eqref{eq:ClosedAction} and \eqref{eq:ChangeBasis}, we find that the operator $\tonde{\hat{1}+ \hat{O} \hat{D}^{-1}}$ acts on the basis ${\bf{v}}_i$ as follows:
\begin{widetext}
\begin{equation}\label{eq:Matrix}
\tonde{\hat{1}+ \hat{O} \hat{D}^{-1}}=
 \left(
\begin{array}{ccc}
1 +(n-1) q^p & -p\frac{(n-1)  \overline{q} (q-1) q^{p-1}}{\sqrt{1-\overline{q}^2}} & 
 p q^{p-1} \sqrt{\frac{A}{n (1-\overline{q}^2)}} \\
 -\frac{(n-1) \overline{q} (q-1) q^{p-1}}{\sqrt{1-\overline{q}^2}} & 1-\frac{(n-1) q^{p-2} \left(\left(p (q-1)^2-1\right) \overline{q}^2+q\right)}{\overline{q}^2-1} &\overline{q} q^{p-2} \sqrt{\frac{A}{n}} \frac{p(1-q)-1}{1-\overline{q}^2}\\
 q^{p-1} \sqrt{\frac{A}{n (1-\overline{q}^2)}}  & \overline{q} q^{p-2} \sqrt{\frac{A}{n}} \frac{p(1-q)-1}{1-\overline{q}^2} &1- \frac{q^{p-2} \quadre{1-\overline{q}^2 
 (n (1-q)+q)-p (1-q) \left(1-n \overline{q}^2+(n-1)
   q\right)} }{1-\overline{q}^2} \\
\end{array}
\right),
\end{equation}
 \end{widetext}
 while the relevant matrix elements of its inverse are:
\begin{equation*}
 \begin{split}\label{eq:ElementsY}
 & Y_{11}=\frac{q^p (p (q-1) ((n-1) q+1)+1)-q^2}{D(q)},\\
 & Y_{12}=\frac{(n-1)  p (1-q) \overline{q} q^{p+1}}{\sqrt{1-  \overline{q}^2} D(q)}=p Y_{21},\\
   \end{split}
 \end{equation*}
  and
\begin{equation*}
 \begin{split}\label{eq:ElementsY2}
  &Y_{22}=\frac{1 }{1+(n-1) q^{p-1}}-\\
  &\frac{(n-1)(1-q) q^{p+2} \left(1+(n-1) q^p-p (1-q)\right)}{\left(1-\overline{q}^2\right) \left(1+(n-1) q^{p-1}\right) D(q)}.
 \end{split}
 \end{equation*}
 with $D(q)$ given in \eqref{eq:D(q)}. The result \eqref{eq:FullGenn} is recovered substituting these expressions into \eqref{eq:quad2}.

\subsection{Conditional distribution of  Hessians}\label{app:Hessian}
\noindent
In this Appendix, we analyze the structure of the $(N-1)n \times (N-1)n$ covariance matrix of the Hessians components $\mathcal{H}^a_{\alpha \beta}$, conditioned to the gradients and energy fields of all the $n$ replicas. We remind that, given \eqref{eq:RelHess2}, the Hessians can be written as
\begin{equation}\label{eq:AgainDecomposHessian}
 {\mathcal{H}}=  {{\mathcal{M}}}+  \theta_{r,k}(\overline{q}) \sum_{\alpha, \beta}  ({\bf e}_\alpha \cdot {\bf w_0}) ({\bf e}_\beta \cdot {\bf w_0}) {\bf e}_\alpha  {\bf e}^T_\beta - \sqrt{2 N} u(\epsilon, \overline{q}) \hat{1},
\end{equation}
where $\mathcal{M}^a$ denotes the $p$-spin part of the Hessian. We compute the conditional law of $\mathcal{M}^a$, and denote  with $\tilde{\mathcal{M}}^a$ the random matrix obeying this law (and similarly for $\tilde{\mathcal{H}}^a_{\alpha \beta}$). 
Since the last two terms in \eqref{eq:AgainDecomposHessian} are deterministic, the covariance matrix of $\tilde{\mathcal{H}}^a$ is the same as the one of $\tilde{\mathcal{M}}^a$, while the averages of the components are shifted by the deterministic terms. \\
We show that, for each $a$, the matrices $\tilde{\mathcal{M}}^a$ are perturbed GOE matrices; in particular, each $ {\tilde{\mathcal{M}}^a}/{\sqrt{N}}$ can be written as a sum of a stochastic matrix $\mathcal{S}^a$ with zero average, and a deterministic matrix $\mathcal{D}^a$, 
\begin{equation}\label{eq:MatSplit}
 \frac{\tilde{\mathcal{M}}^a}{\sqrt{N}} = \frac{\mathcal{S}^a}{\sqrt{N}}+ \mathcal{D}^a.
\end{equation}
The stochastic part $\mathcal{S}^a$ has the block structure:
\begin{equation}\label{eq:MatrixStochastic}
 \left( 
\setlength{\arraycolsep}{.1pt}
\begin{array}{c| c}
   \begin{array}{ccccc}
   m_{11} &  m_{12}& \cdots &\cdots& m_{1M}  \\
  m_{21} &  m_{22}& \cdots&\cdots   &m_{2M}  \\
     \cdots  & \cdots & \cdots & \cdots& \cdots\\
    \cdots    & \cdots &\cdots&  \cdots& \cdots\\ 
            m_{M1}  &  m_{M2}& \cdots&\cdots &m_{MM}\\
  \hline
    n_{M+11} &  n_{M+12}& \cdots & \cdots  &n_{M+1M}\\
       \cdots  &  \cdots & \cdots& \cdots & \cdots\\
          n_{N-11} &  n_{N-12}& \cdots &\cdots &n_{N-1M}
  \end{array}
  &
      \begin{array}{ccc} n_{1M+1} &  \cdots  &n_{1N-1} \\
n_{2M+1} &  \cdots  &n_{2N-1} \\
  \cdots & \cdots & \cdots\\
  \cdots & \cdots & \cdots\\
   
  n_{M M+1} &  \cdots  &n_{M N-1}\\
    \hline
    q_{M+1M+1} &  \cdots  &q_{M+1N-1} \\
\cdots &  \cdots  &\cdots \\
  q_{N-3 M+1} &  \cdots  &q_{N-1 N-1}\\
                     \end{array} 
              \end{array}
      \right)
\end{equation}
where (i)~the larger diagonal block has size $M \times M=(N-n-1) \times (N-n-1)$ and it is made of elements $m_{\alpha \beta}$ that are independent with variance $\sigma^2=p(p-1)$, (ii)~for a generic choice of basis in the subspace $S$, only the elements $n_{\alpha \beta}$ belonging to the same row are correlated, (iii)~the smaller diagonal block has size $n \times n$ and its elements $q_{\alpha \beta}$ are all mutually correlated. 
The deterministic matrix $ \mathcal{D}^a$ is zero everywhere, except in the small $n \times n$ block. 
Form this structure it follows that the reduced density of eigenvalues of $ {\tilde{\mathcal{M}}^a}/{\sqrt{N}}$ is, to leading order in $N$, the one of a GOE matrix with $\sigma^2=p(p-1)$, since the fraction of entries having a modified variance and average is vanishing in the large-$N$ limit. This information suffices to perform the quenched calculation given in the main text, as only the density of eigenvalues is needed to compute the quenched complexity to leading order in $N$. \\
We remark that the partitioning of $\tilde{\mathcal{M}}^a$ into blocks and the properties (i-iii) follow solely from the separation of the coordinates into the subspaces $S^\perp$ and $S$, and are independent on the choice of the basis in both $S^\perp$ and $S$. The covariances of the elements $n_{\alpha \beta}, q_{\alpha, \beta}$ instead depend on the choice of the basis in $S$: in particular, for a particular choice of the basis, that we shall discuss in the following, the elements $n_{\alpha \beta}$ are uncorrelated with eachothers, with variances that differ from the ones of the $m_{\alpha \beta}$.  
As a first step, we discuss how the general structure \eqref{eq:MatSplit} is recovered.

\subsubsection{Block structure of the conditioned Hessian}
\noindent
To compute the averages and covariances of the components $\tilde{\mathcal{M}}_{\alpha \beta}^a$, we group all the independent components of the unconditioned matrices $\mathcal{M}^a$ into an $n N(N+1)/2$-dimensional vector ${\bf M}=({\bf M}_{0}, {\bf M}_{1/2}, {\bf M}_{1})$, where ${\bf M}_\gamma= ({\bf M}_\gamma^{1}, \cdots, {\bf M}_\gamma^{n})$ for $\gamma \in \grafe{0, 1/2,1}$,
\begin{equation*}
 \begin{split}
 & {\bf M}_0^{a}= (\mathcal{M}_{11}^a,\mathcal{M}_{22}^a, \cdots,\mathcal{M}_{MM}^a,\mathcal{M}_{12}^a, \cdots, \cdots,\mathcal{M}_{M-1M}^a)\\
  &  {\bf M}_{1/2}^{a}= (\mathcal{M}_{1M+1}^a,\mathcal{M}_{1M+2}^a, \cdots, \cdots, \cdots,\mathcal{M}_{MN-1}^a)\\
   &   {\bf M}_1^{a}= (\mathcal{M}_{M+1M+1}^a, \cdots,\mathcal{M}_{N-1 N-1}^a,\mathcal{M}_{M+1M+2}^a,  \cdots),
 \end{split}
\end{equation*}
and $M \equiv N-n-1$. The vectors ${\bf M}_{0},  {\bf M}_{1}$ and ${\bf M}_{1/2}$ and have dimension $n M(M+1)/2$, $n^2 (n+1)/2$ and $n^2 M$, respectively. They group the Hessians coordinates along directions that belong both to $S^\perp$, or both to $S$, or one to each subspace, respectively. 
This decomposition reflects the one in \eqref{eq:BareHess}, except that now we consider all the $n$ replicas.  Analogously, we define the $n N$-dimensional vector $\tilde{{\bf g}}=(\tilde{{\bf g}}_{0}, \tilde{{\bf g}}_1)$, with $\tilde{{\bf g}}_{\gamma}= (\tilde{{\bf g}}_\gamma^1, \cdots, \tilde{{\bf g}}_\gamma^n)$, and:
\begin{equation*}
 \begin{split}
\tilde{{\bf g}}_0^a&= (g_1^a,\cdots,g_M^a),\\  
\tilde{{\bf g}}_1^a&=(g_{M+1}^a, \cdots, g_{N-1}^a, \tilde{g}^a_N).
 \end{split}
\end{equation*}
We recall that $\tilde{g}_N^a = {\bm \nabla}h^a \cdot {\bm \sigma}^a= p\, h^a+  \sqrt{2 N} r \tonde{p f_k(\overline{q})- f'_k(\overline{q})\overline{q}}$, and thus for any fixed $\overline{q}$, conditioning to $h^a= \sqrt{2 N} \epsilon$ is equivalent to conditioning to $\tilde{g}_N^a= \sqrt{2 N} \quadre{ p\; \epsilon+  r \tonde{p f_k(\overline{q})- f'_k(\overline{q})\overline{q}}}= \sqrt{2 N} u(\epsilon, \overline{q})$. \\
Before conditioning, the covariance matrices of ${\bf M}$ and $\tilde{{\bf g}}$ have a diagonal structure in this decomposition,
\begin{equation*}
\begin{split}
 \hat{\Sigma}_{{\bf M}{\bf M}}&= \begin{pmatrix}
                \hat{\Sigma}_{{\bf M}{\bf M}}^{0} &&0&&0\\
                0 &&\hat{\Sigma}_{{\bf M}{\bf M}}^{1/2}&&0\\
                0 &&0&&\hat{\Sigma}_{{\bf M}{\bf M}}^{1}
              \end{pmatrix}
               \end{split}
\end{equation*}
and 
\begin{equation*}
\begin{split}
                             \hat{\Sigma}_{\tilde{{\bf g}}\tilde{{\bf g}}}&= \begin{pmatrix}
                \hat{\Sigma}_{\tilde{{\bf g}}\tilde{{\bf g}}}^{0} &&0\\
                             0&&\hat{\Sigma}_{\tilde{{\bf g}}\tilde{{\bf g}}}^{1}
              \end{pmatrix}
              \end{split}
\end{equation*}
as it follows from \eqref{eq:HessTang} and \eqref{eq:CovGrad}. Thus, before conditioning the correlations between Hessians of different replicas preserve the block structure \eqref{eq:BareHess}, in the sense that the components in the block $\mathcal{M}^a_\gamma$ of the replica $a$ are correlated only with the component in the correspondent block $\mathcal{M}^b_\gamma$ of the other replicas $b$. The conditioning to $\tilde{{\bf g}}$ preserves the diagonal form of the covariance matrix as well: indeed, the covariances between ${\bm M}$ and $\tilde{{\bf g}}$, see \eqref{eq:CorelationsHessianGrad}, are of the form:
\begin{equation*}
  \hat{\Sigma}_{{\bf M}\tilde{{\bf g}}}= \begin{pmatrix}
                0 && 0\\
                \hat{\Sigma}_{{\bf M}\tilde{{\bf g}}}^{\frac{1}{2} 0} &&0\\
                0 &&\Sigma_{{\bf M}\tilde{{\bf g}}}^{1 1}
              \end{pmatrix},
\end{equation*}
so that the conditional covariances read:
\begin{equation}\label{eq:CovCart}
  \hat{\Sigma}_{{\bf M}|\tilde{{\bf g}}}=
   \begingroup
\renewcommand*{\arraystretch}{1.5}
  \begin{pmatrix}
                \hat{\Sigma}_{{\bf M}{\bf M}}^{0} &\hspace{-1.5 cm}0&\hspace{-1.5 cm}0\\
                0 &\hspace{-.5 cm}\hat{\Sigma}_{{\bf M}{\bf M}}^{1/2}- \hat{\Sigma}_{{\bf M}\tilde{{\bf g}}}^{\frac{1}{2} 0}  (\hat{\Sigma}_{\tilde{{\bf g}}\tilde{{\bf g}}}^{-1})^{00}\hat{\Sigma}_{\tilde{{\bf g}}{\bf M}}^{0 \frac{1}{2}} &\hspace{-1.5 cm}0\\
                0 &\hspace{-1.5 cm}0&\hspace{-2.3 cm}\hat{\Sigma}_{{\bf M}{\bf M}}^{1}- \hat{\Sigma}_{{\bf M}\tilde{{\bf g}}}^{1 1}  (\hat{\Sigma}_{\tilde{{\bf g}}\tilde{{\bf g}}}^{-1})^{11}\hat{\Sigma}_{\tilde{{\bf g}}{\bf M}}^{11}
              \end{pmatrix}.
                \endgroup
          \end{equation}
As claimed in the main text, the covariances of the largest blocks $\mathcal{M}^a_0$ are left untouched by the conditioning, so that the components of this block form a GOE matrix with variance $\sigma^2=p(p-1)$. \\
We now analyze the structure of $\hat{\Sigma}^{1/{2}}_{{\bf M}|\tilde{{\bf g}}}$, to show that, for generic choices of the basis in the tangent planes, correlations are induced in the blocks $\mathcal{M}^a_{1/2}$ only between elements belonging to the same line. One has:
\begin{equation*}
 \tonde{\hat{\Sigma}^{1/2}_{{\bf MM}}}^{ab}_{\alpha \gamma, \beta \delta}=\langle \mathcal{M}^a_{\alpha \gamma} \mathcal{M}^b_{\beta \delta}\rangle_c=\delta_{\alpha \beta} S^{ab}_{\gamma \delta}, 
\end{equation*}
where $S^{ab}$ is a block of size $n\times n$, equal for every $\alpha$, with components
\begin{equation}
\begin{split}
S^{ab}_{\gamma \delta}&= p(p-1)(p-2)q^{p-3} ({\bf e}^a_\gamma \cdot {\bm \sigma^b}) ({\bf e}^b_\delta \cdot {\bm \sigma^a})\\
&+p(p-1) Q_{ab}^{p-2} ({\bf e}^a_\gamma \cdot {\bf e}^b_\delta)
\end{split}
\end{equation}
and with $Q_{ab}=\delta_{ab}+ (1-\delta_{ab})q$. 
Moreover
\begin{equation}
 \tonde{\hat{\Sigma}^{\frac{1}{2} 0}_{{\bf M} \tilde{{\bf g}}}}^{ab}_{\alpha \gamma, \beta }= \langle \mathcal{M}^a_{\alpha \gamma}\, { g}^b_{\beta} \rangle_c= \delta_{\alpha \beta} p(p-1) q^{p-2} ( {\bf e}^a_\gamma\cdot {\bm \sigma}^b).
 \end{equation}
Finally,
\begin{equation*}
 (\hat{\Sigma}_{\tilde{{\bf g}}\tilde{{\bf g}}}^{0})^{-1}= \frac{1}{p}
 \left( \begin{array}{cccc} 
 \alpha_0 \hat{1}  & \alpha_1 \hat{1} & \cdots & \alpha_1\hat{1} \\
 \alpha_1 \hat{1} &\alpha_0 \hat{1} &\cdots& \cdots\\ 
\cdots  &\cdots  & \cdots  &\cdots\\ 
 \alpha_1\hat{1}  & \alpha_1 \hat{1} & \cdots &\alpha_0 \hat{1} 
\\
 \end{array} 
\right),
\end{equation*}
where the identity matrices have dimension $M \times M$, and
\begin{equation}
\begin{split}
 \alpha_0&= \frac{q+(n-2)q^p}{(1-q^{p-1})(q+(n-1)q^p)},\\
 \alpha_1&=-\frac{q^{p}}{(1-q^{p-1})(q+(n-1)q^p)}.
 \end{split}
\end{equation}
Doing the matrix multiplication, we find
\begin{equation}\label{eq:CorelationsBlock1half}
\begin{split}
 &\tonde{\hat{\Sigma}^{1/2}_{{\bf M|\tilde{{\bf g}}}}}^{ab}_{\alpha \gamma, \beta \delta}=\delta_{\alpha \beta}\; p(p-1)Q_{ab}^{p-2}({\bf e}_\gamma^a \cdot {\bf e}_\delta^b)+\\
 &\delta_{\alpha \beta}\; p(p-1)(p-2) q^{p-3}({\bf e}^a_\gamma \cdot {\bm \sigma}^b)({\bm \sigma}^a \cdot {\bf e}_\delta^b)-\\
 &\delta_{\alpha \beta}\; \frac{p(p-1)^2 q^{2p -4}}{1-q^{p-1}} \sum_{c (\neq a,b)} ({\bf e}^a_\gamma \cdot {\bm \sigma}^c) ({\bf e}^b_\delta \cdot {\bm \sigma}^c)-\\
 &\delta_{\alpha \beta}\;  p(p-1)^2\alpha_1 q^{2p-4} \sum_{c (\neq a)} \sum_{d (\neq b)} ({\bf e}^a_\gamma \cdot {\bm \sigma}^c) ({\bf e}^b_\delta \cdot {\bm \sigma}^d),
 \end{split}
 \end{equation}
 for $\gamma, \delta= M+1, \cdots, N-1$.
 Therefore, correlations arise only between elements $\mathcal{H}^a_{\alpha \beta}$ and  $\mathcal{H}^b_{\alpha \gamma}$, where $\alpha$ is a direction in $S^\perp$ while $\gamma, \delta$ are directions is $S$. For arbitrary $n$ the conditioning does not induce any non-zero average for these components, since such averages are proportional to the elements of $\tilde{{\bf g}}^a_0$, which are all set to zero.\\
We now come to the $n \times n$ blocks $\mathcal{M}^a_1$. From \eqref{eq:CovCart}, one sees that the conditional covariance matrix of these components is in general a dense matrix, meaning that all components are correlated with each others. Furthermore, the conditioning induces non-zero averages for these components. We denote with ${\bm \mu}^{1}_{{\bf M|\tilde{{\bf g}}}}$ the $n^2(n+1)/2$-dimensional vector whose components are the conditional averages $\langle \tilde{\mathcal{M}}^a_{\gamma \delta} \rangle$ of the elements in ${\bf M}_1$. We also introduce the $(n+1)$-dimensional vectors ${\bm \tau}^a=(\sigma^a_{N-n}, \cdots, \sigma^a_N)$ collecting the $(n+1)$ non-zero components of the ${\bm \sigma}^a$, and ${\bm \tau}^0=(0, \cdots, 1)$, as well as the $n(n+1)$-dimensional vectors ${\bm \chi}_1=\tonde{{\bm \tau}^1, \cdots, {\bm \tau}^n}$, $
{\bm \chi}_2=\tonde{{\bm \tau}^0, \cdots, {\bm \tau}^0}$, and ${\bm \chi}_3=\tonde{\sum_{a \neq 1}{\bm \tau}^a, \cdots, \sum_{a \neq n}{\bm \tau}^a}$. 
With this notation, it holds:
\begin{equation}\label{eq:AveragesBlocchetto}
 \begin{split}
{\bm \mu}^{1}_{{\bf M}|\tilde{{\bf g}}}&=  \Sigma_{{\bf M}\tilde{{\bf g}}}^{1}  (\Sigma_{\tilde{{\bf g}}\tilde{{\bf g}}}^{-1})^{1} \tonde{\sqrt{2 N} u(\epsilon, \overline{q})\,{\bm \chi}_2- \langle {\tilde{{\bf g}}\rangle} },
 \end{split}
\end{equation}
where the second term arises because of the signal, that induces non-zero averages to the components of $\tilde{{\bf g}}$.\\
This vector can be determined with the same strategy exploited in Sec. \ref{sec:JointDensity}. In fact, Eq. \eqref{eq:AveragesBlocchetto} can be re-expressed in terms of the inverse correlation matrix $\hat{C}^{-1}$ of the $N$-dimensional vectors ${\bm \nabla}h^a$, or more precisely of its $n(n+1) \times n(n+1)$ block associated to the last $n+1$ components for each replica. We introduce the $n(n+1) \times n(n+1)$ rotation matrix:
\begin{equation}
 \hat{R}=  \left( 
 \begingroup
\begin{array}{cccc} 
  \hat{R}^{1}  & 0 & 0 & 0 \\
0 & \hat{R}^2 &0& 0 \\
\cdots  &\cdots  & \cdots  &\cdots\\
0  &0 & 0 &\hat{R}^n
\\
 \end{array} 
 \endgroup
\right),
\end{equation}
where each block $\hat{R}^{a}$ is $(n+1) \times (n+1)$-dimensional, with columns given by the non-zero components of the vectors ${\bf e}_{M+1}^a, {\bf e}_{M+2}^a, \cdots, {\bm \sigma}^a$. Then $(\Sigma^{-1}_{\tilde{{\bf g}}\tilde{{\bf g}}})^1 = \hat{R}^T (\Sigma^{-1}_{{\bf G G}})^1 \hat{R}$, where $\Sigma^{-1}_{{\bf G G}}$ is the inverse correlation matrix of the $n(n-1)$-dimensional vector ${\bf G}=({\bm \nabla}h^1_{N-n}, \cdots {\bm \nabla}h^1_N, \cdots, \cdots, {\bm \nabla}h^n_{N-n}, \cdots, {\bm \nabla}h^n_N)$. Moreover, $\langle \tilde{{\bf g}}\rangle =-\sqrt{2 N} r f'_k(\overline{q}) \hat{R}^T {\bm \chi}_2$, and $\hat{R} {\bm \chi}_2 = {\bm \chi}_1$. Thus: 
\begin{equation}
 \frac{{\bm \mu}^{1}_{{\bf M}|\tilde{{\bf g}}}}{\sqrt{2N}}=  \Sigma^1_{{\bf M }\tilde{{\bf g}}} \hat{R}^T (\Sigma^{-1}_{{\bf G G}})^1 \tonde{ u(\epsilon, \overline{q}) \,{\bm \chi}_1+ r f'_k(\overline{q}) {\bm \chi}_2}.
\end{equation}
Now, the vectors $(\Sigma^{-1}_{{\bf G G}})^1 {\bm \chi}_i$ can be obtained from the vectors $\hat{C}^{-1} {\bm \xi}_i$ by projecting out the components in $S^\perp$. Using the results
 of Appendix \ref{app:QuadraticForm}, applying the rotation and contracting with the matrix $\Sigma^1_{{\bf M} \tilde{{\bf g}}}$ we obtain:
\begin{equation}\label{eq:Avergaes}
\begin{split}
 &\frac{\langle {\tilde{\mathcal{M}}^a_{\gamma \delta}}\rangle}{\sqrt{2N}}= \zeta_1 \sum_{b \neq a} ({\bm \sigma}^b \cdot {\bf e}^a_\gamma)({\bm \sigma}^b \cdot {\bf e}^a_\delta)+\\
 &\zeta_2 \tonde{( {\bf w_0} \cdot {\bf e}^a_\gamma) \sum_{b \neq a} {\bm \sigma}^b \cdot {\bf e}^a_\delta +  ({\bf w_0} \cdot {\bf e}^a_\delta) \sum_{b (\neq a)} {\bm \sigma}^b \cdot {\bf e}^a_\gamma}+ \\
 &\zeta_3 \;{\sum_{b (\neq a)} {\bm \sigma}^b \cdot {\bf e}^a_\gamma}\: {\sum_{c (\neq a)} {\bm \sigma}^c \cdot {\bf e}^a_\delta},
 \end{split}
\end{equation}
where $\zeta_i= \zeta_i(n, \epsilon, \overline{q}, q;r)$ are linear combinations of the matrix elements $Y_{ij}$, and read:
\begin{equation*}
 \begin{split}
&\zeta_1=\frac{(p-1)[1+ q^{p-2} (1 + 2 (n-1) q)]u(\epsilon, \overline{q})}{p (1 - q) (1 + (n-1) q)-(1-q^{2-p}) (1 + (n-1) q^p)}\\
&+\frac{(p-1)\overline{q}[(p-2)q-(p-1) q^2 + q^{p}] r f'_k(\overline{q})}{q^2 [p (1 - q) (1 + (n-1) q)-(1-q^{2-p}) (1 + (n-1) q^p)]},\\
&\zeta_2=\frac{(p-1) q^{p-1}}{q + (n-1) q^p}r f'_k(\overline{q}),\\
&\zeta_3=\frac{-2(p-1) q^{p-1} u(\epsilon, \overline{q})}{p (1 - q) (1 + (n-1) q)-(1-q^{2-p}) (1 + (n-1) q^p)} \\
&+ \frac{2 (p-1)^2 \overline{q} (1-q) q^{p-1} r f'_k(\overline{q}) / [(1 + (n-1) q^p)]}{p (1 - q) (1 + (n-1) q)-(1-q^{2-p}) (1 + (n-1) q^p)}.
 \end{split}
\end{equation*} 
 
\subsubsection{Explicit covariances in a given basis}
\noindent
 As we have previously remarked, the structure of \eqref{eq:CorelationsBlock1half} and \eqref{eq:Avergaes} is a sole consequence of the separation of the subspaces $S$ and $S^\perp$, and it is independent on the choice of the basis in $S^\perp$ (provided the same choice is made for each tangent plane). Therefore, the correlations and averages depend explicitly only on the last $n$ basis vectors in each tangent plane (having $\alpha= M+1, \cdots, N-1$), which are the ones having non-zero projections on the subspace $S$. We now discuss two possible choices of these basis vectors that strongly simplify the covariances \eqref{eq:CorelationsBlock1half} of the single-replica matrix $\tilde{\mathcal{M}}^{a}$. \\ 
 The first possibility is to choose the basis in such a way that: (i) ${\bf e}_{N-1}^a$ is the projection on the tangent plane at ${\bm \sigma}^a$ of the special vector ${\bf w_0}$, (ii) ${\bf e}_{N-2}^a$ is the projection on the tangent plane at ${\bm \sigma}^a$ of the vector $\sum_{b \neq a}{\bm \sigma}^b$, made orthogonal to ${\bf e}_{N-1}^a$, (iii) the remaining $n-2$ basis vectors are of the form ${\bm \sigma}^{b_1}+ {\bm \sigma}^{b_2}+ \cdots {\bm \sigma}^{b_k}- k {\bm \sigma}^{b_{k+1}} $ for non-repeating indices $b_i \neq a$. For $a=1$, this leads to 
 $$ {\bf e}_{M+k}^1=\frac{1}{\sqrt{(k+1)k (1-q)}} \tonde{\sum_{b=2}^{k+1} {\bm \sigma}^b - k {\bm \sigma}^{k+2}}$$ for $1 \leq k \leq n-2$, while
 \begin{equation*}
  \begin{split}
   {\bf e}_{N-2}^1&=\sqrt{\frac{n (1-\overline{q}^2)}{A}}\sum_{b=2}^n {\bm \sigma}^b-\sqrt{\frac{n (1-\overline{q}^2)}{A}}(n-1)q {\bm \sigma}^1\\
  &-\sqrt{\frac{n}{A (1-\overline{q}^2)}}{(n-1) \overline{q}(1-q)} \tonde{{\bf w_0}- \overline{q} {\bm \sigma}^1}  
  \end{split}
 \end{equation*}
with $A=n(n-1)(1-q) \quadre{1 - n \overline{q}^2 +(n-1) q}$, and 
\begin{equation*}
  {\bf e}_{N-1}^1=\frac{1}{\sqrt{1-\overline{q}^2}}\tonde{{\bf w_0}- \overline{q} {\bm \sigma}^1}.
\end{equation*}
It can be checked that these vectors, together with ${\bm \sigma}^1$, form an orthonormal basis of the subspace $S$. Analogous choices can be made for any replica $a$. \\
Plugging these vectors into \eqref{eq:CorelationsBlock1half} with $a=b$, we find that for any $\gamma=M+1, \cdots, N-3$ it holds $\sum_{c (\neq a)} ({\bf e}^a_\gamma \cdot {\bm \sigma}^c) ({\bf e}^a_\delta \cdot {\bm \sigma}^c)= \delta_{\gamma 	\delta} (1-q)$, and $\sum_{c (\neq a)} ({\bf e}^a_\gamma \cdot {\bm \sigma}^c)=0$. This implies that the components $\mathcal{M}_{\alpha \gamma}^a$ for $\alpha \leq M$ and $M+1 \leq \gamma \leq N-3$ are uncorrelated with each others, and have a modified variance with respect to the one of the larger block, given by:
\begin{equation}\label{eq:VarianzeColonneEasy}
 \sigma_\gamma^2 \equiv \langle \quadre{\tilde{\mathcal{M}}_{\alpha \gamma}^a}^2 \rangle_c= p(p-1) \tonde{1- \frac{(p-1)q^{2p-4}(1-q)}{1-q^{p-1}} }.
\end{equation}
The components $\tilde{\mathcal{M}}^a_{\alpha N-2}$ and  $\tilde{\mathcal{M}}^a_{\alpha N-1}$ are instead correlated, since for $b \neq a$ it holds ${\bf e}^a_{N-1} \cdot {\bm \sigma}^b = {\overline{q} (1-q)}/{\sqrt{1-\overline{q}^2}}$ and ${\bf e}^a_{N-2} \cdot {\bm \sigma}^b =\quadre{{(1-q) \left(1-n \overline{q}^2+(n-1)q\right)}/ {(n-1)(1-\overline{q}^2)}}^{1/2}$, implying:
\begin{equation}\label{eq:MatrixVariancesBlocchetto}
\begin{split}
\Sigma_{12} &\equiv
 \begin{pmatrix}
  \langle \quadre{\tilde{\mathcal{M}}_{\alpha N-2}^a}^2 \rangle_c& \langle \tilde{\mathcal{M}}_{\alpha N-2}^a \tilde{\mathcal{M}}_{\alpha N-1}^a \rangle_c\\
  \langle  \tilde{\mathcal{M}}_{\alpha N-2}^a \tilde{\mathcal{M}}_{\alpha N-1}^a \rangle_c& \langle \quadre{\tilde{\mathcal{M}}_{\alpha N-1}^a}^2 \rangle_c
 \end{pmatrix}\\
 &=
 \begin{pmatrix}
 \sigma^2_{22} & \sigma^2_{12}\\
  \sigma^2_{12} & \sigma^2_{11}
\end{pmatrix},
\end{split}
\end{equation}
with 
\begin{equation*}\label{eq:VarianzeColonneUltime}
\begin{split}
\frac{\sigma^2_{22}}{p(p-1)}&=1- \frac{(p-1) q^{2 p-4} (1-q) \left(1-n \overline{q}^2+(n-1) q\right)}{\left(1-\overline{q}^2\right) \left(1-q^{p-1}\right) \left(1+(n-1)
   q^{p-1}\right)}\\
  \frac{ \sigma^2_{11}}{p(p-1)}&=1-\frac{(n-1) (p-1) q^{2 p-4} \overline{q}^2 (1-q)^2}{\left(1-\overline{q}^2\right) \left(1-q^{p-1}\right) \left(1+(n-1) q^{p-1}\right)}\\
 \frac{\sigma^2_{12}}{p(p-1)} &=-\frac{(p-1)q^{2 p-4}  \overline{q} (1-q)  \sqrt{A/n}}{\left(1-\overline{q}^2\right)\left(1-q^{p-1}\right) \left(1+(n-1) q^{p-1}\right)}.
   \end{split}
\end{equation*}
Thus, with this choice of basis in $S$, the pair of elements in $\mathcal{M}^a_{1/2}$ belonging to the same row and to the last two columns are correlated with each others, and other than that all elements are independent, with variances that depend on the column to which they belongs to.\\
For what concerns the averages \eqref{eq:Avergaes}, we find that in this basis, for $\gamma=M+1, \cdots, N-3$, it holds 
\begin{equation}\label{eq:MedieSimple}
\frac{\langle \tilde{\mathcal{M}}^a_{\gamma \delta} \rangle}{ \sqrt{N}}= \delta_{\gamma \delta}\,\sqrt{2}(1-q) \zeta_1(\epsilon, \overline{q}, q;r),
\end{equation}
while 
\begin{equation*}\label{eq:MatrixAveragesBlocchetto}
 M_{12}\equiv\frac{1}{ \sqrt{N}}\begin{pmatrix}
  \langle \tilde{\mathcal{M}}_{N-2 N-2}^a \rangle& \langle \tilde{\mathcal{M}}_{N-2 N-1}^a \rangle\\
  \langle  \tilde{\mathcal{M}}_{N-2 N-1}^a \rangle& \langle \tilde{\mathcal{M}}_{N-1 N-1}^a \rangle
 \end{pmatrix} =
 \begin{pmatrix}
 \mu_{22} & \mu_{12}\\
  \mu_{12} & \mu_{11}
\end{pmatrix},
\end{equation*}
with the functions $\mu_{ij}=\mu_{ij}(n,\epsilon, \overline{q}, q;r)$ being a linear combination of the $\zeta_i$. \\
For $n \to 1$, one can check that the equality $\mu_{22}(1,\epsilon, \overline{q}, q;r)=(1-q)\zeta_1(1,\epsilon, \overline{q}, q;r)$ holds, while $\mu_{11}(1,\epsilon, \overline{q}, q;r)=0= \mu_{12}(1,\epsilon, \overline{q}, q;r)$. Thus, in this limit there are $(n-1) \to 0$ columns having equal, non-zero average. Similarly, $\sigma_{22} \to \sigma_\gamma$, while $\sigma^2_{11}=p(p-1)$ and $\sigma_{12}=0$. Therefore, the GOE- structure of the unconditioned matrix is recovered, since the number of columns with different average and variance $\sigma^2_\gamma$ is $(n-1)$, and goes to zero. This is the way the annealed limit is recovered.\\
This choice of basis in $S$ is such that there is a unique vector in each tangent plane, ${\bf e}^a_{N-1}$, having a non-zero overlap with the direction of the signal ${\bf w_0}$. 
 An alternative choice can be made, for instance to have the mutual independence of the matrix elements $\tilde{\mathcal{M}}^a_{\alpha N-2}$ and $\tilde{\mathcal{M}}^a_{\alpha N-1}$ for $\alpha \leq M$. To this aim, it is sufficient to choose as basis vectors the linear combinations of the vectors ${\bf e}^1_{N-2}, {\bf e}^1_{N-1}$ that diagonalize the matrix \eqref{eq:MatrixVariancesBlocchetto}. This is given by:
\begin{equation}\label{eq:SmallRotation}
 \begin{split}
{\bf e}'_{N-2}&= \frac{1}{\sqrt{z_{N-2}}}\tonde{\sum_{b \neq a} {\bm \sigma}^b - q (n-1) {\bm \sigma}^a},\\
{\bf e}'_{N-1}&= \frac{1}{\sqrt{z_{N-1}}} \tonde{-\overline{q}\sum_{b=1}^n {\bm \sigma}^b +(1+(n-1)q) {\bf w_0}},
 \end{split}
\end{equation}
with $z_{N-2}=(n-1)(1-q)(1+(n-1)q)$ and $z_{N-1}=(1- n \overline{q}^2+(n-1) q)(1+(n-1)q)$. In this new basis:
\begin{equation}
   \Sigma_{12} \rightarrow \begin{pmatrix}               \tilde{\sigma}_\gamma^2    &0\\
   0&p(p-1)           
   \end{pmatrix} 
 \end{equation}
 with 
 \begin{equation}\label{eq:SigmaTilde}
 \frac{\tilde{\sigma}_\gamma^2}{p(p-1)}=1-\frac{(p-1)(1-q) (1+(n-1) q) q^{2 p-2}}{\left(q-q^p\right) \left(q+(n-1) q^p\right)},
 \end{equation}
meaning that the components of the last two columns of $\tilde{\mathcal{M}}/ \sqrt{N}$ are now independent with variance $\tilde{\sigma}_\gamma^2$ and $p(p-1)$, respectively (the fact that the variance along the direction ${\bf e}'_{N-1}$ is equal to the unconditioned one stems for the fact that  ${\bf e}'_{N-1}$ is orthogonal to any ${\bm \sigma}^a$). Note that in the eigenstates basis, the covariances depend only on the overlap $q$ and are independent on $\overline{q}$: this is natural to expect, since the fluctuating part of the Hamiltonian is the p-spin part, that is blind to the special direction ${\bf w_0}$. The information on the signal is carried by the deterministic part, 
which in the rotated basis reads:
\begin{equation*}
 M_{12} \hspace{-.08 cm}\rightarrow \hspace{-.08 cm}\begin{pmatrix}
 \tilde{\mu}_{22} & \tilde{\mu}_{12}\\
  \tilde{\mu}_{12} & \tilde{\mu}_{11}
\end{pmatrix}\hspace{-.08 cm}=\hspace{-.08 cm}\begin{pmatrix}
               \hspace{-.08 cm}u(\epsilon, \overline{q})\, a_{22}^{(1)}+ r f'_k(\overline{q})\, a_{22}^{(2)}& r f'_k(\overline{q})\, a_{12} \\
  \hspace{-.08 cm}r f'_k(\overline{q})\, a_{12}& 0                 \end{pmatrix} ,
 \end{equation*}
  with 
\begin{equation}
\begin{split}
&\frac{a_{22}^{(1)}}{p-1}\hspace{-.1cm}= \frac{\sqrt{2} (1-q) (1+(n-1) q) \left(1-q^{p-2}\right)}{p (1-q) (1+(n-1)q)-
   \left(1-q^{2-p}\right) \left(1+(n-1) q^p\right)},\\
   &\frac{a^{(2)}_{22}}{p-1}\hspace{-.1cm}={ \sqrt{2}\overline{q} (1-q) q^{p-2} }\times \\
  &\frac{2q-(1-(n-1)q)[q^p+q(p(1-q)+q)]}{(n\hspace{-.05cm}-\hspace{-.05cm}1)q^{2 p}+q^p \left(1-p (1-q)(1+(n\hspace{-.05cm}-\hspace{-.05cm}1) q)-(n\hspace{-.05cm}-\hspace{-.05cm}1) q^2\right)-q^2},\\
&\frac{a_{12}}{p-1}\hspace{-.1cm}=\frac{\sqrt{2 A/n}}{q^p +(n-1)q}.
   \end{split}
\end{equation}
In summary, with this second choice of basis vectors in $S$ we find that the decomposition \eqref{eq:MatSplit} holds, with a deterministic matrix $\mathcal{D}^a$ equal to
\begin{equation}\label{eq:Deterministic}
\mathcal{D}^a=\left( 
\setlength{\arraycolsep}{.1pt}
\begin{array}{c| c}
   \begin{array}{ccccc}
 0 &  0& \cdots &\cdots& 0  \\
 0 &  0& \cdots&\cdots   &0  \\
     \cdots  & \cdots & \cdots & \cdots& \cdots\\
    \cdots    & \cdots &\cdots&  \cdots& \cdots\\ 
           0  & 0& \cdots&\cdots &0\\
  \hline
   0 &  0& \cdots & \cdots  &0\\
       \cdots  &  \cdots & \cdots& \cdots & \cdots\\
          0 & 0& \cdots &\cdots &0\\
                0 & 0& \cdots &\cdots &0\\
                   0 & 0& \cdots &\cdots &0
  \end{array}
  &
      \begin{array}{ccccc} 0 &  \cdots &  \cdots &  \cdots &0 \\
0 &  \cdots &  \cdots&  \cdots  &0 \\
  \cdots & \cdots&  \cdots&  \cdots  & \cdots\\
  \cdots & \cdots &  \cdots&  \cdots & \cdots\\
     0 &  \cdots&  \cdots&  \cdots   &0\\
    \hline
   \mu_\gamma &  0  &  \cdots&  \cdots &0 \\
0 & \mu_\gamma& \cdots&  \cdots  &0 \\
   0 &  0  & \mu_\gamma&  \cdots &0 \\
  0 &  \cdots &  \cdots &\tilde{\mu}_{22} & \tilde{\mu}_{12}\\
    0 & \cdots &  \cdots &\tilde{\mu}_{12} & \tilde{\mu}_{11}\\
                     \end{array} 
              \end{array}
      \right)
\end{equation}
and a stochastic matrix $\mathcal{S}^a$ having the block structure \eqref{eq:MatrixStochastic} with (i) $m_{\alpha \beta}$ gaussian iid with variance $\sigma^2=p(p-1)$, (ii) $n_{ik}$ gaussian iid with variances $\sigma_\gamma^2, \tilde{\sigma}_\gamma^2$ and $\sigma^2$ for $k=M+1, \cdots, N-3$, $k=N-2$ and $k=N-1$, respectively, (iii) $q_{ij}$ Gaussian, in general mutually correlated. 
 We have not determined their covariances since, as we argue in Appendix \ref{app:ThirdOrderEquations}, their expression is not necessary to characterize the lowest order corrections to the density of states of the matrices $\tilde{\mathcal{H}}^a$.\\
 Finally, we point out that with this second choice of basis, the term deriving from the rank-1 perturbation in \eqref{eq:AgainDecomposHessian} is no longer diagonal. For later convenience, we define the constants:
\begin{equation}\label{eq:MuBar}
 \overline{\mu}_{ij}=\tilde{\mu}_{ij}- r \sqrt {2} f''_k(\overline{q}) ({\bf e}'_{N-i} \cdot {\bf w_0}) ({\bf e}'_{N-j} \cdot {\bf w_0}).
\end{equation}

\subsection{Isolated eigenvalues of the conditioned Hessians}\label{app:ThirdOrderEquations}
\noindent
In this Appendix, we derive the equations satisfied by the isolated eigenvalues of the conditioned Hessian matrices $\tilde{\mathcal{H}}$, whenever they exist. We focus of the spectrum of the centered matrix $\mathcal{A}$ defined by
\begin{equation}
 \frac{\tilde{H}}{\sqrt{N}}\equiv \mathcal{A}- \sqrt{2} u(\epsilon, \overline{q})  \hat{1},
\end{equation}
and having itself the block structure
\begin{equation}\label{eq:newMatrix} 
 \mathcal{A}= \begin{pmatrix}
               A_0 & A_{1/2}\\
               A_{1/2}^T & A_{1}
              \end{pmatrix},
\end{equation}
where the largest $(N-1-n) \times (N-1-n)$ block $A_0$ is a GOE with $\sigma^2=p(p-1)$, and $A_{1/2}$ and $A_1$ have the statistics described in Appendix~\ref{app:Hessian}. In particular, we choose the basis in the subspace $S$ to be equal to the second one discussed in the Appendix.\\
In the large-$N$ limit, the bulk of the density of eigenvalues of $\mathcal{A}$ is controlled by the largest block $A_0$, and is thus a centered semicircle. We aim at determining the poles of the resolvent of \eqref{eq:newMatrix} that lie on the real axis outside the support of the semicircle, meaning that they are smaller than $ -2 \sqrt{p(p-1)}$. From the block structure of $\mathcal{A}$ it follows that the trace of $(z- \mathcal{A})^{-1}$ has two contributions, one coming from the largest $(N-1-n) \times (N-1-n)$ block, and one given by the small $n \times n$ block. We focus on this second contribution, since the corresponding matrix elements lie in the subspace $S$, and have therefore a non-zero overlap with the signal ${\bf w}_0$. The poles of the part  of the resolvent coming from this block correspond to isolated eigenvalues having an eigenvector with a non-zero component in the direction of the signal.\\
The quantity to determine are thus the poles of $ \langle \text{Tr} \grafe{1/N \cdot D(z)} \rangle$, where 
\begin{equation}\label{eq:Dz}
 D(z) \equiv z \hat{1}- {A_1}- A^T_{1/2} \tonde{z \hat{1}- {A_0}}^{-1} A_{1/2},
\end{equation}
and where now the average is over the distribution of the entries of the matrix $\mathcal{A}$. To compute these poles, we exploit the fact that in the large $N$ limit:
\begin{equation}\label{eq:MegaApprox}
\left \langle \text{Tr} \grafe{ \frac{1}{ N\, D(z)}} \right \rangle=\text{Tr} \grafe{  \frac{1}{N \langle D(z) \rangle}} .\end{equation}
This can be shown setting $D= \langle D \rangle + \delta D$ and making use of the expansion:
\begin{equation}
 D^{-1}= \frac {1}{\langle D \rangle} -  \frac {1}{\langle D \rangle} \delta D  \frac {1}{\langle D \rangle}+ \frac {1}{\langle D \rangle}\delta D  \frac {1}{\langle D \rangle}\delta D  \frac {1}{\langle D \rangle}+ \cdots
\end{equation}
Taking the average of the trace, we find that the corrections to the leading order term in \eqref{eq:MegaApprox} are given by:
\begin{equation}\label{eq:SecOrdCorrections}
\sum_{ijklm} \langle D \rangle^{-1}_{ij} \langle D \rangle^{-1}_{kl}\langle D \rangle^{-1}_{mi} \; \langle \delta D_{jk}  \delta D_{lm}\rangle,
\end{equation}
where the sum is over indices taking $n$ distinct values. The fluctuating part $\delta D$ of \eqref{eq:Dz} is contributed by two independent terms: the first one is made by the fluctuating components $q_{ij}/\sqrt{N}$ of the block $A_1$, while the second term is made by the fluctuating part of 
\begin{equation}
 X \equiv A^T_{1/2} \tonde{z \hat{1}- {A_0}}^{-1}\hspace{-.1 cm} A_{1/2}= \langle X \rangle + \delta X
\end{equation}
 around its mean value. Since the covariances of the $q_{ij}$ are $O(1)$, the first term contributes to the sum \eqref{eq:SecOrdCorrections} with $O(1/N)$. We now consider the contribution of the second term. For large $N$:
\begin{equation}
\left \langle \tonde{A^T_{1/2} \tonde{z \hat{1}- {A_0}}^{-1}\hspace{-.1 cm} A_{1/2}}_{ij} \right \rangle= G_{\sigma}(z) \frac{\sum_{\alpha} \langle n_{i \alpha} n_{\alpha j}\rangle}{N},
\end{equation}
where
\begin{equation}\label{eq:ResolventGOE}
 G_\sigma(z)=\frac{z+ \sqrt{z^2-4 \sigma^2}}{2 \sigma^2}
\end{equation}
is the resolvent of a GOE matrix with variance $\sigma^2$, while, given the results of Appendix \ref{app:Hessian}, we have
\begin{equation}
 \frac{\sum_{\alpha} \langle n_{i \alpha} n_{\alpha j}\rangle}{N}= \delta_{ij} \begin{cases}
 \sigma^2_\gamma &\mbox{ if } N-n \leq i \leq N-3\\
 \tilde{\sigma}^2_\gamma &\mbox{ if }  i = N-2\\
 \sigma^2 &\mbox{ if } i = N-1.
 \end{cases}
\end{equation}
Then:
\begin{equation}
\begin{split}
 \langle \delta X_{jk} \delta X_{lm} \rangle &= \frac{C_{jklm}^{(1)}}{N^2} \sum_{\alpha \gamma} \text{cov}\quadre{\tonde{\frac{1}{z - A_0}}_{\alpha \alpha}, \tonde{\frac{1}{z - A_0}}_{\gamma \gamma}} \\
 &+\frac{C_{jklm}^{(2)}}{N^2} \sum_{\alpha \gamma} \langle\tonde{\frac{1}{z - A_0}}^2_{\alpha \gamma} \rangle,
 \end{split}
\end{equation}
where the constants are of $O(1)$ in $N$. Thus, the behavior in $N$ of this second contribution is controlled by the decay of the covariances of the matrix elements of the resolvent of a GOE matrix; since the latter go to zero with $N$ as it can be readily checked in perturbation theory, it follows that this is a subleading correction to the leading term in \eqref{eq:MegaApprox}.
Therefore, in the large-$N$ limit:
\begin{equation}
 \langle D(z) \rangle = \left(
  \setlength{\arraycolsep}{.1pt}
 \begin{array}{ccccc}
  d&0&0 &\cdots & \cdots\\
  0&d&0&\cdots&\cdots\\
 0&\cdots& d&\cdots & \cdots\\
  \cdots & \cdots &\cdots&z\hspace{-.05cm}-  \hspace{-.05cm}\overline{\mu}_{22}\hspace{-.05cm}-  \hspace{-.05cm} \tilde{\sigma}^2_\gamma G_\sigma(z)&- \overline{\mu}_{12}\\
 \cdots&\cdots&\cdots& - \overline{\mu}_{12} &z\hspace{-.05cm}-  \hspace{-.05cm} \overline{\mu}_{11}\hspace{-.05cm}-  \hspace{-.05cm}\sigma^2 G_\sigma(z)\\
 \end{array}
\right),
\end{equation}
with $d=z-\mu_\gamma-\sigma_\gamma^2 G_\sigma(z)$.\\
The poles of the RHS of \eqref{eq:MegaApprox} can be found as zeros of the determinant of $\langle D(z) \rangle$, and are therefore solutions of:
\begin{equation}
\quadre{z-\mu_\gamma-\sigma_\gamma^2 G_\sigma(z)}^{n-2} \Pi_n(z)=0,
\end{equation}
with 
\begin{equation}
 \Pi_n(z)=\text{det}\grafe{ \begin{pmatrix}
  z- \overline{\mu}_{22}- \tilde{\sigma}^2_\gamma G_\sigma(z) & - \overline{\mu}_{12}\\
  -\overline{\mu}_{12} & z- \overline{\mu}_{11}- \sigma^2 G_\sigma(z)
 \end{pmatrix}}.
\end{equation}
In the following, we focus on the solutions of $\Pi_n(z)=0$, since the corresponding eigenvectors have a non-zero component with the signal. Before doing that, it is instructive to consider the stability criterion which is obtained within the annealed approximation: besides giving some indications on what happens qualitatively also in the quenched case, it turns out to be the right criterion for the stationary points that are at the equator, $\overline{q}=0$. 
\subsubsection{Isolated eigenvalue: annealed approximation}
\noindent
As stated in the main text, the annealed approximation is obtained setting $n=1$. In this case, since $\overline{\mu}_{12} \to 0$, $\overline{\mu}_{22} \to \mu_\gamma$ and $\tilde{\sigma}^2_\gamma \to \sigma^2_\gamma$, and given that $\overline{\mu}_{11} \to \mu=- \sqrt{2} r f''_k(\overline{q})(1-\overline{q}^2)$, one is left with the equation:
\begin{equation}\label{eq:EqationAnnealed}
z -\mu-\sigma^2 G_\sigma(z)=0.
\end{equation}
Substituting \eqref{eq:ResolventGOE} into \eqref{eq:EqationAnnealed} we get
\begin{equation}\label{eq:EqWithSquareAnnealed}
 \frac{z}{2} -\mu= \frac{\sqrt{z^2- 4 \sigma^2}}{2}.
\end{equation}
Taking the square of the resulting equation leads to the solution
\begin{equation}\label{eq:EigenAnnealedNonShifted}
z= - \sqrt{2} r f''_k(\overline{q})(1-\overline{q}^2) - \frac{p(p-1)}{\sqrt{2} r f''_k(\overline{q})(1-\overline{q}^2)}.
\end{equation}
This solution is defined for arbitrary values of $\mu$; however, it has to be considered only whenever it leads to a LHS of \eqref{eq:EqWithSquareAnnealed} that is positive. This holds provided that $\mu < -\sigma$, as one easily finds by substitution. Solving for $r$, one finds that this corresponds to:
\begin{equation}\label{eq:ConditionErre}
 r \geq \sqrt{\frac{p(p-1)}{2}}\frac{1}{(k-1)  \overline{q}^{k-2} \left(1-\overline{q}^2\right)}.
 \end{equation}
In particular, this implies that for $k=1$ there is no solution (i.e., no isolated eigenvalue exists), as well as for $k \geq 3$ and $\overline{q}=0$. We find that this remains true also within the quenched calculation.\\
The result \eqref{eq:EigenAnnealedNonShifted} is consistent with the fact that, for $n=1$, the conditioned Hessian coincides with the non-conditioned one (modulo the shift by $\sqrt{2N} u(\epsilon, \overline{q})$), and therefore it reduces to a GOE matrix perturbed with a rank-1 perturbations with negative eigenvalue equal to $\mu$. Eq. \eqref{eq:EigenAnnealedNonShifted} follows then from a general result holding for matrices of the form $\hat{M}=\hat{M}_0+ R(\mu)$, where $\hat{M}_0$ is a random matrix with eigenvalues density $\rho(\lambda)$ with compact support in $\quadre{a, b}$ and $R(\mu)$ is a rank-1 perturbation with negative eigenvalue $\mu$. In this case, it is known \cite{BenaychGeorges, Edwards} that an isolated eigenvalue exists whenever $\mu < 1/ G_{0}(a^-)$, where 
\begin{equation}\label{eq:ResZero}
 G_{0}(z)= \int \frac{\rho(\lambda)}{z-\lambda} d\lambda
\end{equation}
is the resolvent associated to the unperturbed random matrix, and it equals to $z(\mu)=G_0^{-1}(1/\mu)$, where $G_0^{-1} (\cdot)$ is the functional inverse of $G_0(\cdot)$. Applying this result to the GOE case ~\cite{footnote9}, one recovers Eq. \eqref{eq:EigenAnnealedNonShifted}.\\
 We point out that the eigenvalue of the full Hessian $\tilde{H}/ \sqrt{N}$ is obtained through an additional shift by the factor $\sqrt{2} u (\epsilon, \overline{q})$. For fixed $\overline{q}$, one finds that the latter vanishes at a value of energy given by:
\begin{equation}\label{eq:StabilityEnergy}
\begin{split}
&\epsilon_{st}^{(\text{ann})}(\overline{q},r) \equiv \frac{1}{\sqrt{2}p} \quadre{\mu + \frac{p(p-1)}{\mu} - \tonde{\frac{p}{k} -1}r \overline {q}^k}.
\end{split}
\end{equation}
This condition can be recovered within the replica framework, in the RS setting, see Eq.~\eqref{eq:ConditionIsolatedEv}.

\subsubsection{Isolated eigenvalue: quenched calculation}
\noindent
We now perform the quenched calculation of the isolated eigenvalue, which accounts for the correlations between minima at fixed, quenched realization of the random Gaussian field. This requires to determine the zeros of $\Pi_n(z)$ in the limit $n \to 0$, which are solutions of the equation:
\begin{equation}\label{eq:EqEigenQuenched}
 \tonde{z- \overline{\mu}_{22}- \tilde{\sigma}^2_\gamma G_\sigma(z)} \tonde{z- \overline{\mu}_{11}- \sigma^2 G_\sigma(z)}- \overline{\mu}_{12}^2=0,
\end{equation}
where all the functions appearing in \eqref{eq:EqEigenQuenched} are evaluated at $n=0$. Substituting \eqref{eq:ResolventGOE} into \eqref{eq:EqEigenQuenched} we obtain the equation:
\begin{equation}\label{eq:CubicEquation}
\begin{split}
&\frac{z^2}{2}+ z \tonde{-(\overline{\mu}_{11}+ \overline{\mu}_{22})+ \frac{\tilde{\sigma}^2_\gamma \overline{\mu}_{11}}{2 \sigma^2} + \frac{\overline{\mu}_{22}}{2}}\\
&+\tonde{\overline{\mu}_{22}\overline{\mu}_{11} - \overline{\mu}^2_{12} - \tilde{\sigma}^2_\gamma}=\\
&- \frac{\sqrt{z^2- 4 \sigma^2}}{2 \sigma^2} \tonde{- \sigma^2 z + \sigma^2 \overline{\mu}_{22} + \tilde{\sigma}^2_\gamma \overline{\mu}_{11}}.
\end{split}
\end{equation}
Taking the square of this equation and rearranging the components we find the third order equation
\begin{equation}\label{eq:ThirdOrder}
 F_3 z^3+ F_2 z^2 + F_1 z+ F_0=0,
\end{equation}
with coefficients
\begin{equation}
\begin{split}
 F_3&=- \overline{\mu}_{11} \tonde{1- \frac{\tilde{\sigma}^2_\gamma}{\sigma^2}},\\
 F_2&=\tonde{1- \frac{\tilde{\sigma}^2_\gamma}{\sigma^2}} \quadre{\overline{\mu}_{11}^2 + \overline{\mu}_{22}\overline{\mu}_{11}+ \sigma^2}+\overline{\mu}_{11} \overline{\mu}_{22}- \overline{\mu}_{12}^2,\\
 F_1&=-2 (\overline{\mu}_{22}\overline{\mu}_{11}- \overline{\mu}_{21}^2)\quadre{\overline{\mu}_{22}+\overline{\mu}_{11}- \frac{1}{2}\tonde{\frac{\tilde{\sigma}^2_\gamma}{\sigma^2}\overline{\mu}_{11}+ \overline{\mu}_{22}}}\\
 &-\tonde{1- \frac{\tilde{\sigma}^2_\gamma}{\sigma^2}}\sigma^2 \overline{\mu}_{22}-\sigma^2 \tonde{\overline{\mu}_{22}+ \frac{\tilde{\sigma}^4_\gamma}{\sigma^4}\overline{\mu}_{11} },\\
 F_0&= \tonde{\overline{\mu}_{22}\overline{\mu}_{11}-\overline{\mu}_{12}^2 - \tilde{\sigma}^2_\gamma}^2+ \sigma^2 \tonde{\overline{\mu}_{22}+ \frac{\tilde{\sigma}^2_\gamma}{\sigma^2}\overline{\mu}_{11}}^2.
 \end{split}
\end{equation}
Note that for $k=1$, $\overline{\mu}_{11}=0$ and thus \eqref{eq:ThirdOrder} reduces to a second order equation. For $k>1$, the Eq.\eqref{eq:ThirdOrder} has three solutions: for the values of parameters that we are considering, we find that at most one out of these three solutions is real and, for some values of parameters, exits the support of the semicircle. We denote this solution with $z_{I}$. To determine its domain of existence, we require the consistence with the equation \eqref{eq:CubicEquation}, i.e., we ask that, when evaluated at $z= z_I$, the RHS of \eqref{eq:CubicEquation} has a sign that is opposite to the sign of the expression $- \sigma^2 z_I + \sigma^2 \overline{\mu}_{22} + \tilde{\sigma}^2_\gamma \overline{\mu}_{11}$.
This condition has to be imposed separately, since \eqref{eq:ThirdOrder} is obtained from \eqref{eq:CubicEquation} by taking a square, and thus it is insensitive to the sign in front of the square root in the definition of the resolvent \eqref{eq:ResolventGOE} (it is a generalization to the quenched case of the condition $\mu <-\sigma$ found in the annealed approximation).
For those $z_I$ that meet this condition, the eigenvalue of the full Hessian $\tilde{H}/ \sqrt{N}$ is given by $z_I-\sqrt{2} u (\epsilon, \overline{q})$: imposing this expression to be zero gives the critical energy $\epsilon_{\text{st}}(\overline{q}, r)$ discussed in the main text.\\
The quenched calculation of the isolated eigenvalue gives results that are in general quantitatively different (although qualitatively very similar) with respect to the annealed one. The annealed limit is exact only at the equator $\overline{q}=0$, since in this case $q_{SP}(\epsilon, 0)= 0$ and thus the equation $\Pi_n(z)=0$ reduces to \eqref{eq:EqationAnnealed}. In this case, one finds that no eigenvalue exists for $k=1$ and $k \geq 3$, since the denominator in \eqref{eq:ConditionErre} vanishes, while it exists for $r \geq r_c$ for $k=2$.

\subsection{Kac-Rice calculation: additional results}\label{app:Results}
\begin{figure*}[!htbp]
\subfloat[][]{\includegraphics[width=.5\linewidth]{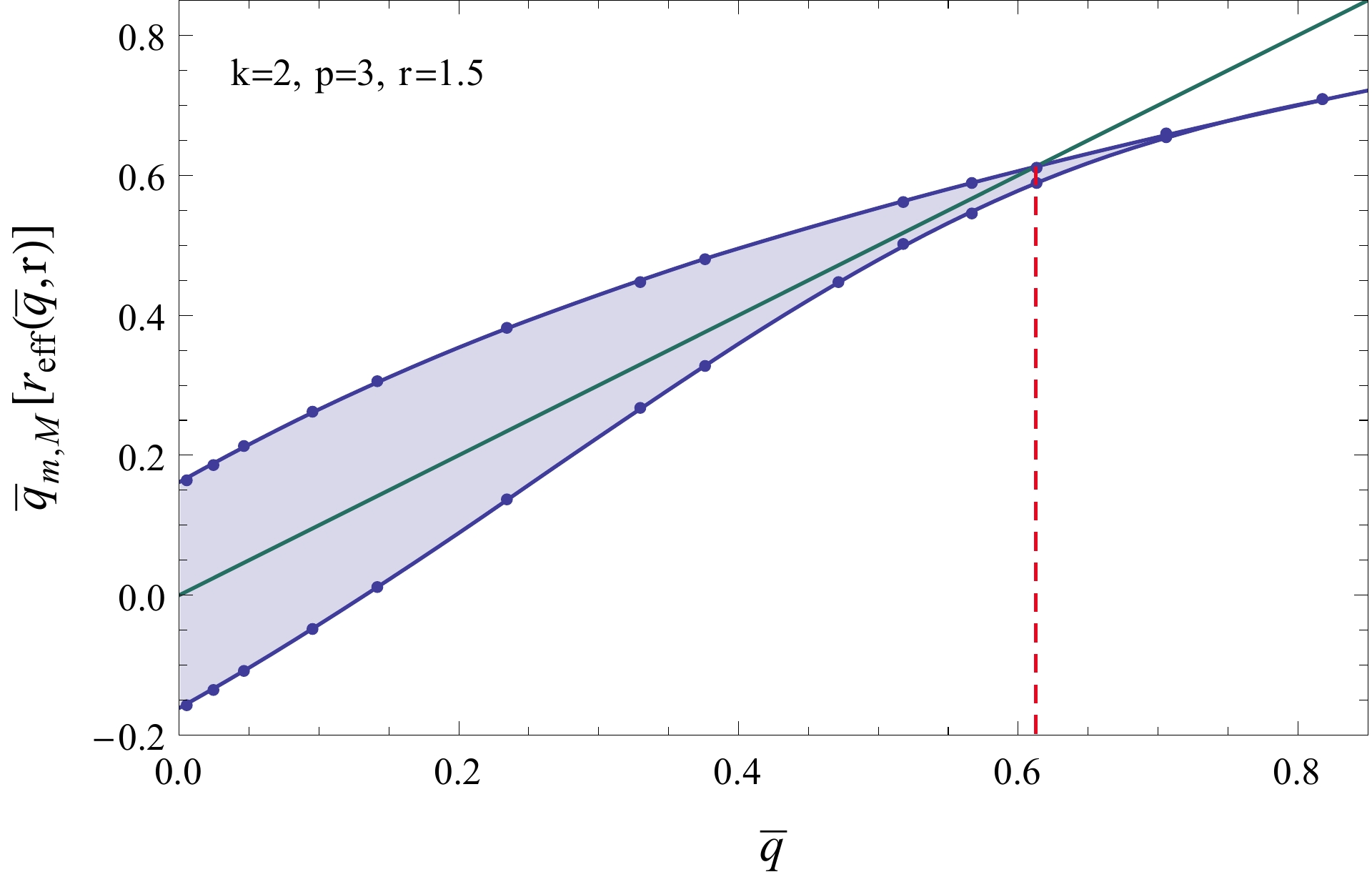}}
\subfloat[][]{ \includegraphics[width=.5\linewidth]{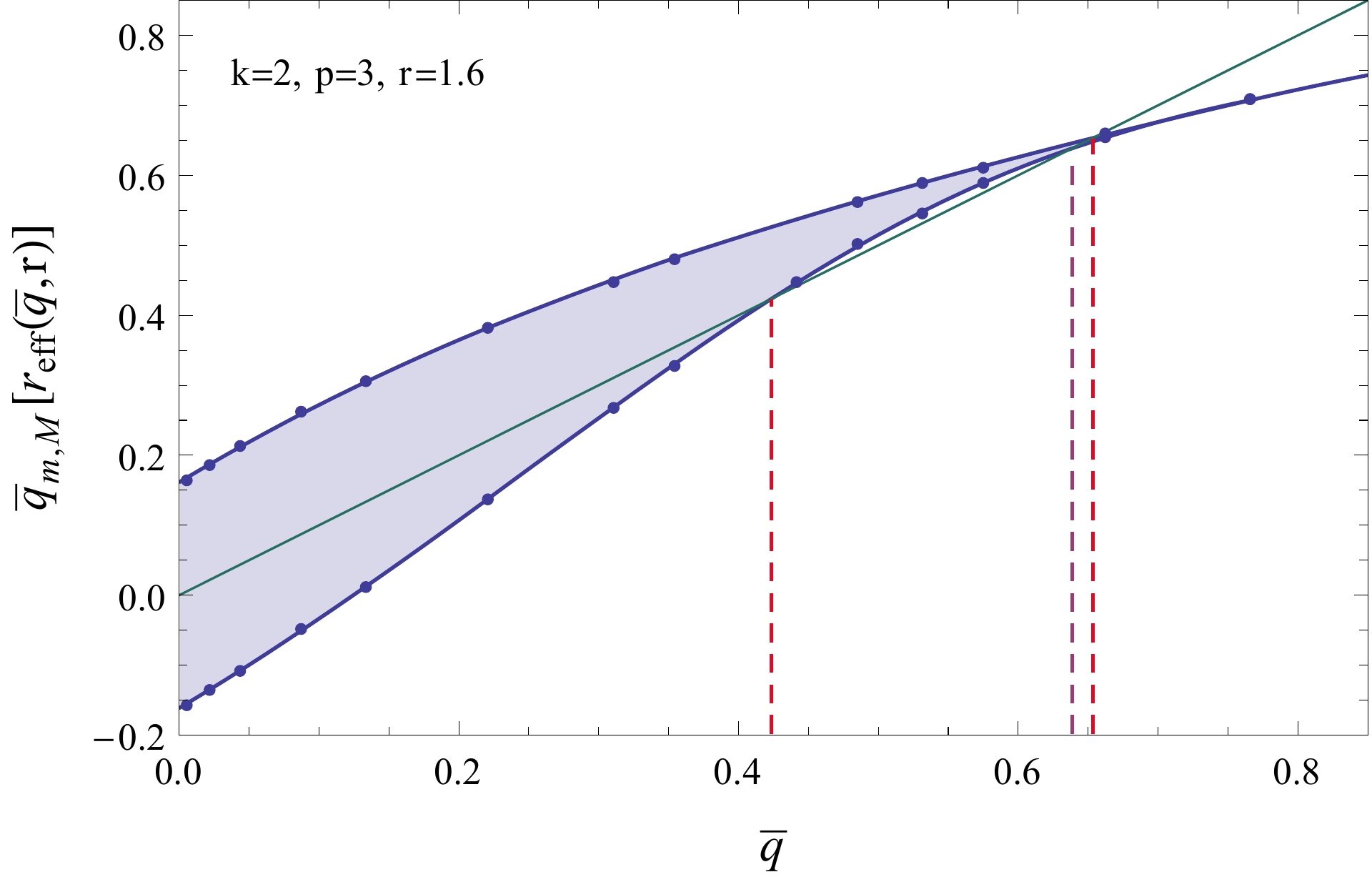}}\\
\subfloat[][]{ \includegraphics[width=.5\linewidth]{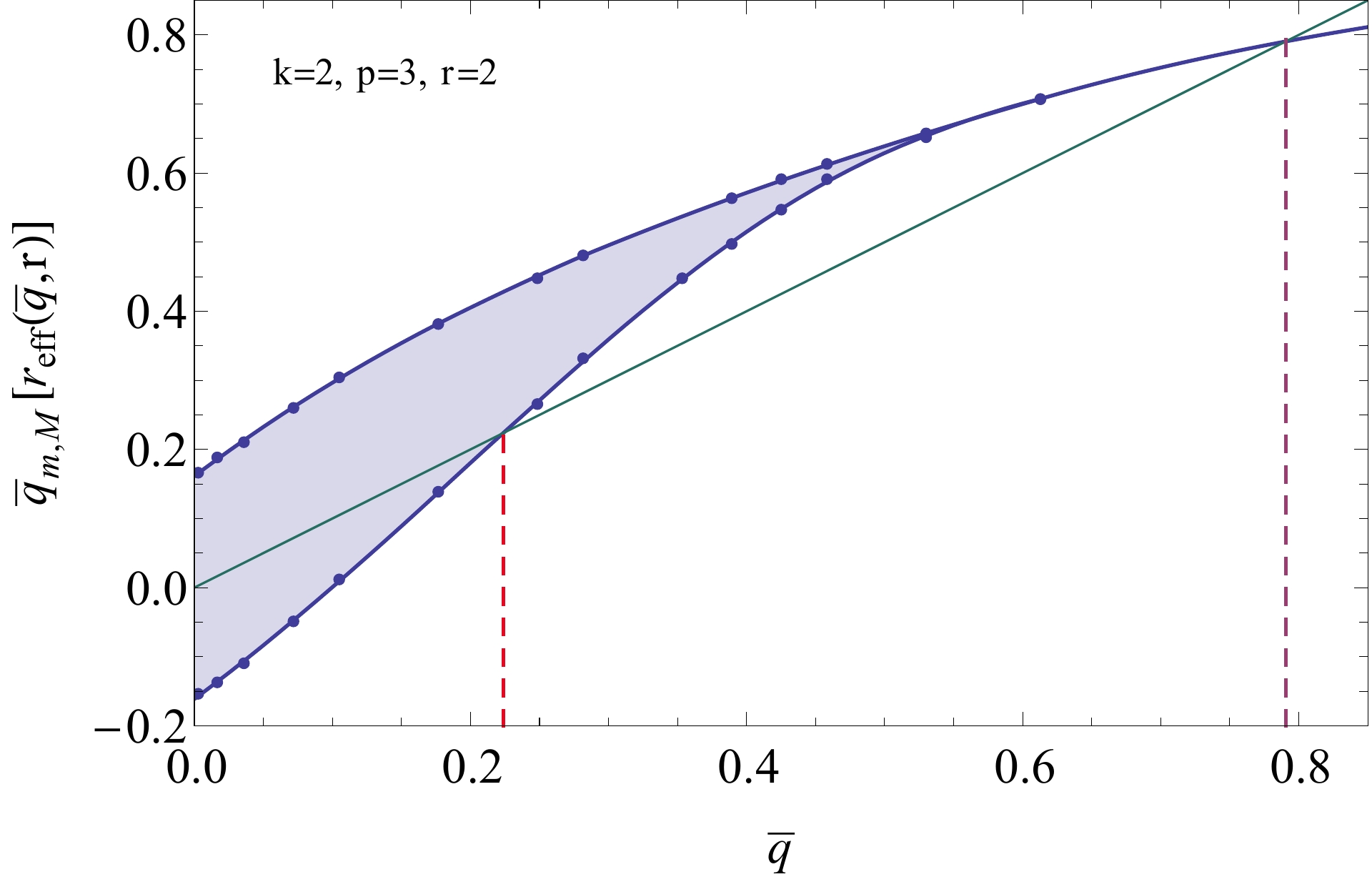}}
\subfloat[][]{\includegraphics[width=.5\linewidth]{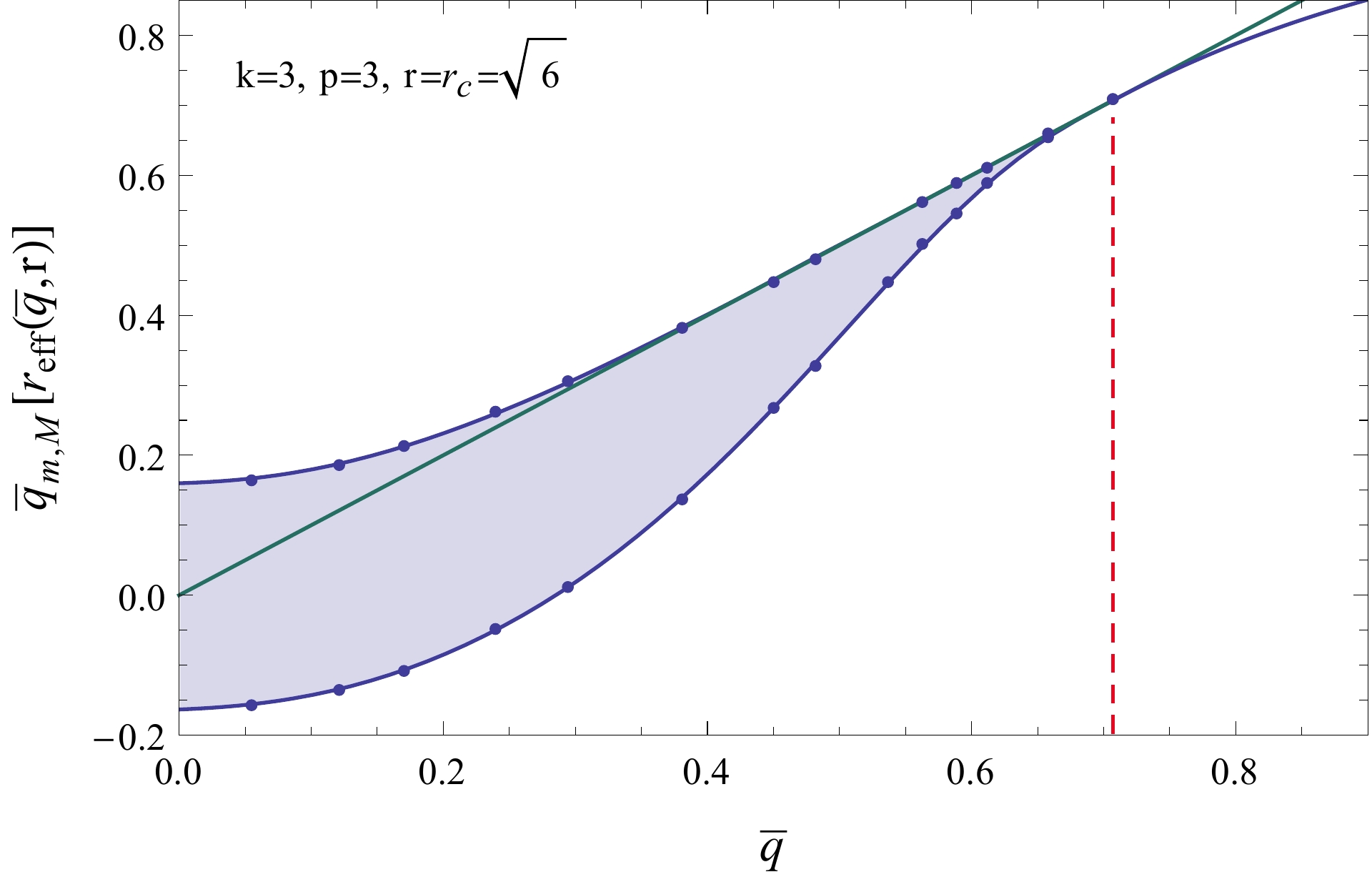}}
\caption{The blue curves are the functions $\overline{q}_m(r)$ and $\overline{q}_M(r)$ of Fig.~\eqref{fig:Bandek1}, plotted as a function of $r_1^{\text{eff}}(r, \overline{q})= r \overline{q}$ for $k=2$, and  $r_1^{\text{eff}}(r, \overline{q})= r \overline{q}^2$ for $k=3$. The green curve is a line with slope $1$. The values of $\overline{q}$ for which the straight line lies within the colored bands give the latitudes at which exponentially many stationary points are found. (a-c) Plots for $k=2$ and $p=3$, and fixed values of $r$. The intersection points between the green line and the blue curves define the boundaries $\overline{q}_m(r), \overline{q}_M(r)$ of the bands in Fig.~\eqref{fig:Bandek2}. (d) Plot for $k=3=p$ and $r=r_c$. The straight line lies within the colored bands for all $ \overline{q} \leq \overline{q}_c \approx 0.71$. }\label{fig:BandeMappatek2p3}
\end{figure*}

\begin{figure*}[!htbp]
\subfloat[][]{
\includegraphics[width=.5\linewidth]{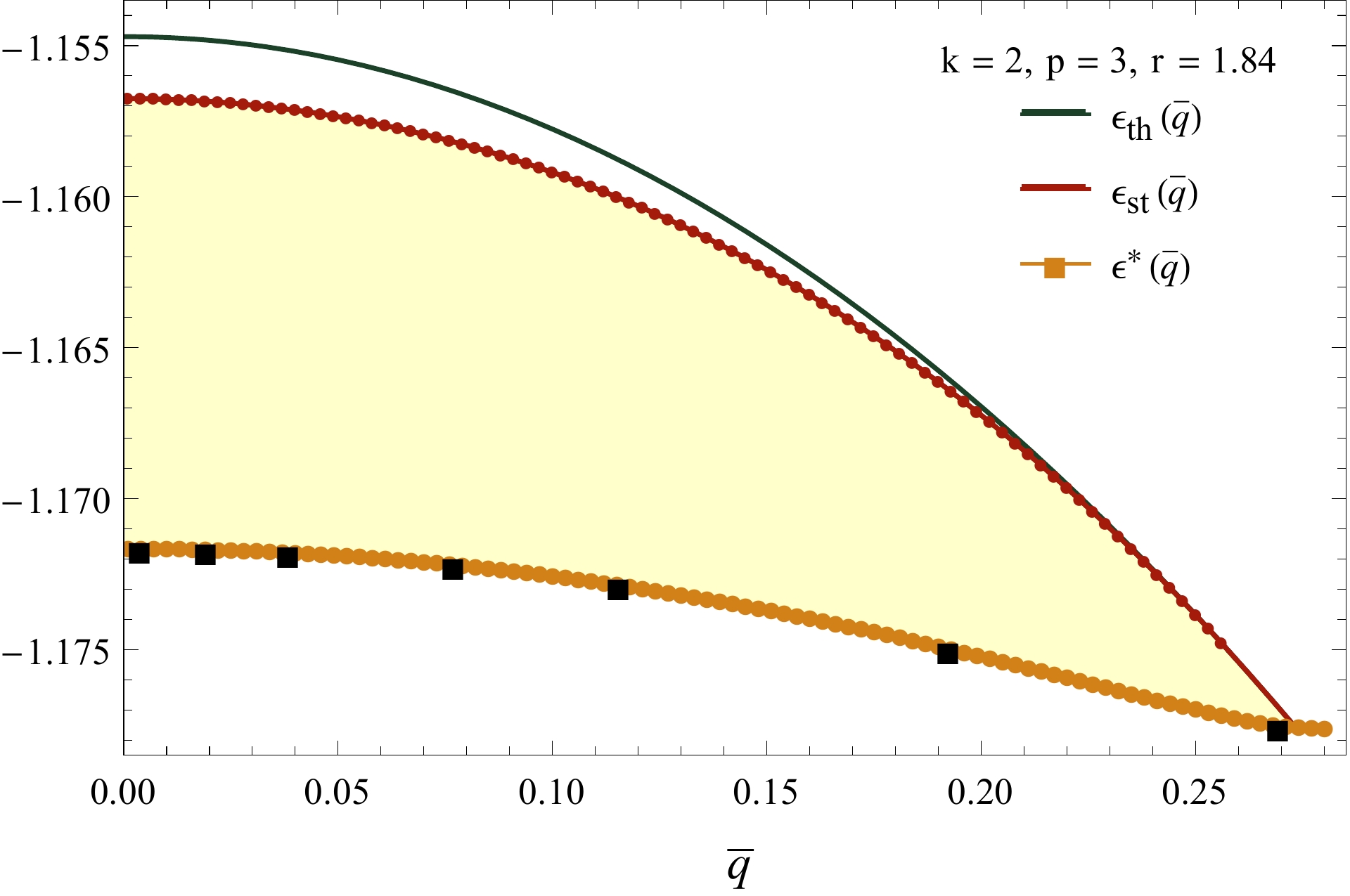}}
\subfloat[][]{\includegraphics[width=.5\linewidth]{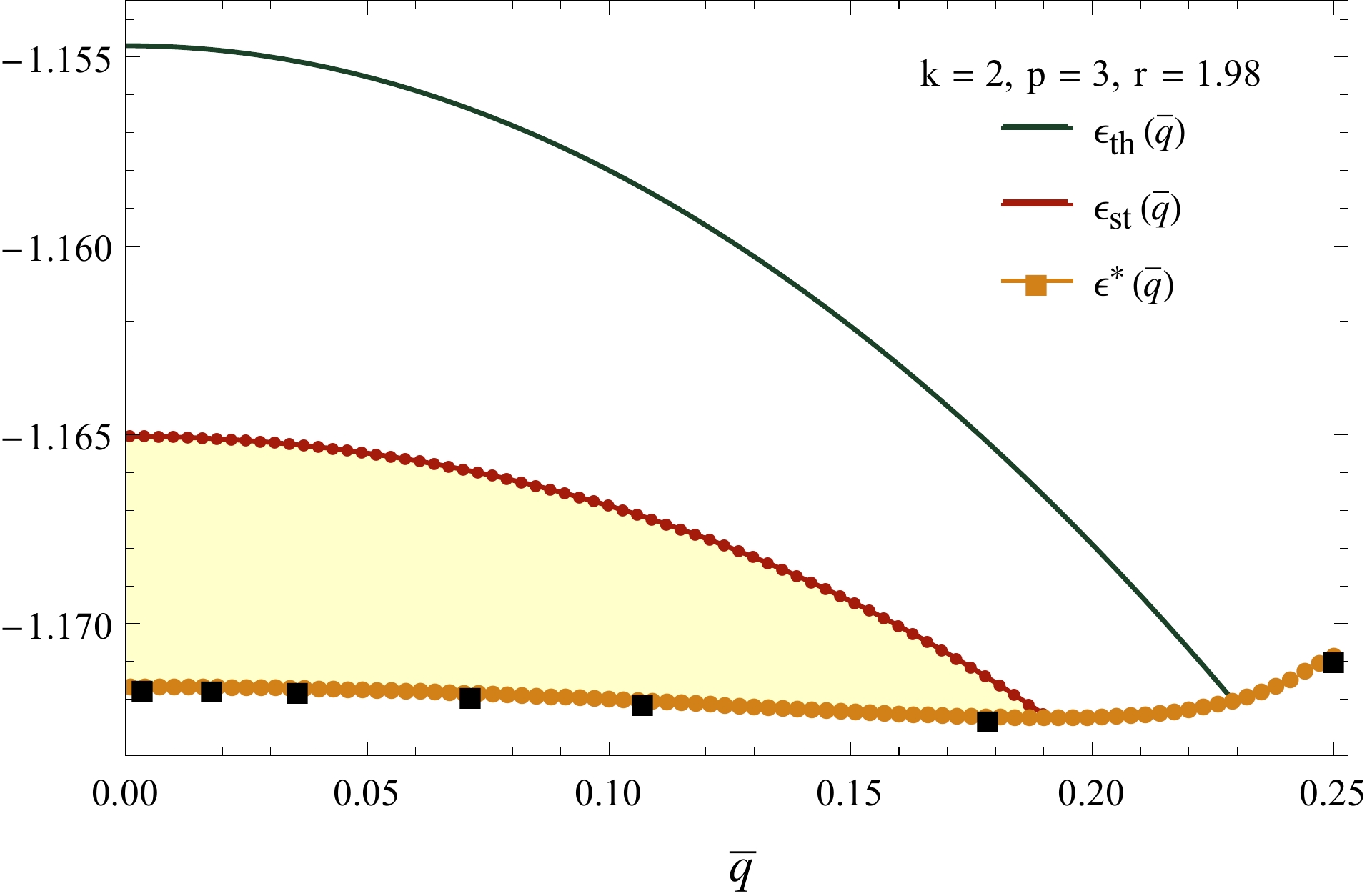}}
\caption{Comparison between the energies $\epsilon_{\text{st}}(\overline{q}),\epsilon_{\text{th}}(\overline{q})$ and $\epsilon^*(\overline{q})$ for $k=2$ and two values of $r>r_c$. The orange points are obtained from the direct solution of the saddle point equation for $k=2$, the black squares exploiting the mapping from $k=1$. The intersection points between $\epsilon_{\text{th}}$ and $\epsilon^*$ correspond to the dashed lines in Fig. \ref{fig:Bandek2}, the ones between $\epsilon_{\text{st}}$ and $\epsilon^*$ to the solid lines. The yellow strip identifies the energies of the \emph{stable} stationary points. (a) For $r= 1.84$ and $\epsilon< \epsilon_{\text{th}}$, the isolated eigenvalue renders unstable the higher energy points at $\overline{q}< 0.26$, while it does not exist for the larger latitudes $0.26 < \overline{q} < 0.27 \approx \overline{q}_M$. (b) For $r =1.98$ and $\epsilon< \epsilon_{\text{th}}$, the stationary points in the interval $\overline{q}_M \approx 0.19 < \overline{q} < 0.23$ are all unstable irrespective of their energies, while for the smaller latitudes only the points at higher energy are unstable. For these values of $r$, the absolute minima of the energy landscape (not in the figure) are at $\overline{q}^*\approx 0.746$ for $r=1.84$ and  $\overline{q}^*\approx 0.786$ for $r=1.98$, and have energy density $\epsilon^*\approx -1.327$ and $\epsilon^*\approx -1.369$, respectively.
} \label{fig:Isolatedk2p3}   
\end{figure*}

\begin{figure*}[!htbp]
\subfloat[][]{
\includegraphics[width=.5\linewidth]{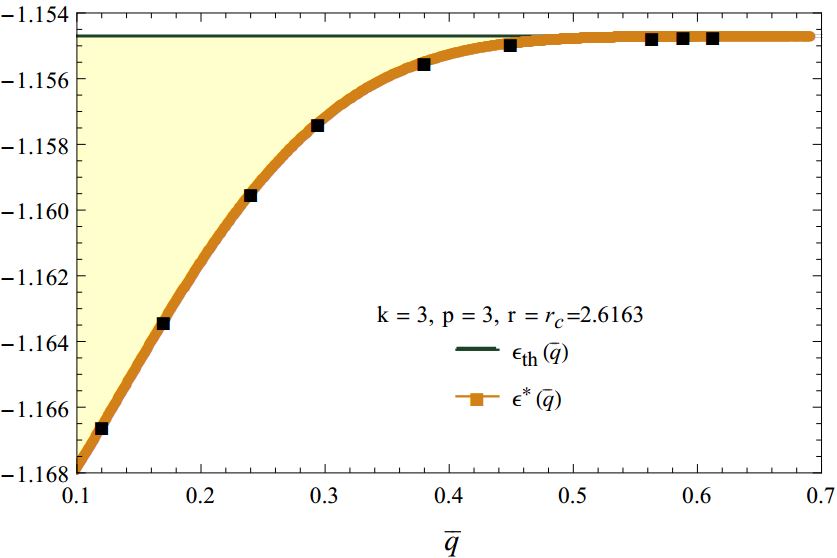}}
\subfloat[][]{\includegraphics[width=.5\linewidth]{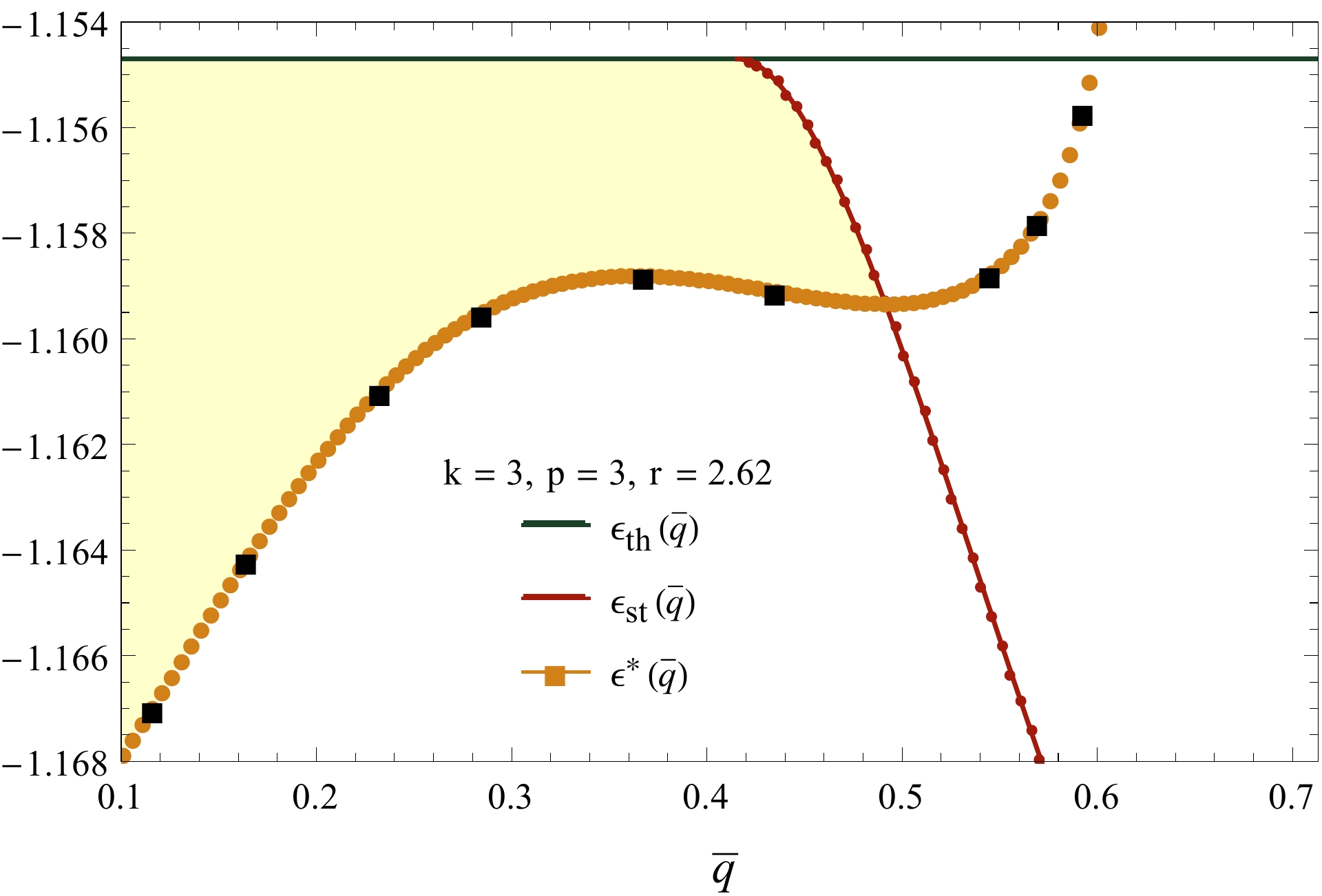}}
\caption{Comparison between the energies $\epsilon_{\text{st}}(\overline{q}),\epsilon_{\text{th}}(\overline{q})$ and $\epsilon^*(\overline{q})$ for $k=3=p$. The orange points are obtained from the solution of the saddle point equations for $k=3$, the black squares are obtained exploiting the mapping from $k=1$. The intersection points between $\epsilon_{\text{th}}$ and $\epsilon^*$ correspond to the dashed lines in Fig. \ref{fig:Bandek3}, while the intersection between $\epsilon_{\text{st}}$ and $\epsilon^*$ gives $\overline{q}_M$. The yellow strip identifies the energies of the \emph{stable} stationary points. (a) Exactly at $r =r_c$, all states at the latitudes $\overline{q}_m<\overline{q} < \overline{q}_c=0.707$ are stable. The isolated eigenvalue exists only for the states $q=\overline{q}_c$, which are at their threshold energy (the eigenvalue is attached to the lower edge of the semicircle, that touches zero for the points at this latitude). 
(b) For $r \approx 2.62 >r_c$, some of the stationary points in the interval $ 0.415 \lesssim \overline{q} <\overline{q}_M \approx 0.49 $ are unstable because of the eigenvalue (the ones at higher energy). The stationary points in the interval $\overline{q}_M  < \overline{q} <0.6$ are all unstable irrespective of their energies. For this value of $r$, the deepest stable minimum (not in the figure) is the isolated one at $\overline{q}^*\approx 0.88$.
} \label{fig:Isolatedk3p3}   
\end{figure*}

\noindent 
This Appendix contains some additional results related to the content of Sec.~\ref{sec:Results}: we discuss how the mapping \eqref{eq:ExplicitMapping} is exploited to derive the bands in Figs.~\ref{fig:Bandek2} and \ref{fig:Bandek3}, comment on some impliecations on the thermodynamical transitions, and provide some details on the isolated eigenvalue of the Hessian of the stationary points.\\
For $k=1$, the band containing the stable stationary points is delimited by the curves $\overline{q}_m(r)$ and $\overline{q}_M(r)$ plotted in Fig.~\eqref{fig:Bandek1}. To determine the analogous curves for $k=2$ (and fixed $r$), it is sufficient to consider the functions $\overline{q} \to \overline{q}_m(r_1^{\text{eff}})$ and $\overline{q} \to  \overline{q}_M(r_1^{\text{eff}})$, with $r_1^{\text{eff}} (r, \overline{q})= r \overline{q}$. Indeed, the latitudes $\overline{q}$ satisfying $\overline{q}_m(r_1^{\text{eff}}) \leq \overline{q} \leq  \overline{q}_M(r_1^{\text{eff}})$ are such that the complexity $\Sigma_{p,2}(\epsilon, \overline{q})$ is positive over a finite energy interval. In Fig.~\ref{fig:BandeMappatek2p3}, we give an example of this mapping for different values of $r$: for the smaller $r$, there is a connected strip containing exponentially many stationary points, which encloses the equator. For the intermediate $r$, the strip is instead separated into a larger band enclosing to the equator, and a thinner one at larger overlap $\overline{q}$ (the thinner band at larger overlap has its counterpart at negative overlap). The landscape phase transition in which the band splits into disconnected components occurs between these values of $r$. The largest $r$ in Fig.~\ref{fig:BandeMappatek2p3} corresponds to $r>r_c$; in this case, the band of states enclosing the equator has shrunk but it is still finite, while the strip closer to the \emph{North Pole} has collapsed to a single state. The bands at $k=3$ can be obtained with an analogous procedure, using $r_1^{\text{eff}}= r \overline{q}^2$. In the same figure we show the case $p=3$ and $r=r_c=\sqrt{6}$, to illustrate that at the critical point there is a unique connected band containing the equator, with maximal latitude $\overline{q}_M(r_c)= \overline{q}_c$. 
\noindent We now re-examine the thermodynamic properties of the system at $k>1$, and clarify how they can be deduced from the case $k=1$. Consider the $k=1$ curves $\epsilon^*_1(\overline{q}, r_1)$ satisfying $\Sigma_{p,1}(\epsilon_1^*(\overline{q},r_1), \overline{q}; r_1)=0$. For $k>1$ and fixed $r$, it follows from \eqref{eq:ExplicitMapping} that $
 \Sigma_{p,k}(\epsilon_k^*(\overline{q},r),\overline{q}; r)=0$ with $\epsilon^*_k(\overline{q},r)= \epsilon_1^*(\overline{q}, r \overline{q}^{k-1})+ r (1-1/k) \overline{q}^k$. This allows one to reconstruct the full spectrum on minima $\epsilon^*_k(\overline{q},r)$ from its $k=1$ counterpart; the minimization of this function over $\overline{q}$ gives the thermodynamic energy $\epsilon_k^*(r)$, and the corresponding latitudes $\overline{q}_k^*(r)$ plotted as yellow squares in Figs.~\ref{fig:Bandek2} and \ref{fig:Bandek3}. It happens that the latitudes $\overline{q}_k^*(r)$, whenever they are not equal to zero, coincide with the image under the mapping \eqref{eq:ExplicitMapping} of the corresponding ones $\overline{q}^*_1$ at $k=1$, i.e., $\overline{q}_k^*(r)= \overline{q}_1^*(\tilde{r}_1)$, where $\tilde{r}_1=\tilde{r}_1(r)$ solves $\tilde{r}_1 - r \quadre{\overline{q}_1^*(\tilde{r}_1)}^{k-1}=0$. The fact that $\overline{q}^*(\tilde{r}_1)$ are stationary points of $\epsilon^*_k(\overline{q}, r)$ is easily checked, as the derivative 
  \begin{equation*}
  \frac{\partial \epsilon^*_k(\overline{q},r)}{\partial \overline{q}}
  = \tonde{\frac{\partial \epsilon_1^*}{\partial \overline{q}} + \frac{\partial \epsilon_1^*}{\partial r} r \overline{q}^{k-2} + r \overline{q}^{k-1}}
 \end{equation*}
 equals to zero at $\overline{q}=\overline{q}^*(\tilde{r}_1)$, since
 \begin{equation*}
  \frac{\partial \epsilon_1^*(\overline{q}_1^*(r_1), r_1)}{\partial \overline{q}} =0, \quad
  \frac{\partial \epsilon_1^*}{\partial r}(\overline{q}^*(\tilde{r}_1), \tilde{r}_1)=- \overline{q}^*(\tilde{r}_1).
  \end{equation*}
  The second equality follows from the fact that $\epsilon^*_k(\overline{q},r)= E(\overline{q})-r \overline{q}^k/k$ with $E(\overline{q})$ the energy of the deepest states of the $p$-spin Hamiltonian at fixed overlap $\overline{q}$ with the \emph{North Pole}, see Sec. \ref{sec:ReplicaAnalysis}. We now characterize both $r_{\text{2ND}}$ and the spinodal point $r_{1\text{SP}}$ in terms of the function $\overline{q}^*_1(r_1)$ or, more precisely, of its inverse $r_1^*(\overline{q})$. Notice that, since $\overline{q}^*_1(r_1)$ is defined by the condition $\partial \epsilon^*_1(\overline{q}^*, r_1)/\partial \overline{q}=0$, it holds $r_1^*(\overline{q})= \partial E(\overline{q})/\partial \overline{q}$. \\  
  For general $k$, we define the function $r_k(\overline{q}) \equiv  r_1^*(\overline{q})/ \overline{q}^{k-1}$, which associates to each $\overline{q}$ the value of $r_k$ for which  $\overline{q}^*_k(r_k)= \overline{q}$, i.e., for which $\overline{q}$ is the latitude of the deepest minimum. For $k=2$, $r_2(\overline{q})$ is monotone increasing, and takes a finite minimum value at $\overline{q}=0$, which is precisely $r_{\text{2ND}}$\cite{footnote10}. For  smaller $r$, $\overline{q}_2^*(r)$ is frozen to $0$, and the corresponding energy $\epsilon^*(r)$ is frozen to the ground state energy of the $p$-spin model with $r=0$. \\  
  For $k=3$ and larger, the function $r_k(\overline{q})$ is non-monotone, and two latitudes are associated to each fixed, large enough $r$: the larges of these latitudes is the one of the local minimum $\overline{q}^*_2(r)>0$ of $\epsilon^*_3(\overline{q},r)$, while the smaller is the one of the local maximum. The function $r_k(\overline{q})$  has minimum at a point $\overline{q}_{\text{SP}}$, defined by
  \begin{equation}\label{eq:rSpinodal}
  \frac{d r_k(\overline{q})}{d \overline{q}}=0 \longrightarrow \overline{q} \frac{d  r^*_{1}(\overline{q})}{d \overline{q}}-(k-1) r^*_{1}(\overline{q})=0.
 \end{equation}
 At this point, the local maximum and minimum merge, and thus $r_k(\overline{q}_{\text{SP}})=r_{1\text{SP}}$. For general $k$, it holds $r_{1\text{SP}}=r_c$ whenever $p=k$, (see for instance Fig.~\ref{fig:Bandek3}~(b)). This can be seen in the following way: for $\overline{q} \geq \overline{q}_c$, the function $\epsilon^*_1(\overline{q},r)$ is obtained from the annealed complexity, or, equivalently, from the solution of the RS equation in Sec.~\ref{sec:ReplicaAnalysis}. This gives $
   \epsilon^*_1(\overline{q}, r)= -\sqrt{{p (1-\overline{q}^2)}/{2}}- r \overline{q}$,
and minimizing and solving for $r$ we get $
   r^*_{1}(\overline{q})= \sqrt{p/[2 (1-\overline{q}^2)]}\overline{q} \text{  for  } \overline{q}> \overline{q}_c.$
 The solution of Eq. \eqref{eq:rSpinodal} then reads
 \begin{equation}
  \overline{q}_{\text{SP}}= \sqrt{\frac{k-2}{k-1}},
 \end{equation}
 which is consistent (i.e., larger than $\overline{q}_c$) for $k >p$. In this case, 
 \begin{equation}
 r_{1\text{SP}}= \sqrt{\frac{p(k-2)}{2}}f'_k \tonde{\frac{k-2}{k-1}}^{-1} \text{  for  } k \geq p.
 \end{equation}
 At $k=p$, one recovers $\overline{q}_{\text{SP}}=\overline{q}_c$ and $r_{1\text{SP}}=r_c$, see Eqs.~\ref{eq:rCritico} and  \eqref{eq:qc}. For $k<p$, $r^*_{1}(\overline{q})$ has to be computed using the solutions to the RSB equations.\\
We conclude this Appendix with some details on how the bands of minima are modified when accounting for the instability due to the isolated eigenvalue of the Hessians, focusing on the cases $k=2$ and $p=3$, and $k=3=p$.\\
For $k=2$,we find that for $r \gtrsim r_c$ the first stationary points that are affected by the eigenvalue are the ones at smaller overlap $\overline{q}$: among them, the isolated eigenvalue renders unstable the ones at higher energy, see Fig.~\ref{fig:Isolatedk2p3}~(a). Therefore, its effect is to diminish from above the width of the energy interval in which stable points are found. As $r$ increases, the instability propagates to the largest latitudes $\overline{q}$, until eventually for these larger latitudes the energy $\epsilon_{\text{st}}(\overline{q}, r)$ becomes smaller that $\epsilon^*(\overline{q},r)$, see Fig.~\ref{fig:Isolatedk2p3}~(b); at these intermediate values of $r$, there are still stable stationary points at small overlap $\overline{q}$ and energy strictly smaller than the \emph{threshold}, while the ones at larger overlap are all unstable, irrespective of their energy. The band is thus narrowed. The last points that become unstable are the ones at the equator; the instability of these points can be computed within the annealed approximation, as illustrated in Appendix \ref{app:ThirdOrderEquations}. Note that in both the cases considered in Fig.~\ref{fig:Isolatedk2p3}, since $r>r_c$, the deepest minimum in the landscape is not in the band , but it is rather the minimum of the annealed complexity, which is stable as its energy density is below the \emph{threshold}.\\
\noindent For $k=3=p$, the isolated eigenvalue appears at $r=r_c$, see  Fig.~\ref{fig:Isolatedk3p3}~(a): at the critical point $\overline{q}_c$, all the energies $\epsilon_{\text{st}}, \epsilon_{\text{th}}, \epsilon^*$ coincide, and coincide with the energy $\epsilon_c$: exactly at this latitude, the annealed complexity is equal to zero at $\epsilon_c$, and negative otherwise. The corresponding stationary point is marginally stable, with the minimal eigenvalue of the Hessian being right at the lower boundary of the support of the semicircle, which is at zero.
For larger $r$, the band close to the equator becomes affected by the instability due to the eigenvalue, see Fig.~\ref{fig:Isolatedk3p3}~(b). In particular, within the numerical accuracy we find that $\epsilon_{\text{st}}(\overline{q})$ intercepts $\epsilon^*(\overline{q})$ at a latitude that corresponds to the local minimum of $\epsilon^*(\overline{q})$, meaning that exactly at the local minimum the isolated eigenvalue is equal to zero. The isolated minimum of the annealed complexity at higher overlap is instead always stable. 

\subsection{Stability of metastastable minima found with replicas}\label{app:StabilityMonasson}
\noindent
The stability of the metastable minima (counted by \eqref{Sc}) with respect to fluctuations in the structure of the overlap matrix $Q_{ab}$ is probed by the replicon eigenvalue of the matrix
$M_{\alpha \beta; \gamma \delta}=\partial^2 (n S[Q_{\alpha, \beta}])/ \partial Q_{\alpha \beta} \partial Q_{\gamma \delta}$, evaluated at the saddle point. This can be determined from the $m (m-1) \times m(m-1)$ block $M_{a b; c d}$ of $M_{\alpha \beta; \gamma \delta}$, which corresponds to  indices $a,b,c,d$ of replicas belonging to the same group with mutual overlap $q_{ab}=q_1$. The latter is given by
\begin{equation}
\begin{split}
 &M_{a b; c d}= -\frac{\beta^2}{2} p (p-1) q_{ab}^{p-2} \tonde{\delta_{ac} \delta_{bd}+ \delta_{ad} \delta_{bc}}\\
 &+{Q^{-1}_{ac} Q^{-1}_{bd}+Q^{-1}_{ad}Q^{-1}_{bc}},
 \end{split}
\end{equation}
where $Q^{-1}$ is the inverse of the overlap matrix. When evaluated at the saddle point and for $n \to 0$, is has the structure \cite{crisantisommers}
\begin{equation}
\begin{split}
&M_{ab;cd}=M_1 \frac{\delta_{ac} \delta_{bd}+ \delta_{ad} \delta_{bc}}{2}+M_2 \frac{\delta_{ac}+ \delta_{bd}+ \delta_{ad}+ \delta_{bc}}{4}\\
&+ M_3,
\end{split}
\end{equation}
with 
\begin{equation}
 \begin{split}
 M_1&=-{\beta^2} p(p-1) q_1^{p-2}+  2\tonde{\frac{1}{1- q_1}}^2,\\
 M_2&=\frac{4}{(1-q_1)^2} \frac{m (q_0-q_1)^2 + (1-q_1)(q_1-\overline{q}^2)}{(1-q_1+m (q_1-q_0))^2},\\
 M_3&=\frac{2}{(1-q_1)^2}\quadre{\frac{m (q_0-q_1)^2 + (1-q_1)(q_1-\overline{q}^2)}{(1-q_1+m (q_1-q_0))^2}}^2.
 \end{split}
\end{equation}
The replicon eigenvalue is given by $M_1$, and vanishes whenever
\begin{equation}\label{eq:RepliconZero}
\frac{1}{\beta(1-q_1)}=  \sqrt{\frac{ p(p-1)}{2}}. 
\end{equation}
The stability condition obtained in the annealed approximation (see Eq. \ref{eq:ResZero} and below) can be recovered within the replica setting, in the RS framework. It indeed corresponds to the vanishing of the longitudinal eigenvalue, and is obtained setting to zero the eigenvalues of the matrix of second derivatives (with respect to the order parameters $\overline{q}$ and $\beta(1-q_1)$) of the  replica-symmetric limit of the action, which is given by:
$$
\frac{S_{\rm RS}}{\beta}=\frac{p}{4}\beta(1-q)+r f_k(\overline{q})+\frac{1}{2}\frac{1-\overline{q}^2}{\beta(1-q)}.
$$  
The matrix of second derivatives reads 
\begin{equation}
\begin{pmatrix}
  r f''_k(\overline{q})- \frac{1}{\beta(1-q_1)}& \frac{\overline{q}}{\quadre{\beta(1-q_1)}^2}\\
  \frac{\overline{q}}{\quadre{\beta(1-q_1)}^2}& \frac{1- \overline{q}^2}{\quadre{\beta(1-q_1)}^3}
 \end{pmatrix},
\end{equation}
and has two eigenvalues that both vanish  whenever 
\begin{equation}\label{eq:ConditionIsolatedEv}
 r f''_k(\overline{q})(1-\overline{q}^2)=\frac{1}{\beta(1-q_1)}.  
 \end{equation}
 This criterion can be re-written in terms of the resolvent $G(z)$ associated to the Hessian of the $p$-spin Hamiltonian in absence of the signal (at $r=0$), since  $\beta(1-q_1)$ is the spin susceptibility of the $p$-spin model, which is related to the inverse of the Hessian matrix. More precisely, $\beta(1-q_1)= - G(0)$, so that the condition in Eq. \eqref{eq:ConditionIsolatedEv} is equivalent to $
  r f''_k(\overline{q})(1-\overline{q}^2)= - 1/G(0)$. This is precisely the condition of vanishing eigenvalue obtained in the annealed Kac-Rice calculation, as it equals to $z(\mu)=0$ where $z(\mu)= G^{-1}(1/\mu)$ and $\mu=- r f''_k(\overline{q})(1-\overline{q}^2)$. As remarked in Appendix~\ref{app:ThirdOrderEquations}, this condition is exact at the equator $\overline{q}=0$; in particular, it allows to obtain the value of $r$ where the equator band disappears in the case $k=2$.


\begin{thebibliography}{99}
\bibitem{footnote1}
We know now that more general and complex cases are possible but nevertheless FRSB and 1RSB still stand as the two paradigms to describe 
mean-field glassy systems.
\bibitem{footnote5} In absence of the signal ($r=0$), the expression \eqref{eq:FinalGradientsDistribution} becomes independent of $\overline{q}$, and it is maximal at $q=0$. When optimizing Eq. \eqref{eq:Full3} over $\overline{q}$ as well, one finds that the complexity is dominated by $\overline{q}=q=0$. This is consisent with the fact that for the $p$-spin Hamiltonian, the quenched calculation for the complexity reproduces the annealed result, that is indeed recovered setting $\overline{q}=q=0$.
\bibitem{footnote6}
Note nevertheless that they are correlated in blocks: the blocks $\tilde{M}_{\gamma}^a$ correlate only with the correspondent blocks $\tilde{M}_{\gamma}^b$, for $a \neq b$ and $\gamma=0, 1/2, 1$.
\bibitem{footnote8}
Note however that no polynomial algorithms allowing to initialize the system at finite overlap with the signal is yet known \cite{reviewkrzakalazdeborova, krzakala, montanari}.
\bibitem{footnote10}
 Indeed, using  $r_1^*(\overline{q})= \partial E(\overline{q})/\partial \overline{q}$ and $E(\overline{q})= E_{GS}+ \frac{C_p}{2}\overline{q}^2+ O(\overline{q}^3)$, we get $r_2(0)= C_p$, in agreement with Sec.~\ref{sec:ReplicaAnalysis}.
\end{thebibliography}
\end{document}